\documentclass[a4paper,11pt]{article}
\pdfoutput=1 
\usepackage{jheppub}
\usepackage[T1]{fontenc} 
\usepackage[all]{xy}
\usepackage{rotating}
\usepackage{float}
\usepackage{tikz}
\usetikzlibrary{calc,decorations.markings}

\let\a=\alpha \let\b=\beta   
 \let\h=\eta   
  \let\n=\nu  \let\p=\pi

\def\diag{\mathop{\rm diag}\nolimits}

\def\a{\alpha}
\def\b{\beta}

\def\CF{{\cal F}}

\def\CH{{\cal H}}
\def\CI{{\cal I}}

\def\CM{{\cal M}}
\def\CN{{\cal N}}
\def\CO{{\cal O}}
\def\CP{{\cal P}}

\def\CS{{\cal S}}
\def\CT{{\cal T}}

\def\CW{{\cal W}}

\def\CZ{{\cal Z}}

\def\U{U}
\def\SU{SU}

\def\beq#1\eeq{\begin{align}#1\end{align}}

\makeatletter
\newcommand*{\rom}[1]{\expandafter\romannumeral #1}
\makeatother

\title{\boldmath Non-unitary  TQFTs from  3D $\mathcal{N}=4$   rank 0 SCFTs }

\abstract{We propose a novel procedure of assigning a pair of non-unitary topological quantum field theories (TQFTs), TFT$_\pm [\CT_{\rm rank \;0}]$, to a (2+1)D interacting $\CN=4$ superconformal field theory (SCFT) $\CT_{\rm rank \;0}$ of rank 0, i.e.\ having no Coulomb and Higgs branches. The topological theories arise from particular degenerate  limits of the SCFT.  Modular data of the non-unitary TQFTs are extracted from the  supersymmetric partition functions   in the degenerate limits. As a non-trivial dictionary, we propose that $F = \max_\alpha \left(- \log |S^{(+)}_{0\alpha}| \right) =  \max_\alpha \left(- \log |S^{(-)}_{0\alpha}|\right)$, where $F$ is the round three-sphere free energy of $\CT_{\rm rank \;0 }$ and $S^{(\pm)}_{0\alpha}$ is the first column in  the modular S-matrix of  TFT$_\pm$.  From the dictionary, we derive the lower bound on $F$, $F \geq -\log \left(\sqrt{\frac{5-\sqrt{5}}{10}} \right) \simeq 0.642965$, which holds for any  rank 0 SCFT. The bound is saturated by the minimal $\CN=4$ SCFT proposed by Gang-Yamazaki, whose associated topological theories are both the Lee-Yang TQFT. We explicitly work out the (rank 0 SCFT)/(non-unitary TQFTs) correspondence for infinitely many examples.  }

\author[a,b,c]{Dongmin Gang,}
\author[c]{Sungjoon Kim,}
\author[d]{Kimyeong Lee,}
\author[e]{Myungbo Shim,}
\author[f]{Masahito Yamazaki}

\affiliation[a]{
Department of Physics and Astronomy $\&$ Center for Theoretical Physics,
\\
Seoul National University, 1 Gwanak-ro, Seoul 08826, Korea}
\affiliation[b]{Asia Pacific Center for Theoretical Physics (APCTP),
	Pohang 37673, Korea}
\affiliation[c]{Department of Physics, Pohang University of Science and Technology (POSTECH), \\Pohang 37673, Republic of Korea}

\affiliation[d]{Korea Institute for Advanced Study, 85 Hoegiro, Dongdaemun-Gu, Seoul 02455, Korea}
\affiliation[e]{Department of Physics and Research Institute of Basic Science, \\ Kyung Hee University, Seoul 02447, Korea}
\affiliation[f]{Kavli Institute for the Physics and Mathematics of the Universe (WPI),\\
	University of Tokyo, Kashiwa, Chiba 277-8583, Japan}

\emailAdd{arima275@snu.ac.kr}
\emailAdd{sjkim0305@postech.ac.kr}
\emailAdd{klee@kias.re.kr}
\emailAdd{mbshim1213@khu.ac.kr}
\emailAdd{masahito.yamazaki@ipmu.jp}

\begin{document} 
\maketitle
\flushbottom


\section{Introduction}

Supersymmetric quantum field theories with 8 supercharges (8 $Q$s) provide a fertile ground for many interesting research topics connecting various areas in theoretical and mathematical physics. For example, 
Seiberg-Witten's approach \cite{Seiberg:1994rs} to 4-dimensional (4D) $\CN=2$ supersymmetric  gauge theories provides analytic understanding of  non-perturbative phenomena, such as confinement,  in strongly coupled  gauge theories.  4D $\CN=2$ superconformal field theories can be geometrically constructed using wrapped M5-branes in M-theory, and the 4D-2D correspondence connects  physics of 4D  supersymmetric field theories with  mathematical structures on 2D Riemann surfaces in an unexpected way \cite{Gaiotto:2009we,Alday:2009aq,Gaiotto:2009hg}. Interestingly, there exist non-trivial superconformal field theories with 8 supercharges in higher dimensional space-time, 5D and 6D,  as predicted by String/M-theory \cite{Witten:1995zh,Seiberg:1996bd,Intriligator:1997pq}. More recently, it is found that 2D chiral algebras (resp.\ 1D topological quantum mechanics)  appear as  protected subsectors of 4D $\CN=2$ (resp.\ 3D $\CN=4$) superconformal field theories \cite{Beem:2013sza,Chester:2014fya,Chester:2014mea, Beem:2016cbd,Dedushenko:2016jxl,Dedushenko:2017avn,Panerai:2020boq}. 

Extended SUSY gauge theories have rich structures in their vacuum moduli space, and one natural invariant is the rank, i.e.\ the complex dimension of the Coulomb branch.  
There have been numerous efforts in classifying  SCFTs with 8 $Q$s for a given low rank in various space-time dimensions \cite{Chang:2019dzt,Argyres:2020nrr,Heckman:2013pva,Heckman:2015bfa,Bhardwaj:2015xxa,Jefferson:2017ahm,Jefferson:2018irk,Bhardwaj:2018yhy,Bhardwaj:2018vuu,Apruzzi:2019opn, Bhardwaj:2020gyu}. 
For  3D $\CN=4$ SCFTs, however, the rank is in general not a
duality-invariant concept since 
the Coulomb and Higgs branches are exchanged under the 3D mirror symmetry \cite{Intriligator:1996ex}.  
For this reason, we hereby modify the definition of rank as the maximum of the dimension of Coulomb branch and  
that of the Higgs branch. 
Another peculiar fact about 3D $\CN=4$ theories  is that there exist non-trivial interacting SCFTs of rank 0, as studied 
by two of the authors of this present paper in \cite{Gang:2018huc}. This is in contrast with the case of $D\geq 4$, where it is often implicitly assumed that there is no non-trivial interacting rank 0 SCFTs with 8 $Q$s, so that the classification program starts with rank 1. (Recently, 4D/5D rank 0  SCFTs  were found through a geometrical engineering but it is yet unclear if they are interacting SCFTs \cite{Closset:2020scj}.)  
Note that most of the classification schemes in previous studies do not work for rank 0 cases since the existence of Coulomb or Higgs branch operators is an crucial assumption in the analysis. 

In this paper, we initiate the classification of rank 0 3D $\CN=4$ SCFTs by establishing the following correspondence:
\begin{align}
\begin{split}
&\textrm{3D  $\CN=4$ superconformal field theories of rank 0 } 
\\
&\longleftrightarrow \textrm{A pair of 3D non-unitary topological quantum field theories (TQFTs)}\;.
\end{split}
\end{align}
The non-unitary TQFTs emerge at particular choices of non-superconformal R-symmetry, $\nu=\pm 1$ in \eqref{R-symmetry mixing}, of rank 0 $\CN=4$ SCFTs. In  the  limits, due to huge Bose/Fermi cancellations the unrefined superconformal index gets contributions only from Coulomb-branch  or Higgs-branch operators and their descendants.  For rank 0 theories, the index  becomes trivial (i.e.\ $1$) since there are no non-trivial Coulomb/Higgs branch operators.    Other partition functions on various rigid supersymmetric Euclidean backgrounds also drastically simplify in the degenerate limits for rank 0 theories. Our correspondence says that the simplified partition functions  are actually equal to the partition functions of corresponding  non-unitary TQFTs on the same 3D spacetime. (See \eqref{rank 0 to TFT} for a precise statement.)  Concrete dictionaries of the correspondence  are given  in Table \ref{Table : Dictionaries}.  In the degenerate limits, as seen in the superconformal index case,  contributions from local operators become unimportant for rank 0 theories and only non-local loop operators become relevant physical observables. Similarly,  loop operators are  the only physical observables in general TQFTs.
Using the correspondence, one can map the problem of  classifying  rank 0 SCFTs  to the   classification of non-unitary TQFTs, which is much easier to handle. 
Mathematically, TQFTs are described by modular tensor categories (MTCs) and classification of MTCs has been studied intensively in the literature  \cite{turaev1992modular,rowell2006quantum, rowell2009classification,wen2016theory, bruillard2016rank,Cho:2020ljj}.  The most basic quantity characterizing a 3D CFT is the round three-sphere free energy, usually denoted as $F$.  The  $F$ always monotonically decreases under the renormalization group (RG) flow and thus is regarded as a proper measure of the degrees of freedom of  3D CFT \cite{Jafferis:2011zi,Liu:2012eea,Casini:2012ei}. In one of the most interesting and surprising dictionaries of the correspondence,  the $F$ of a rank 0 CFT is related  to the modular S-matrix of  a non-unitary TQFT in a very simple way as given in Table \ref{table:non-unitary/SCFT}.     Combining the dictionary and universal algebraic properties of the S-matrix, we derive following lower bound on $F$
\begin{align}
F \geq -\log \left( \sqrt{\frac{5-\sqrt{5}}{10}}\right),
\end{align}
which should hold for any rank 0 SCFTs. Interestingly, the lower bound is saturated by the minimal theory studied in \cite{Gang:2018huc}.

The correspondence is  similar in spirit with the  (4D $\CN=2$ SCFT)/(2D chiral algebra) correspondence mentioned above.  In both correspondences, non-unitary algebraic structures, chiral algebras on the one hand and modular tensor category on the other, appear as protected subsectors of unitary superconformal field theories. But there are several crucial differences. First, two theories  in our correspondence are defined on  the same  3D space-time while the 2D chiral algebra lives in the 2D subspace of 4D space-time of the SCFT. 
 Secondly, basic physical objects are BPS local operators in the (4D SCFTs)/(2D chiral algebra) story while BPS non-local loop operators are basic objects in our correspondence. 
That non-local loop operators play crucial roles can be a great advantage of our classification approach over the conventional conformal bootstrap approaches, since the latter are based on correlation functions of local operators.   
We note that the 3D non-unitary TQFTs are sensitive to the global structure of the 3D rank 0 SCFTs and two theories in the correspondence share the same one-form symmetry as well as their 't Hooft anomalies. 
 In 3D, the quantity $F$ (unlike the stress-energy tensor central charge $C_T$) is sensitive to the global structure of CFT and the conformal bootstrap approach never give a constraint on $F$ but only on $C_T$, which is not a proper measure of the degrees of freedom in a strict sense \cite{Nishioka:2013gza}. 
 
The remaining part of paper is organized as follows. In  section \ref{sec : Rank 0/(non-unitary) correspondence}, we present  the precise statement of the correspondence with several concrete dictionaries. As an application of the correspondence, we derive  interesting lower bounds on $F$ for rank 0 SCFTs. In section \ref{sec : examples}, we explicitly work out the correspondence in detail with several classes of infinitely many rank 0 SCFTs.  The results are summarized in Table \ref{table:non-unitary/SCFT}. In Appendix \ref{App : some reviews}, we give brief reviews on supersymmetric partition functions of 3D $\CN\geq 3$ gauge theories and modular data of 3D TQFTs which are basic ingredients of the dictionaries. In other Appendices,  we collect technical details and supplementary materials. 

\section{(Rank 0  SCFT)/(Non-unitary  TQFTs) correspondence } \label{sec : Rank 0/(non-unitary) correspondence}

In this section, we establish a correspondence between
\begin{align}
\begin{split}
&\CT_{\rm rank \;0} \; :\; \textrm{a 3D $\mathcal{N}=4$ interacting SCFT with empty Coulomb and Higgs branches}
\\
& \xrightarrow{ \qquad }  {\rm TFT}_{\pm } [\CT_{\rm rank\; 0}]\;:\; \textrm{a pair of 3D non-unitary TQFTs}\;.
\end{split}
\end{align}
The basic dictionaries for the correspondence are summarized in Table \ref{Table : Dictionaries}. 

\subsection{\texorpdfstring{Non-unitary TQFTs from  $\CN=4$ SCFTs of rank 0 }{Non-unitary TQFTs from  N=4 SCFTs of rank 0}}

\paragraph{3D Rank 0 $\CN=4$ SCFTs }  In this paper, 3D  $\CN=4$ rank 0 SCFT is defined as 
\begin{align}
(\textrm {Rank 0 theory}):= (\textrm {Theory with no Coulomb and Higgs branches})\;.
\end{align}
3D $\mathcal{N}=4$ SCFTs have $SO(4)\simeq SU(2)_L \times SU(2)_R$ R-symmetry. 
The Coulomb (Higgs) branch is parametrized by chiral primary operators charged under the $SU(2)_R$ ($SU(2)_L$) symmetry while neutral under  the $SU(2)_L$ ($SU(2)_R$) symmetry. In our definition, the rank 0 theory can have mixed branches parametrized by chiral primaries charged under both of $SU(2)_L$ and $SU(2)_R$. Rank 0 SCFTs  in Section \ref{sec : examples} with $\CN=5$ supersymmetry actually have the mixed branches. Rank 0 theories cannot have a continuous flavor symmetry commuting with the $SO(4)$ R-symmetry, since a flavor current multiplet contains Higgs- or Coulomb- branch operators. By the same reasoning, the rank 0 theories cannot have SUSY more than $\CN=5$. 

\paragraph{Axial $U(1)$ symmetry and R-symmetry mixing} Let $R_\nu$ and $A$ be the two Cartan generators of the $SO(4)$ R-symmetry:
\begin{align}
\begin{split}
&R_{\nu=0} := R +R'\;, \quad A :=   R-R'\;,
\\
& R_{\nu} := R_{\nu=0}+ \nu A\;. \label{R-symmetry mixing}
\end{split}
\end{align}
Here $R$ and $R'$ are the Cartans of $SU(2)_L$ and $SU(2)_R$ respectively. In our convention, they are normalized as $R,R' \in \frac{1}2 \mathbb{Z}$. In terms of an $\mathcal{N}=2$ subaglebra, $R_{\nu}$ is the R-charge while $A$ is the charge of a $U(1)$ flavor symmetry (commuting with the $\CN=2$ supersymmetry) called the axial $U(1)$ symmetry. The mixing between the $U(1)$ R-symmetry and the axial flavor  symmetry is parametrized by $\nu$.  
The IR superconformal R-symmetry corresponds to $\nu=0$. 

\paragraph{Supersymmetric partition functions }  Generally, the partition function $\CZ^{\mathbb{B}}_{\CT} (b^2, m,\nu;s)$  of a 3d $\CN=4$ SCFT $\CT$ on  a rigid supersymmetric  background $\mathbb{B}$ depends on the followings:
\begin{align}
\begin{split}
 \CM &:\; \textrm{topology of $\mathbb{B}$}\;,
\\
b^2\;(\textrm{or } q) &: \; \textrm{squashing  (or $\Omega$-deformation) parameter }\;,
\\
m \;(\textrm{or } \eta=e^m) \;& :\; \textrm{real mass (or fugacity variable) for axial $U(1)$ symmetry}\;,
\\
\nu\;&  : \; \textrm{R-symmetry mixing parameter in  \eqref{R-symmetry mixing}}\;,
\\
s \in H^1 (\CM, \mathbb{Z}_2)\;&  : \;\textrm{SUSY-compatible spin-structure}\;.
\end{split}
\end{align}
We consider supersymmetric backgrounds $\mathbb{B}$ whose topologies are given as $\CM_{g,p}$, a degree $p$ bundle over a genus $g$ Riemann surface $\Sigma_g$:
\begin{align}
\xymatrix{
S^1  \ar[r]^{\; p \quad  } &\CM_{g,p}    \ar[d]  &  \\
&\Sigma_g & 
} 
\end{align}
We refer readers 
to Appendix \ref{App : review on localization} for a brief review on  localizations on supersymmetric backgrounds. We can turn on the $\Omega$-deformation parameter (sometimes called squashing parameter) only for $g=0$. For even $p$, one can consider two supersymmetric backgrounds depending on the choice of the spin structure along the fiber $[S^1]\in H_1 (\CM,\mathbb{Z}_2)$.  We denote $s=+1$ ($s=-1$) for the periodic (anti-periodic) boundary  condition for fermionic fields along the $S^1$.  For $p=0$, the partition function can be regarded as a version of the BPS index $\CI (s)$ and its spin structure dependence can be interpreted as
\begin{align}
\CI(s)= \begin{cases} 
\textrm{Tr}_{\CH_{\rm BPS}} (-1)^{R_\nu}, \quad  \textrm{for } s=-1\;,
\\ 
\textrm{Tr}_{\CH_{\rm BPS}} (-1)^{2j_3},  \quad \textrm{for } s=1\;. \label{indices in differenent s}
\end{cases}
\end{align}
As BPS indices, they can be defined without overall phase factor ambiguities. 
For $p\neq 0$, on the other hand,  local counterterms  affect the phase factor of the partition function \cite{Closset:2012vg} and  it is non-trivial to keep track of the local counterterms. Throughout this paper, we are for the most part interested in the absolute value of partition functions.  For $g=1$ and $p=0$, the supersymmetric partition function is independent of all the parameters and is simply an integer number called the Witten index. 

\paragraph{Emergence of non-unitary TQFT in the limits $\nu \rightarrow \pm 1$} As main result of the paper, we propose  that for any rank 0 $\CN=4$ SCFT $\CT_{\rm rank \;0}$ we can associate a pair of non-unitary TQFTs denoted by  $ \textrm{TFT}_\pm [\CT_{\rm rank\;0}]$, satisfying the following relation
\begin{align}
\begin{split}
\textbf{Main proposal : }&\CZ^{\mathbb{B}}_{\CT_{\rm rank \;0}} \left(b^2, m\; (\textrm{or}\; \eta),\nu;s \right) \xrightarrow{\quad m \rightarrow 0 \;(\textrm{or}\; \eta \rightarrow 1),\; \nu\rightarrow \pm 1 \quad }  \CZ^{\CM_{g,p}}_{{\rm TFT}_\pm [\CT_{\rm rank\;0}]} (s) \;.  \label{rank 0 to TFT}
\end{split}
\end{align}
The partition function $\CZ^{\CM_{g,p}}_{{\rm TFT}}$ of the topological theory  depends  only on the topological structure $\CM_{g,p}$ of $\mathbb{B}$ and  (possibly) a choice of a spin-structure $s$ on it.\footnote{The overall phase factor of the partition function depends also on the choice of the framing. As with supersymmetric partition functions, there is no canonical choice of the framing for non-zero $p$ and we mostly focus on the absolute values of the partition functions. }   We claim that in the degenerate limits  \rom{1}) the rigid supersymmetry partition function becomes independent on the squashing  parameter $b^2$ (or $\Omega$-deformation parameter $q$) and \rom{2}) it becomes the partition function of a non-unitary topological quantum field theory. 

We call a topological quantum field theory a non-spin (or bosonic) TQFT when its partition function is independent on the choice of the spin structure, and a spin (or fermionic) TQFT otherwise. We  propose that 
\begin{align}
\begin{split}
&\textrm{ The non-unitary  $\textrm{TFT}_{\pm} [\CT_{\rm rank\; 0}]$ is a spin (fermionic) TQFT}\;, 
\\
&\qquad \textrm{if $R_{\nu =\pm 1} +2 j_2 \in 2\mathbb{Z}+1$ for some BPS local operators \;.  }
\\
&\qquad\textrm{or equivalently,}\;  
\\
&\qquad \textrm{if $\CI^{\rm sci}(q,\eta, \nu=\pm 1;s)$ contains   $q^{\frac{1}2(\textrm{odd integer})}$ terms\;.} 
\end{split} \label{criterion of spin/non-spin}
\end{align}
Here $\CI^{\rm sci}$ is the superconformal index defined in \eqref{Def : superconformal index} and the index at $\nu=\pm 1$ is in general a power series in  $q^{1/2}$ since $R_{\nu=\pm 1}+2j_3 \in \mathbb{Z}$.  From \eqref{indices in differenent s}, we expect that the supersymmetric partition on $\CM_{g,p=0}$ depends on the spin-structure $s$ if  the above condition is met. This is because that the  condition $R_{\nu =\pm 1} +2 j_3 \in 2\mathbb{Z}+1$ implies that $(-1)^{R_{\nu = \pm 1}} \neq (-1)^{2j_3}$ for some BPS local operators which acts on the Hilbert-space $\CH_{\rm BPS}$ non-trivially.  The above condition gives  sufficient but not necessary condition for $\textrm{TFT}_{\pm}$ to be fermionic. To see this, consider a rank 0 SCFT $\CT_{\rm rank \;0}$ not satisfying the above condition whose associated non-unitary topological field theories, $\textrm{TFT}_{\pm}[\CT_{\rm rank \; 0}]$, are bosonic. Then,    $\CT_{\rm rank \;0} \otimes \CT_{\rm spin\;top}$ with an unitary fermionic topological field theory $ \CT_{\rm spin\;top}$ is still a rank 0 SCFT not satisfying the above condition since the decoupled topological sector does not contribute to the superconformal index.  But its associated non-unitary TQFTs $\textrm{TFT}_{\pm}[\CT_{\rm rank \;0} \otimes \CT_{\rm spin\;top}] = \textrm{TFT}_{\pm}[\CT_{\rm rank \;0} ] \otimes  \CT_{\rm spin\;top} $ are fermionic due to the decoupled unitary spin TQFT $ \CT_{\rm spin\;top} $.
 
The proposal in \eqref{rank 0 to TFT} can be easily proven for the case when  $\CZ^{\mathbb{B}}$ is the superconformal index \eqref{Def : superconformal index}.  In the degenerate limit, $\nu\rightarrow 1$ and $\eta=1$, the index becomes
\begin{align} 
\mathcal{I}^{\rm sci}(q,\eta ,\nu;s ) \xrightarrow{\;\; \eta =1, \nu\rightarrow  1 \;\; }  \begin{cases}
\textrm{Tr}_{\mathcal{H}_{\rm rad}(S^2)} (-1)^{2j_3} q^{R + j_3}\; \; , \quad s=+1\;,\\
\textrm{Tr}_{\mathcal{H}_{\rm rad}(S^2)} (-1)^{R} q^{R + j_3} \; \;, \quad s=-1\;.
\end{cases}  
\end{align}
All  unitary multiplets of 3D $\CN=4$ superconformal algebra are listed in \cite{Cordova:2016emh}. From the classification, it is not difficult to check that the index above gets contributions only from operators in a short multiplet denoted by $B_1[0]^{(2R,2R')}_{\Delta}$  with $R=0$ in \cite{Cordova:2016emh}. The  bottom state inside the multiplet corresponds to a  Coulomb branch operator parametrizing the Coulomb branch. From the superconformal index in the degenerate limit, one can actually compute the Hilbert-series of the Coulomb branch \cite{Razamat:2014pta}. 
Since we are considering a rank 0 theory $\CT_{\rm rank\; 0}$ with empty Coulomb branch, the index gets contributions only from the identity operator and becomes simply ($q$-independent) $1$. Similarly, one can also confirm that the index  becomes  $1$ in the other degenerate limit, $\nu\rightarrow -1$ and $\eta=1$, since there is no Higgs branch. In summary, from the superconformal multiplet analysis, we conclude that
\begin{align}
\mathcal{I}_{\CT_{\rm rank \;0}}^{\rm sci}(q,\eta ,\nu;s )   \xrightarrow{\;\; \eta = 1, \nu\rightarrow \pm 1 \;\; }   1,  \;\;  \textrm{for any rank 0 theory } \CT_{\rm rank\; 0}\;. \label{triviality of SCI from supermultiplet analysis}
\end{align}
This proves the proposal in \eqref{rank 0 to TFT} for the case when $\CZ^{\mathbb{B}}$= (superconformal index) since $\CZ^{S^2\times S^1}=1$ for all topological theories. 

We currently do not know the full proof of the proposal \eqref{rank 0 to TFT} for other supersymmetric partition functions $\CZ^{\mathbb{B}}$. As noticed in \cite{Cho:2020ljj}, however, the triviality of superconformal index for a non-conformal choice of R-symmetry is a strong evidence for the appearance of non-unitary TQFT, 
while the triviality of the index at the superconformal R-symmetry implies an emergence of a unitary TQFT at the infra-red fixed point. We  explicitly  test the proposal with infinitely many rank 0 SCFTs and various supersymmetric partition functions in  section \ref{sec : examples}. 
 
\subsection{Dictionaries}
In Table \ref{Table : Dictionaries}, we summarize basic dictionaries of the correspondence. 
The dictionary in the first line is the most basic one and other dictionaries (except for the last one for $F$) follow from it. On the one hand, partition functions (with insertion of loop operators along the fiber $S^1$) of a topological field theory  on the geometries $\CM_{g,p}$ are determined by the modular data, i.e.\ $S$ and $T$ matrices, of the topological theory.  On the other hand, the supersymmetric partition function on $\mathbb{B}$ with topology $\CM_{g,p}$ can be computed using localization technique as briefly summarized in Appendix \ref{App : review on 3D TQFT}.  

\begin{table}[ht]
	\begin{center}
		\begin{tabular}{|c|c|}
			\hline
			$\quad \textrm{TFT}_\pm [\CT_{\rm rank\;0}]$ \quad &  $\CT_{\rm rank \;0}$ 
			\\
			\hline
			\hline
			$\CZ^{\CM_{g,p}}_{\rm TFT_\pm } (s)$& BPS partition function $\CZ^{\mathbb{B}}_{\CT_{\rm rank\; 0 }}\big{|}_{\nu\rightarrow \pm 1, m=0} (s)$
			\\
			& with  (\textrm{topology of }$\mathbb{B}$) = $\CM_{g,p}$ 
			\\
			\hline
			Spin or non-spin & Equation \eqref{criterion of spin/non-spin}
			\\
			\hline
			Rank   $N$ & Witten index
			\\
			\hline
			& \textrm{Bethe vacua } $\{\vec{z}_\alpha\}_{\a=0}^{N-1}$
			\\
			\textrm{Simple objects } & \textrm{ or }
			\\
			& \textrm{BPS loop operators $\{ \CO^{\pm}_\alpha (\vec{z}) \}_{\alpha=0}^{N-1}$}
			\\
			\hline
			$(S^{\pm}_{0\alpha})^{-2}$ &    $\CH_\alpha (m=0, \nu\rightarrow \pm 1;s=-1)$
			\\
			\hline
			$T^\pm_{\alpha\beta}$ (only for non-spin)&  $ \delta_{\alpha \beta} (\frac{\CF_\alpha}{\CF_{\alpha=0}}) \big{|}_{\nu \rightarrow \pm 1, m=0} $ 
			\\
			\hline
			$(T^2)^\pm_{\alpha\beta}$&  $ \delta_{\alpha\beta} (\frac{\CF_\alpha}{\CF_{\alpha=0}})^2\big{|}_{\nu \rightarrow \pm 1, m=0,s=-1} $ 
			\\
			\hline
			$S^{\pm}_{0 0}$ &   $\bigl|\CZ^{S^3_b}_{\CT_{\rm rank\;0}}(m=0, \nu \rightarrow \pm 1)\bigr|$ 
			\\[0.5ex] 
			\hline
			$W^\pm_\beta(\alpha)$ & $  \CO^\pm_\alpha  (\vec{z}_\beta)|_{\nu \rightarrow\pm 1, m=0}$ 
			\\
			\hline
			$\max_\alpha (-\log |S^{\pm}_{0\alpha}|)$& $F$ (three-sphere free energy)
			\\
			\hline
		\end{tabular}
	\end{center}
	\caption{Basic dictionaries in (rank 0 SCFT)/(non-unitary TQFTs) correspondence. $S^\pm_{\alpha\beta}$ and $T^\pm_{\alpha\beta} $  are modular matrices of ${\rm TFT}_{\pm} [\CT_{\rm rank\;0}]$.  We define $W_\beta(\alpha):=\frac{S_{\alpha \beta}}{S_{0 \beta}} = \langle \beta | \CO^A_\alpha | \beta\rangle$, see \eqref{action of loop operators},  from which one can compute S-matrix $S_{\alpha \beta} = W_\beta (\alpha) W_0 (\beta) S_{00}$. $\CH_\alpha$ and $\CF_\alpha$ are handle gluing and fibering operator at the $\a$-th Bethe-vacuum respectively, see \eqref{Handle gluing} and \eqref{Fibering}. $\CO_{\alpha=0}$ corresponds to the trivial loop and $W_\beta (0) =1$. The rank of a TFT  (not to be confused with the rank of its associated 3D $\CN=4$ SCFT) is defined as the dimension of $\CH(\mathbb{T}^2)$ for  non-spin case while  is defined as the dimension of  $\CH_{--}(\mathbb{T}^2)$, i.e.\ Hilbert-space in  NS-NS sector, for spin case. Similarly, the $S$ and $T$ matrices  (resp. $S$ and $T^2$) for non-spin TQFT case  (resp. spin TQFT case) are modular matrices acting on $\CH(\mathbb{T}^2)$ (resp. $\CH_{--}(\mathbb{T}^2)$). Simple objects are in one-to-one with a basis of $\CH(\mathbb{T}^2)$ (resp. $\CH_{--}(\mathbb{T}^2)$) for non-spin TQFT case (resp. spin TQFT case). We refer readers to Appendix \ref{App : review on 3D TQFT} for a general  review on the modular data of non-spin and spin TQFTs. }
	\label{Table : Dictionaries}
\end{table}
 For a ${\rm TQFT}$ its rank, i.e.\ the size of modular matrices, is equal to the ground state degeneracy on the two torus. For a supersymmetric field theory, the degeneracy is equal to the Witten index. The dictionaries for $S_{0\alpha}^2, T_{\alpha \beta}$ and $W_\beta (\alpha)$ simply follow from comparison between \eqref{Zgp from S and T} and \eqref{Supersymmetric loop in Zgp}. The Bethe-vacuum corresponding to the trivial simple object, $\alpha=0$, can be determined by requiring that 
\begin{align}
\textrm{Trivial object $\alpha=0$ : } \frac{1}{\sqrt{|\CH_{\alpha=0} (m=0, \nu \rightarrow \pm 1)|}} = S^{\pm}_{00}=|\CZ^{S^3_b}(m=0, \nu\rightarrow \pm 1)|\;.  \label{Consistency for trivial vacuum}
\end{align}
In topological field theories simple objects  (labeled by $\alpha$) are  loop operators, while 
in supersymmetric field theories $\alpha$ labels the types (gauge and flavor charge) of loop operators.  Therefore, we expect that for each Bethe-vacuum $\alpha$ there is a corresponding supersymmetric loop operator $\CO_\alpha(\vec{z})$, see around  \eqref{Supersymmetric loop in Zgp}:
\begin{align}
\textrm{(Bethe vacua)-to-(loop operators) map : } \vec{z}_\alpha \rightarrow \CO^\pm_\alpha \;.
\end{align}
 The trivial object $\alpha=0$ corresponds to the identity operator. The map  will be determined by requiring the following consistency conditions
 \begin{align}
 	\begin{split}
 	&W^\pm_{0}(\alpha) = \frac{S^\pm_{\alpha 0}}{S^\pm_{00}}  = \sqrt{\frac{\CH_{\alpha=0}(m=0, \nu \rightarrow \pm 1)}{\CH_{\alpha}(m=0, \nu \rightarrow \pm 1)}}\; \textrm{ and }W^\pm_{0}(\alpha) = \CO^\pm_\alpha (\vec{z}_{\alpha=0}) \big{|}_{\nu \rightarrow \pm 1, m=0}\;,
 	\\
 	&\Rightarrow  \CO^\pm_\alpha (\vec{z}_{\alpha=0})  =  \sqrt{\frac{\CH_{\alpha=0}(m=0, \nu \rightarrow \pm 1)}{\CH_{\alpha}(m=0, \nu \rightarrow \pm 1)}}\;.
 	\label{Consistency for Bethe-to-loop}
 	\end{split}
 \end{align}

The  dictionary for $F$ in the last line is one of the most non-trivial and interesting statements in this paper.  It says that
\begin{align}
\begin{split}
&F[\CT_{\rm rank \;0}]
\\
&=-\log  |S_{0 \alpha_*}| {\rm\;  of\; TFT}_\pm[\CT_{\rm rank \; 0}] \quad  \textrm{($\alpha_*$ is chosen such that $|S_{0\alpha_*}|\leq |S_{0\alpha}|$ for other $\a$)}\;. \label{S-alpha}
\end{split}
\end{align}
$F$ is the free energy on round three-sphere, which is the quantity appearing in the F-theorem and is a proper measure of degree of freedom. The relation is surprising since it relates the quantity ($F$) at the superconformal point,  $\nu=0$, to the  quantity ($S_{0\alpha_*}$) in the degenerate limits $\nu =\pm 1$. 

One possible explanation for the dictionary above is as follows. In general,  $S_{0\alpha_*}$ in a topological field  theory computes the three-sphere partition function with an insertion of loop operator $\CO_{\alpha_*}^{\Gamma = \textrm{(unknot)} }$ of type $\alpha_*$ along the unknot in $S^3$. In the rank 0 theory $\CT_{\rm rank \; 0 }$, on the other hand, there is a flavor vortex loop operator associated to the $U(1)$ axial flavor symmetry. The loop operator is known to act on the  three-sphere partition as a difference operator shifting the parameter $\nu$ \cite{Dimofte:2011ju,Kapustin:2012iw,Drukker:2012sr}
\begin{align}
\textrm{(flavor vortex loop)$_\pm$ of charge $\pm 1$ } \;  \longleftrightarrow  \; \exp (\pm \partial_{\nu})\;.
\end{align}
This means that the $S^3$ partition function at the conformal point can be identified with the $S^3$ partition function with an insertion of the vortex loop operators of charge $+ 1$($-1$)  in TFT$_-$ (TFT$_+$)\footnote{If we turn on the squashing parameter $b$, the partition function at $\nu=-1$ with the flavor vortex loop is given as $\mathcal{Z}^{S^3_b+ \CO_{\rm flavor\; vortex}}(m=0, \nu=-1) =  \mathcal{Z}^{S^3_b}\left(m= i \pi (b-\frac{1}b), \nu=0\right)  $. Interestingly, the partition function is actually independent $b$  \cite{Gaiotto:2019mmf,Bullimore:2020jdq} and $|\mathcal{Z}^{S^3_b}\left(m= i \pi (b-\frac{1}b), \nu=0\right)| = |\mathcal{Z}^{S^3_b}\left(m= i \pi (b-\frac{1}b), \nu=0\right)|_{b=1}= |\mathcal{Z}^{S^3_b}\left(m= 0, \nu=0)\right|= e^{-F} $.   }
\begin{align}
\begin{split}
&|\CZ^{S^3} ( \nu=0) |=  \exp (\partial_\nu )\cdot |\CZ^{S^3} ( \nu=-1) | =  |\CZ^{S^3+ \CO_{\rm flavor \;vortex}} (\nu=-1) |
\\
& \Rightarrow  |\CZ^{S^3} ( \nu=0) |= |S_{0{\alpha=\textrm{(flavor vortex)}_+}} \textrm{ of TFT}_-|\; .
\end{split}
\end{align}
Hence if one  identify $\alpha =(\textrm{flavor vortex})_+$ ($\alpha =(\textrm{flavor vortex})_-$) of  TFT$_-$ (TFT$_+$) with $\alpha = \alpha_*$ in \eqref{S-alpha}, then the dictionary follows. Actually, according to F-maximization \eqref{F-maximization}, we expect that
\begin{align}
\begin{split}
&|\mathcal{Z}^{S^3}(\nu=0)| = |S_{0 \alpha=\textrm{(flavor vortex)}_+} \textrm{ of TFT}_-| < |S_{00} \textrm{ of TFT}_-| = |\mathcal{Z}^{S^3}(\nu=-1)|\;,
\\
&|\mathcal{Z}^{S^3}(\nu=0)| = |S_{0 \alpha=\textrm{(flavor vortex)}_-} \textrm{ of TFT}_+| < |S_{00} \textrm{ of TFT}_+| = |\mathcal{Z}^{S^3}(\nu=1)|\;.
\end{split}
\end{align}
It is compatible with the desired identification, $(\textrm{flavor vortex loop}) = (\alpha_* \textrm{in }\eqref{S-alpha} )$. The  property above is also compatible with the fact  that TFT$_{\pm}$ are non-unitary since they violate the unitarity condition \eqref{Unitarity in S}. 

The argument above gives circumstantial evidence but not a full proof for the dictionary on $F$. 
The dictionary will be confirmed  explicitly in section \ref{sec : examples} with infinitely many examples. We leave general proof or disproof of the dictionary for future work.

\subsection{\texorpdfstring{Application: lower bounds on $F$}{Application: lower bounds on F}}
Here we derive interesting lower bounds on $F$ for rank 0 SCFTs  using the correspondence introduced in the previous subsection. 

One immediate and interesting consequence of the  dictionaries in Table \ref{Table : Dictionaries} is 
\begin{align}
	F >    -\log \frac{1}{\sqrt{\textrm{Witten index}}}  \;, \;\;  \textrm{for all interacting $\CN\geq 4$ SCFT $\CT_{\rm rank \; 0}$ of rank $0$}\;. 
\end{align}
This follows from the following fact
\begin{align}
\sum_{\alpha =0}^{N-1} (S_{0\alpha})^2 = 1\;\Rightarrow  \; \textrm{min}_\alpha |S_{0\alpha}| = |S_{0\alpha_*} \textrm{ in \eqref{S-alpha}} |<\frac{1}{\sqrt{N}}\;, \label{bound for general rank}
\end{align}
combined with the dictionaries for Witten index and $F$.
More interestingly, using the dictionary,  one can prove following
\begin{align}
F \geq    -\log \sqrt{\frac{5-\sqrt{5}}{10}} = 0.642965 \;, \;\;  \textrm{for any interacting $\CN\geq 4$ SCFT $\CT_{\rm rank \; 0}$ of rank $0$}\;.  \label{lower bound on F}
\end{align}
The bound is saturated by the minimal $\CT_{\rm min}$ theory which will be introduced in section \ref{sec : examples}. The inequality above follows from the following fact combined with the dictionary for $F$
\begin{align}
|S_{0\alpha_*} \textrm{in \eqref{S-alpha}}|  \leq    \sqrt{\frac{5-\sqrt{5}}{10}}  = 0.525731 \quad \textrm{for all non-unitary TFT}\;. \label{lower bound on S-alpha for non-unitary TFT}
\end{align}
Thanks to \eqref{bound for general rank}, we only need to check the inequality for non-unitary TQFTs up to rank 3 since  and $\frac{1}{\sqrt{4}} =0.5 <  \sqrt{\frac{5-\sqrt{5}}{10}}$. Let us first consider rank 2 case. Let $x = S_{00}^2$ and $y=S_{01}^2$ where $S_{\alpha \beta}$ is the S-matrix  of a non-unitary TQFT.   Then the two positive real numbers $x$ and $y$ should satisfy followings
\begin{align}
\textrm{GSD}_{g=0} = x+y=1\;, \quad  \textrm{GSD}_{g=2} = \frac{1}x+\frac{1}y = k \in \mathbb{Z}_{>0}\;. \
\end{align}
Here $\textrm{GSD}_g$ denotes the ground state degeneracy on genus $g$ Riemann surface, see \eqref{GSDg from S}.  
One can solve the equations and we have two solutions
\begin{align}
	x= \frac{1}2 \pm \frac{1}2 \sqrt{\frac{k-4}{k}}\;, \quad y= \frac{1}2 \mp \frac{1}2 \sqrt{\frac{k-4}{k}} \;.
\end{align} 
Imposing the conditions, $x,y \in \mathbb{R}_{>0}$ and $x>y$ (non-unitarity condition, see \eqref{Unitarity in S}),  we have only one solution for each $k> 4$. As the natural number $k$ increases, $y^{1/2} = |S_{0\alpha_*}|$ in the solution  decreases. Thus, we have  
\begin{align}
\begin{split}
& |S_{0\alpha_*}|  = y^{1/2} \leq (y^{1/2} \textrm{ at } k=5) = \sqrt{\frac{5-\sqrt{5}}{10}}\;, 
 \\
& \textrm{ for any rank 2 non-unitary TQFT}\;.  \label{bound for rank 2}
 \end{split}
\end{align}
Now let us move on to the rank 3 case. Let $x= S_{00}^2, y=S_{01}^2$ and  $z=S_{02}^2 = S_{0\alpha_*}^2$. Then, the three positive real numbers should satisfy 
\begin{align}
\begin{split}
&\textrm{GSD}_{g=0} = x+y+z=1\;, \quad \textrm{GSD}_{g=2} = \frac{1}x+\frac{1}y+\frac{1}z=k_1 \in \mathbb{Z}_{>0}\;, 
\\
&\textrm{GSD}_{g=3} = \frac{1}{x^2}+\frac{1}{y^2}+\frac{1}{z^2} = k_2 \in \mathbb{Z}_{>0}\;.
 \end{split}
\end{align}
One can confirm that  any solution to equations above satisfying the non-unitarity conditions, $\textrm{min}(x,y)\geq z$ and $\textrm{max}(x,y)> z$, have following property\footnote{Since $\frac{1}x+\frac{1}y+\frac{1}z \leq \frac{3}z$, $\frac{1}{x^2}+\frac{1}{y^2}+\frac{1}{z^2} \leq \frac{3}{z^2}$ and thus $z\leq \textrm{min}(\frac{3}{k_1}, \sqrt{\frac{3}{k_2}})$, it is enough to check that $z^{1/2}\leq \frac{1}2$ for only finitely many cases $(0<k_1\leq 12$ and $0<k_2 \leq 48)$. }
\begin{align}
\begin{split}
&|S_{0\alpha_*}|  = z^{1/2} \leq (z^{1/2} \textrm{ at } k_1=10 \textrm{ and } k_2 = 36)  = \frac{1}{2} < \sqrt{\frac{5-\sqrt{5}}{10}}\;,
\\
&\textrm{ for any rank 3 non-unitary TQFT}\;.  \label{bound for rank 3}
\end{split}
\end{align}
We are not certain if there is a rank 3 non-unitary TQFT saturating the bound $|S_{0\alpha_*}|  =\frac{1}2$.  
The results in \eqref{bound for general rank}, \eqref{bound for rank 2} and \eqref{bound for rank 3} imply the bound in \eqref{lower bound on S-alpha for non-unitary TFT}, from which the conclusion in \eqref{lower bound on F} follows.

\section{Examples } \label{sec : examples}
In this section, we introduce infinitely many examples of (2+1)D $\mathcal{N}=4$ interacting rank 0 superconformal field theories $\CT_{\rm rank \;0}$. Using the dictionary  in Table \ref{Table : Dictionaries}, we compute the set $\{|S^{\pm}_{0\alpha}|\}$ for non-unitary TQFTs $\textrm{TFT}_\pm [\CT_{\rm rank \;0}]$. We also independently compute the three-sphere free energy $F$ by performing the localization integral for $\CZ^{S^3_b}$ at $b=1, m=0,\nu=0$ and confirm the non-trivial dictionary for $F$. See  Table \ref{table:non-unitary/SCFT} for the summary. 

\begin{table}[ht]
	\centering
	\begin{tabular}{|c| c| c |c| c|}
		\hline
		$\CT_{\rm rank \;0 }$& TFT$_\pm [\CT_{\rm rank\;0}]$ &  \textrm{Set of $\{|S^{\pm }_{0\alpha}|\}$}& $\exp (-F)$ \\ [0.5ex] 
		\hline \hline
		$\CT_{\rm min}$&(Lee-Yang)   &   \{$\sqrt{\frac{5+\sqrt{5}}{10}}$ , $\sqrt{\frac{5-\sqrt{5}}{10}}$ \} &$\sqrt{\frac{5-\sqrt{5}}{10}}$
		\\
		\hline
		$(U(1)_1+H)$&  $\textrm{Gal}_{d} (SU(2)_{6})/\mathbb{Z}_2^f$  &  $\{2 \zeta^1_{6}, 2 \zeta^3_6 \}$ & $2 \zeta^1_6$
		\\
	      &$(\textrm{with }d=\zeta_{6}^{3})$   & &
		\\
		\hline 
		$SU(2)_{k}^{\frac{1}2 \oplus \frac{1}2}$&$\textrm{Gal}_{d} (SU(2)_{4|k|-2})/\mathbb{Z}_2^f$ & $\big{\{}  2 \zeta^{2n-1}_{4|k|-2} \big{\}}_{n=1}^{|k|}$ & $2\zeta_{4|k|-2}^1$ 
		\\
		($|k|>1$)&$(\textrm{with }d=\zeta_{4|k|-2}^{2|k|-1})$   & &
		\\
		\hline
		$T[SU(2)]_{k_1,  k_2}$  &  \textrm{See the caption} &  $\big{\{} (\frac{1}{\sqrt{2}}\zeta_{|k_1 k_2-1|-2}^{n})^{\otimes 2} \big{\}}_{n=1}^{|k_1 k_2 -1|-1}$  & $\frac{1}{\sqrt{2}}\zeta_{|k_1 k_2-1|-2}^{1}$
		\\[0.5ex] 
		\hline
		$\frac{T[SU(2)]}{SU(2)^{\rm diag}_{|k|=3}}$  & (Lee-Yang)$^{\otimes 2} \otimes U(1)_{2}$&  $\big{\{} \frac{1}{\sqrt{10}}^{\otimes 4}, \frac{5+\sqrt{5}}{10 \sqrt{2}}^{\otimes 2} , \frac{5-\sqrt{5}}{10 \sqrt{2}}^{\otimes 2} \big{\}}$& $\frac{5-\sqrt{5}}{10\sqrt{2}} $
		\\[1.5ex] 
		\hline
		$\frac{T[SU(2)]}{SU(2)^{\rm diag}_{|k|=4}}$  & $\frac{\textrm{Gal}_{\zeta_{10}^7}(SU(2)_{10}) \times SU(2)_2}{\mathbb{Z}_2^{\rm diag}} $ & $\big{\{} \frac{1}{2}, \frac{1}{2 \sqrt{3}}^{\otimes 5} ,\frac{3+\sqrt{3}}{12 }^{\otimes 2} , \frac{3-\sqrt{3}}{12}^{\otimes 2} \big{\}}$    & $\frac{3-\sqrt{3}}{12 }$ 
		\\[1.5ex] 
		\hline
		$\frac{T[SU(2)]}{SU(2)^{\rm diag}_{|k|=5}}$  &Gal$_{d} \big{(}(G_2)_3\big{)} \otimes U(1)_{-2} $    & $ \big{\{}\frac{1}{\sqrt{6}}^{\otimes 2} , \frac{1}{\sqrt{14}}^{\otimes 6}, $ &  $\sqrt{\frac{5}{84}-\frac{1}{4\sqrt{21}}}$  
		\\
	     &$ \left(d=\sqrt{\frac{5}{84}+\frac{1}{4\sqrt{21}}}\,\right)$  & $ \sqrt{\frac{5}{84}\pm \frac{1}{4\sqrt{21}}}^{\,\otimes 2} \big{\}}$  &  
		\\
		\hline
	     &  & $ \big{\{}\frac{1}{\sqrt{2|k|-4}}^{\otimes (|k|-3)} , \frac{1}{\sqrt{2|k|+4}}^{\otimes (|k|+1)} $ &  $\frac{1}{\sqrt{8|k|-16}} $ \;
		\\
		$\frac{T[SU(2)]}{SU(2)^{\rm diag}_{|k|\geq 6}}$   & {\bf ?} &$(\frac{1}{\sqrt{8|k|-16}} + \frac{1}{\sqrt{8|k|+16}})^{\otimes 2}$\;,\;  & \;$- \frac{1}{\sqrt{8|k|+16}}  $  
		\\
		& &  $(\frac{1}{\sqrt{8|k|-16}} - \frac{1}{\sqrt{8|k|+16}})^{\otimes 2}$\big{\}} & 
	    \\[1ex]
		\hline
	\end{tabular}
	\caption{Non-unitary TQFTs from rank 0  $\CN=4$ SCFTs. $\textrm{Gal}_{d}(\textrm{TFT})$ denotes a Galois conjugate of an unitary topological field theory TFT with $S_{00} (\textrm{Gal}_{d}[\textrm{TFT}])=d$. We define $\zeta_k^n :=\sqrt{\frac{2}{k+2}} \sin \frac{n \pi}{k+2}$. 
		For  the rank 0 SCFT $\CT_{\rm rank \; 0}=T[SU(2)]_{k_1, k_2}:= \frac{T[SU(2)]}{SU(2)^C_{k_1}\times SU(2)^H_{k_2}}$ with $|k_1 k_2-1|>3$ and $\textrm{min}(|k_1|,|k_2|)>2$, the corresponding non-unitary TQFTs are $\textrm{TFT}_+ = \big{[}\textrm{Gal}_{\zeta_{|k_1 k_2 -1|-2}^{|k_2|}} (SU(2)_{|k_1 k_2 -1|-2})  \big{]}\otimes U(1)_2$  and $\textrm{TFT}_- = \big{[}\textrm{Gal}_{\zeta_{|k_1 k_2 -1|-2}^{|k_1|}} (SU(2)_{|k_1 k_2 -1|-2})  \big{]}\otimes U(1)_2$. For $\CT_{\rm rank \;0} = \frac{T[SU(2)]}{SU(2)^{\rm diag}_{|k|\geq 6}}$, we could not identify their associatec TFTs with previously  known non-unitary TQFTs.  }
	\label{table:non-unitary/SCFT}
\end{table}

\subsection{\texorpdfstring{The minimal $\mathcal{N}=4$ SCFT $\CT_{\rm min}$}{The minimal N=4 SCFT T(min)}}

\subsubsection{SUSY enhancement}
In \cite{Gang:2018huc}, it was claimed that
\begin{align}
\begin{split}
&\textrm{(3D $\mathcal{N}=2$ gauge theory, $U(1)_{k=-3/2}$ coupled to a chiral multiplet $\Phi$ of charge $+1$)} 
\\
&\xrightarrow{\quad \textrm{at IR} \quad }(\textrm{3D $\mathcal{N}=4$ superconformal field theory $\mathcal{T}_{\rm min}$})\;.
\end{split}
\end{align}
As a quick evidence for the SUSY enhancement,  there are  following two gauge invariant   monopole operators  in the theory 
\begin{align}
\hat{\phi} \hat{\phi}|\mathfrak{m}=+1\rangle \;, \quad \hat{\psi}^*|\mathfrak{m}=-1\rangle\;,
\end{align}
with following quantum numbers \cite{Borokhov:2002cg,Benini:2011cma,Gang:2018huc}
\begin{align}
\begin{split}
&A = -\mathfrak{m}=-1, \; R= \frac{1-R_\Phi}{2} |\mathfrak{m}| +2R_\Phi =1, \; j=1 \;, \; \Delta=2 \;, 
\\
&A = -\mathfrak{m} =1, \; R = \frac{1-R_\Phi}{2} |\mathfrak{m}|   + (1-R_\Phi) =1, \; j=1\;, \; \Delta=2\;.  \label{monopole ops in Tmin}
\end{split}
\end{align}
Here $A$ is the charge of $U(1)_{\rm top}$ and R is the charge of the superconformal $U(1)$ R-symmetry;  $j \in \frac{\mathbb{Z}}2$ and  $\Delta$ are the Lorentz spin and the conformal dimension respectively. 
We use the fact that  \cite{Gang:2018huc}
\begin{align}
R_\Phi := (\textrm{Superconformal $U(1)$ R-symmetry charge of $\Phi$}) = \frac{1}3\;,  \label{superconformal R-charge of minimal theory}
\end{align}
which can be determined by the F-maximization \cite{Jafferis:2010un}. Here $|\mathfrak{m}\rangle \in \CH_{\rm rad}(S^2)$ (the radially quantized Hilbert-space on $S^2$) is a half BPS bare monopole operator. The bare monopole operator has a $U(1)$ gauge charge $-\frac{3}2 \mathfrak{m}+\frac{1}{2}|\mathfrak{m}|$ and should be dressed by  excitations ($\hat{\phi}, \hat{\psi}$ and their complex conjugations) of matter fields  to be gauge-invariant. The dressed  monopole operators are $1/4$ BPS local operators. According to the classification in \cite{Cordova:2016emh}, the monopole operators above, if they survive  at the IR  superconformal point,  must  belong to an extra SUSY-current multiplet of the 3D $\CN=2$ superconformal algebra. The multiplet consists of conformal primaries of the following quantum numbers as well as their conformal descendants,
\begin{align}
	[(R, j , \Delta) = (0,\frac{1}2,\frac{3}2)] \xrightarrow{\quad Q, \bar{Q}\quad } 	[(R, j , \Delta) = (\pm 1,1,2)]  \xrightarrow{\quad Q, \bar{Q}\quad }  [(R, j , \Delta) = (0,\frac{3}2,\frac{5}2)] \;.
\end{align}
Here $Q :=Q_1 +iQ_2$ and $\bar{Q}:=Q_1- i Q_2$ are the $\CN=2$ supercharges. 
The local operators in the top component with $(R, j, \Delta)= (0, \frac{3}2,\frac{5}2)$ correspond to the conserved current for the extra supersymmetry, whose existence guarantees the SUSY enhancement. 

Further, it was claimed in \cite{Gang:2018huc} that the infra-red (IR)  superconformal field theory (SCFT) $\mathcal{T}_{\rm min}$ is the minimal 3D $\mathcal{N}=4$ SCFT having the smallest  three-sphere free-energy $F$ and the smallest non-zero stress-energy tensor central charge $C_T$  whose exact values are \cite{Gang:2019jut,Gang:2018huc}
\begin{align}
\begin{split}
F(\CT_{\rm min}) & = -\log \sqrt{\frac{5-\sqrt{5}}{10}}
\\
& \simeq 0.642965\;,
\\
\frac{C_T (\mathcal{T}_{\rm min})}{C_T (\textrm{free theory with single $\Phi$})} &= \frac{8}{26}\left( 8- \frac{5\sqrt{5+ 2\sqrt{5}}}\pi \right)
\\
&\simeq 0.992549\;. \label{F and CT of minimal theory}
\end{split}
\end{align} 
 There is no vacuum moduli space in the minimal  theory and thus the minimal SCFT is  of rank 0.  
 
 \paragraph{Superconformal index} Applying the  localization results in  \cite{Kim:2009wb,Imamura:2011su} (see also Appendix A of \cite{Aharony:2013dha}), the superconformal index for $\CT_{\rm min}$ can be written as
 \begin{align}
 \begin{split}
 &	\CI_{\CT_{\rm min}}^{\rm sci} \big{(}q, \eta,\nu;s=1 \big{)} = \sum_{ \mathfrak{m} \in \mathbb{Z}} \oint_{|a|=1} \! \frac{da}{2\pi i a} \, q^{\frac{|\mathfrak{m}|}6} (a(-1)^\mathfrak{m})^{-\frac{3\mathfrak{m}}2 - \frac{|\mathfrak{m}|}2}(\eta  q^{\frac{\nu}2 })^{-\mathfrak{m}} \textrm{P.E.}[f_{\rm single}(q,a;\mathfrak{m})]
 \\	
 & \textrm{with } f_{\rm single} (q,a;\mathfrak{m}) := \frac{q^{\frac{1}6 + \frac{|\mathfrak{m}|}2 }a }{1-q} - \frac{q^{\frac{5}6 + \frac{|\mathfrak{m}|}2}a^{-1} }{1-q}\;.
 \end{split}
 \end{align}
 In the above, we use the superconformal $R$ charge of $\Phi$  given in \eqref{superconformal R-charge of minimal theory}.
 At the conformal point, $\nu=0$, the index becomes
 \begin{align}
 	\CI_{\CT_{\rm min}}^{\rm sci} \big{(}q, u,\nu=0;s=1 \big{)} =  1 -q + \left(\eta+\frac{1}\eta\right) q^{3/2}-2 q^2 +\left(\eta+\frac{1}\eta \right) q^{5/2} -2 q^3 +\ldots \;. \label{sci for Tmin}
 \end{align}
 The terms in $q^{3/2}$ come from the monopole operators in \eqref{monopole ops in Tmin} and the index is compatible with the claimed SUSY enhancement \cite{Evtikhiev:2017heo}. 
On the other hand, the index at the non-conformal point $\nu = \pm 1$ is 
 \begin{align}
 	\begin{split}
 		&\CI_{\CT_{\rm min}}^{\rm sci} \big{(}q, \eta,\nu,s=1 \big{)}\big{|}_{\nu \rightarrow \pm 1}  
 		\\
 		&= 1+ \left(-1+\eta^{\mp 1}\right) q+\left(-2+\eta +\frac{1}\eta\right) q^2 +\left(-2+\eta+\frac{1}\eta\right) q^3 +\ldots \;.
 	\end{split}
 \end{align}
 As anticipated from the superconformal multiplet analysis  in \eqref{triviality of SCI from supermultiplet analysis}, the index becomes ($q$-independent) 1 in the degenerate limits, $\nu\rightarrow \pm 1$ and $\eta \rightarrow 1$.  This reconfirm that the $\CN=4$ theory is of rank 0 and gives a strong signal that  a topological field theory emerges   in the limits.  

 \subsubsection{Lee-Yang TQFT in degenerate limits}
 Here we claim that
 \begin{align}
{\rm TFT}_{\pm }[\CT_{\rm min}] = (\textrm{Lee-Yang TQFT})
 \end{align}
 with explicit checks of the dictionaries in Table \ref{Table : Dictionaries}.  The non-unitary TQFT has following modular data
 \begin{align}
  S =  \begin{pmatrix}
   \sqrt{\frac{1}{10} \left(\sqrt{5}+5\right)} &  -\sqrt{\frac{1}{10} \left(5-\sqrt{5}\right)} \\
  -\sqrt{\frac{1}{10} \left(5-\sqrt{5}\right)}  & -\sqrt{\frac{1}{10} \left(\sqrt{5}+5\right)} 
  \end{pmatrix}\;, \quad  T =  \begin{pmatrix}
   1 &  0 \\
 0  & \exp (-\frac{2 \pi i }5)
  \end{pmatrix}\;. \label{S,T of Lee-Yang}
 \end{align}

\paragraph{Squashed three-sphere partition function}
 The squashed three-sphere partition function for the minimal theory is ($\hbar = 2\pi i b^2 , b \in \mathbb{R}$)
 \begin{align}
 \begin{split}
 &\mathcal{Z}^{S^3_b}_{\CT_{\rm min}}  (b,m,\nu)= \int\! \frac{dZ}{\sqrt{2\pi \hbar}} \,\, e^{-\frac{Z^2 + 2Z \left(m+ (i \pi +\frac{\hbar}2)\nu \right)}{2\hbar}} \psi_{\hbar}(Z)\;.
 \end{split} \label{S^3_b-ptn of minimal theory}
 \end{align}
 Here $\psi_\hbar$ is a special function called the quantum dilogarithm, for which readers are referred to Appendix \ref{App : QDL}.  The partition function in the limit $b\rightarrow 1$ was studied in \cite{Gang:2019jut} and one can check that
 \begin{align}
 \begin{split}
 &\Big{|}\mathcal{Z}^{S^3_b}_{\CT_{\rm min}}  (b,m=0,\nu\rightarrow \pm 1) \Big{|}  \xrightarrow{\quad b\rightarrow 1 \quad }  \sqrt{\frac{1}{10} \left(\sqrt{5}+5\right)}+\sum_{n=2}^\infty s_n (1-b)^n\;,
 \\
 & \textrm{with } s_n =0 \textrm{ up to arbitrary higher order $n$}\;.
 \end{split}
  \label{ZS3 for minimal theory in  limits}
 \end{align}
That the partition function becomes independent on the squashing parameter $b$ in the degenerate limit is a strong signal  that the theory  becomes topological. Furthermore, the partition function is identical to the $S^3$ partition function ($S_{00}$, see \eqref{S,T of Lee-Yang}) of the Lee-Yang TQFT,
 \begin{align}
  \Big{|}\mathcal{Z}^{S^3_b}_{\CT_{\rm min}}  (b,m=0,\nu = \pm 1) \Big{|}  = \Big{|}\mathcal{Z}^{S^3}_{\rm Lee-Yang}\Big{|}\;.
 \end{align}
The free energy $F = -\log   \big{|}\mathcal{Z}^{S^3_b}_{\CT_{\rm min}}  (b=1,m=0,\nu=0) \big{|}$ of the minimal theory, given in \eqref{F and CT of minimal theory}, nicely matches with $\max_{\alpha}  (-\log |S_{0\alpha}|)$, see \eqref{S,T of Lee-Yang} for the S-matrix,
 \begin{align}
 \left(F   \textrm{  of }\CT_{\rm min} \right)  =\left(  \max_{\alpha}  (-\log |S_{0\alpha}|) \textrm{  of Lee-Yang TQFT} \right)\;,
 \end{align}
 which confirms the dictionary for $F$ in Table \ref{Table : Dictionaries}.
 
 \paragraph{Perturbative invariants $\CS_n^\alpha$} The integrand in  \eqref{S^3_b-ptn of minimal theory} can be expanded as
 \begin{align}
 \begin{split}
 &\log \CI_{\hbar}(Z, m,\nu) =  \log \left( e^{-\frac{Z^2 + 2Z \left(m+ (i \pi +\frac{\hbar}2)\nu \right)}{2\hbar}} \psi_{\hbar}(Z)\right) 
 \\
 &\xrightarrow{\quad \hbar\rightarrow 0 \quad } \frac{1}\hbar \CW_0 (Z, m, \nu) +\CW_1 (Z, m, \nu)  + \ldots \quad \textrm{with }
 \\
 & \CW_0 = \textrm{Li}_2 (e^{-Z}) - \frac{Z^2}2  - Z (m+i\pi \nu)\;, \quad \CW_1 = -\frac{1}2 \log (1-e^{-Z}) - \frac{Z \nu}2\;.
 \end{split}
 \end{align}
 There are two Bethe-vacua \eqref{Bethe-vacua} determined by the following algebraic equation
 \begin{align}
 \textrm{Bethe-vacua of $\CT_{\rm min}$} \;:\; \Big{\{}z\;:\;  \frac{(z-1) e^{-m-i \pi  \nu }}{z^2} =1 \Big{\}}\;.
 \end{align}
 In the degenerate limits, $m=0$ and $\nu \rightarrow \pm 1$, the two Bethe-vacua approach following values 
 \begin{align}
  z_{\alpha =0} \rightarrow \frac{1}{2} \left(\sqrt{5}-1\right) \;, \quad z_{\alpha =1}  \rightarrow \frac{1}{2} \left(-\sqrt{5}-1\right)\;.  
 \end{align}
 Perturbative invariants  \eqref{perturbative invariants} of two saddle points associated to the two Bethe-vacua in the degenerate limits are
 \begin{align}
 \begin{split}
 & \CS^{\alpha =0 }_0 \rightarrow \frac{7 \pi ^2 }{30} \;, \quad \CS_1^{\alpha =0} \rightarrow -\frac{1}2 \log \left(\frac{5-\sqrt{5}}2 \right)\;, \quad \CS_2^{\alpha =0}  \rightarrow - \frac{7}{120}\;, 
 	\\
 &\CS^{\alpha =1 }_0 \rightarrow  -\frac{17 \pi ^2 }{30} \;, \quad \CS_1^{\alpha =1} \rightarrow -\frac{1}2 \log \left(\frac{5+\sqrt{5}}2 \right)\;, \quad \CS_2^{\alpha =1}  \rightarrow - \frac{7}{120}\;, 
 \\
 & \CS^\alpha_{n\geq 3}\rightarrow 0\;. 
 \end{split}
 \end{align}
That this is compatible with the expected properties in \eqref{perturbative invariants in the degenerate lmits} is  a highly non-trivial evidence for emergence of topological theory in the degenerate limits. 
 
 \paragraph{Fibering and Handle gluing } Using the formulae in   \eqref{Fibering} and \eqref{Handle gluing} combined with the above computation of $\CS^\alpha_n$,  we have
 \begin{align}
 \begin{split}
& \bigg{\{} \CF_{\alpha} (m=0, \nu \rightarrow  \pm 1, s=-1) \bigg{ \}}_{\alpha=0,1} \longrightarrow \;\left\{\exp \left(-\frac{7 i \pi  }{60} \right), \exp \left(\frac{17 i \pi }{60} \right) \right\}\;,
 \\
&\bigg{\{} \CH_{\alpha} (m=0, \nu \rightarrow  \pm 1, s=-1)  \bigg{\}}_{\alpha=0,1}\longrightarrow  \;\bigg{\{} \frac{5-\sqrt{5}}2 ,  \frac{5+\sqrt{5}}2 \bigg{\}}\;.
\end{split}
 \end{align}
 Since the $1/\sqrt{\CH_{\a=0}}$ is equal to  $|\CZ^{S^3_b}|$ at $\nu=\pm 1 $ in \eqref{ZS3 for minimal theory in  limits},  the $z_{\alpha=0}$ is indeed the Bethe-vacuum corresponding to the trivial object according to the criterion in  \eqref{Consistency for trivial vacuum}.  The computations above also confirm the dictionary for $S_{0\a}^{-2}$ and $T_{\alpha\beta}$ in Table \ref{Table : Dictionaries}, see \eqref{S,T of Lee-Yang} for modular matrices of Lee-Yang TQFT. 
 \paragraph{Supersymmetric loop operator} For a $U(1)$ gauge theory, the supersymmetric dyonic loop operator $\CO_{(p,q)}$ of (electric charge, magnetic charge)=$(p,q)$ is 
 \begin{align}
 	\CO_{(p,q)} = z^p(1-z^{-1})^{q}\;. 	
 \end{align}
The consistency condition in \eqref{Consistency for Bethe-to-loop} is met  when we choose the (Bethe vacua)-to-(loop operators) map as follow
\begin{align}
&\CO_{\alpha=0} = (\textrm{identity operator})\;, \quad  \CO_{\alpha=1} =\CO_{(p,q)=(1,0)} \;.
\end{align}
Then, using the dictionary in Table \ref{Table : Dictionaries}
\begin{align}
\begin{split}
&W_{\beta=0,1} (0)=1 \;, \quad W_{\beta=0} (1) = z_{0} = \frac{1}2 (\sqrt{5}-1)\;,  \quad  W_{\beta=1} (1) = z_{1} = \frac{1}2 (-\sqrt{5}-1)\;.
\end{split}
\end{align}
From $\CW_{\beta}(\alpha)$, one can compute the S-matrix using the formula $S_{\a \b }= S_{00} W_\beta (\a) W_{0}(\beta)$, and confirm that it is identical to the S-matrix of the Lee-Yang TQFT given in \eqref{S,T of Lee-Yang}.

\subsection{\texorpdfstring{$U(1)_1 + H$ : SUSY enhancement $\CN=3 \rightarrow \CN=5$}{U(1){1} + H : SUSY enhancement N=3 to N=5}}

\subsubsection{SUSY enhancement}
We define
\begin{align}
\begin{split}
(U(1)_k +H) &:= (\textrm{3D $\CN=4$ $U(1)$ gauge theory with CS level $k$}
\\
&\qquad \;  \textrm{ coupled to a hypermultiplet of charge $+1$})\;.
\end{split}
\end{align}
For non-zero $k$, the theory has  $\CN=3$ supersymmetry instead of $\CN=4$ since the CS term breaks some of the $\CN=4$ supersymmetry. The $\CN=3$ theory has the $U(1)_{\rm top}$ flavor symmetry associated to the dynamical $U(1)$ gauge theory. As pointed out in \cite{Garozzo:2019ejm,Beratto:2020qyk}
\begin{align}
\begin{split}
\textrm{For $k=1$},\; &\textrm{the $(U(1)_k +H)$ has enhanced $\CN=5$ supersymmetry at IR}
\\
& \textrm{and the resulting IR SCFT  is of rank 0 }.
\end{split}
\end{align}
The $U(1)_{\rm top} = SO(2)_{\rm top}$ symmetry becomes the $U(1)$ axial symmetry, which is a subgroup of  $SO(4) \subset SO(5)$ R-symmetry, in the supersymmetry enhancement. 
\begin{align}
SO(3)_R \times SO(2)_{\rm top} \xrightarrow{\quad \textrm{RG} \quad } SO(5)_R\;.
\end{align}
For $|k|=1$, there are two BPS monopole operators whose quantum numbers are
\begin{align}
	\begin{split}
	&A=+1, \;  j = \frac{1}2, \;R =1, \; \Delta  = \frac{3}2\;\textrm{ and}
	\\
	&A=-1, \;  j = \frac{1}2, \;R =1,\;  \Delta  = \frac{3}2\;. \label{monopole op in U(1)+H}
	\end{split}
\end{align}
Here $A$ is the charge of $U(1)_{\rm top}$ and $R \in \frac{\mathbb{Z}}2$ is the spin  of $SO(3)$ R-symmetry.  $j$ and  $\Delta$ are the Lorentz spin and the conformal dimension respectively. The    BPS operators belong to   extra SUSY-current multiplet \cite{Cordova:2016emh}  of $\CN=3$ superconformal algebra, which  consists of the following conformal primaries and their descendants
\begin{align}
\begin{split}
	&[(R, j , \Delta) = (0,0,1)] \xrightarrow{\quad \mathbf{Q} \quad } 	[(R, j , \Delta) = ( 1,\frac{1}2,\frac{3}2)]  \xrightarrow{\quad \mathbf{Q} \quad } 
	\\
	& [(R, j , \Delta) = (1,1,2)] \oplus  [(R, j , \Delta) = (0,0,2)] \xrightarrow{\quad \mathbf{Q} \quad }  [(R, j , \Delta) = (0,\frac{3}2,\frac{5}2)] \;.
\end{split}
\end{align}
Here $\mathbf{Q}=(Q_1, Q_2, Q_3)$ are the $\CN=3$ supercharges. The local operators in the top component of the supermultiplet with $(R, j, \Delta)= (0, \frac{3}2,\frac{5}2)$ correspond to the conserved current for extra supersymmetry, and thus  the existence of the multiplet implies the supersymmetry enhancement \cite{Garozzo:2019ejm,Beratto:2020qyk}.
 
\paragraph{Superconformal index } Using the localization summarized in \ref{App : review on localization}, the index is given as 
\beq
 &\CI^{\rm sci}_{\U(1)_{k}+H} (q,\eta,\nu;s=1)\nonumber
\\
= &\sum_{\mathfrak{m} } \oint_{|a|=1}\frac{da}{2\pi i a}q^{\frac{|\mathfrak{m}|}{4}}((-1)^{\mathfrak{m}}a)^{k \mathfrak{m}}(\h q^{\frac{1}{2}\n})^{-\mathfrak{m}}\textrm{P.E.}[f_{\rm single} (q, \mathbf{a}, \eta ; \mathfrak{m})]\;, \label{index for U(1)+H}
\eeq
where $\h$ is a fugacity of $\U(1)_{\rm top}$ and
\beq
f_{\rm single}(q,\mathbf{a}, \eta;\mathfrak{m}):= \frac{q^{\frac{1}{4}+\frac{1}2 | \mathfrak{m}|}(a+\frac{1}a)}{1-q}  - \frac{q^{\frac{3}{4}+\frac{1}2  | \mathfrak{m}|}(a+\frac{1}a)}{1-q} \;.
\eeq
Using the above expression, the index can be evaluated and we find\footnote{The same computation was done in \cite{Beratto:2020qyk,Garozzo:2019ejm}. }
\beq
 &\CI^{\rm sci}_{\U(1)_{k}+H} (q,\eta,\nu;s=1)\nonumber
\\
&=\begin{cases}
1+q^{1/2}+\left(-\eta -\frac{1}{\eta }-1\right) q+\left(\eta +\frac{1}{\eta }+2\right) q^{3/2}+\ldots , \quad   |k|=1
\\
1+q^{1/2}- q+\ldots \;, \quad |k|>1
\end{cases}  
\eeq
The term $(-\eta -\frac{1}\eta ) q$ comes from the monopole operators  \eqref{monopole op in U(1)+H} in extra SUSY-current multiplet and implies the SUSY enhancement. Note that the SUSY enhancement occurs only at $|k|=1$. 

In the degenerate limit $\n=\pm1$ and $\h=1$, the index becomes
\beq
 &\CI^{\rm sci}_{\U(1)_{k}+H} (q,\eta=1,\nu=\pm1;s=1)\nonumber
\\
&=\left\{\begin{array}{lr}
1+\left(1-\eta^{\mp 1}\right) q^{1/2}+\left(\eta^{\mp 1}-1\right) q +\ldots \big{|}_{\eta =1} = 1\;, &\ \quad  (|k|=1)\\
\textrm{Non-trivial power series in }q^{1/2}\;, & \quad (|k|>1) \\
\end{array}\right..
\eeq
It is compatible the expectation that the theory is a $\CN=4$ (actually $\CN=5$) SCFT of rank 0  when $|k|=1$. It also implies that  there emerge  non-unitary TQFTs in the degenerate limits only when $|k|=1$.  The non-unitary TQFTs are expected to be fermionic according to \eqref{criterion of spin/non-spin}. 

\subsubsection{Non-unitary TQFTs in  degenerate limits}
Here we extract modular data of $\textrm{TFT}_{\pm}[U(1)_{k=1} +H]$ by computing various  supersymmetric partition functions. 
\paragraph{Squashed three-sphere partition function} It can be written as (see Appendix \ref{App : review on localization})

\beq
\begin{split}
&\CZ^{S^3_b}_{\U(1)_k+H} (b, m, \nu) =  \int \!\frac{\text{d}Z}{\sqrt{2\pi \hbar}}\,\, \CI_\hbar (Z,m,\nu)\; \textrm{ with }
\\ 
& \CI_\hbar (Z,m,\nu) = \exp \left( \frac{k Z^2}{2 \hbar} \right)  \exp \left( -\frac{ZW}{ \hbar} \right)\bigg{|}_{W = m +(\pi i + \frac{\hbar}2)\nu} 
\!\!
\prod_ {\epsilon_1\in   \{\pm 1\}}\Psi_{\hbar}\Big[ \epsilon_1 Z+\frac{\pi i}{2}+\frac{\hbar}{4}\Big] \;.
\end{split} \label{S3b for U(1)+H}
\eeq
Here $\Psi_\hbar  (X):= \psi_\hbar( X)\exp (\frac{X^2}{4\hbar})$ as defined in \eqref{Bigpsi}. 
In a round sphere limit ($b=1$)  with $k=1$ and $m=0$, the integral reduces to (using Appendix \ref{App : QDL})

\beq
\begin{split}
\CZ^{S^3}_{\U(1)_1+H} (b=1, m=0, \nu)
=
 \frac{e^{i \delta}}{4\p} \int \text{d}Z \frac{ e^{\frac{k Z^2}{4\p i}} e^{- Z \nu} }{\cosh(Z/2)}\;.
\end{split}
\label{U(1)_1+H round shpere}
\eeq
Here $e^{i \delta}$ is an unimportant phase factor sensitive to  local counterterms. For $\nu=0$, this integration can be exactly evaluated  by applying the residue theorem to the  integral along the  contour depicted in  Fig \ref{U(1)_1+H contour}:
\begin{figure}[H]
    \centering
    \begin{tikzpicture}
[decoration={markings,
	mark=at position 0.13 with {\arrow[line width=1pt]{>}},
	mark=at position 0.3 with {\arrow[line width=1pt]{>}},
	mark=at position 0.5 with {\arrow[line width=1pt]{>}},
	mark=at position 0.7 with {\arrow[line width=1pt]{>}},
	mark=at position 0.9 with {\arrow[line width=1pt]{>}}
}
]
\draw[help lines,->] (-5,0) -- (5,0) coordinate (xaxis);
\draw[help lines,->] (0,-0.5) -- (0,1.5) coordinate (yaxis);

\path[draw,line width=0.8pt,postaction=decorate] 
(4,0) node[below] {$\infty$} -- (4,1) -- (0.3,1)  arc (0:180:0.3) -- (-4,1) -- (-4,0) node[below] {$-\infty$} -- (4,0);

\node at (0,1) {$\times$};

\node[below] at (xaxis) {Re$Z$};
\node at (0,1.8) {Im$Z$};
\node at (0.3,0.7) {$\pi i$};
\node at (0.7,-0.3) {$C_1$};
\node at (2,1.3) {$C_2$};
\node at (-0.6,1.4) {$A_1$};
\node at (-2,1.3) {$C_3$};
\end{tikzpicture}
    \caption{A contour for the evaluation of \eqref{U(1)_1+H round shpere}. There is a simple pole at $Z = \pi i$ inside the contour.}
    \label{U(1)_1+H contour}
\end{figure}
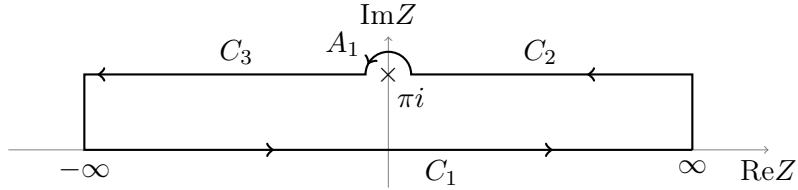
\begin{align}
    \int_{-\infty}^{\infty} \text{d} Z \frac{e^{\frac{Z^2}{4 \pi i}}}{\cosh(Z/2)}
&+  \int_{\infty}^{-\infty} \text{d} Z \frac{e^{\frac{Z^2}{4\pi i}} e^{\frac{Z}{2} e^{\frac{\pi i}{4}}}}{i \sinh (Z/2)} + \pi i (-2 i e^{\frac{\pi i }{4}}) = 2\pi i (-2 i e^{\frac{\pi i }{4}})\;,
\nonumber\\
\Rightarrow
\int_{-\infty}^{\infty} \text{d} Z \frac{e^{\frac{Z^2}{4 \pi i}}}{\cosh(Z/2)}
& = 2\pi e^{\frac{\pi i}{4}} + e^{-\frac{\pi i}{4}}\int_{-\infty}^{\infty} \text{d}Z \frac{e^{\frac{Z^2}{4\pi i}}e^{\frac{Z}{2}}}{\sinh(Z/2)}
\nonumber\\
& = 2\pi e^{\frac{\pi i}{4}} + 
\frac{e^{-\frac{\pi i}{4}}}{2}\Bigg( \int_{-\infty}^{\infty} \text{d}Z \frac{e^{\frac{Z^2}{4\pi i}}e^{\frac{Z}{2}}}{\sinh(Z/2)}
-\int_{-\infty}^{\infty} \text{d}Z \frac{e^{\frac{Z^2}{4\pi i}}e^{-\frac{Z}{2}}}{\sinh(Z/2)}
\Bigg)
\nonumber\\
& = 2\pi e^{\frac{\pi i}{4}} 
+ e^{-\frac{\pi i}{4}} \int_{-\infty}^{\infty} \text{d}Z
 e^{\frac{Z^2}{4\pi i}}
\nonumber\\
& = 2\pi ( e^{\frac{\pi i}{4}} + e^{-\frac{\pi i}{2}}) = 4\pi e^{-\frac{\pi i}{8}} \sin\Big(\frac{\pi}{8}\Big)\;.
\end{align}
Here, the residue at the simple pole $Z=\pi i$ is $-2 i e^{\frac{\pi i}{4}}$. The first, second, and third term in the first line comes from the path $C_1$, $C_2 + C_3$, and an arc $A_1$ respectively. At the third equality, we have used the changing variable as $Z \rightarrow -Z$ from the last term in the second line.

Likewise, the integration for $\nu=\pm 1$ can be computed exactly in a similar way by using the same contour and we found
\begin{align}
    \int_{-\infty}^{\infty} \text{d}Z \frac{e^{\frac{Z^2}{4\pi i}} \, e^{-Z}}{\cosh(Z/2)}
    =
    4\pi e^{-\frac{5\pi}{8}}\sin\Big(\frac{3\pi}{8}\Big)\;.
\end{align}
Restoring the overall factor $1/(4\pi)$  in \eqref{U(1)_1+H round shpere}, we finally  have
\begin{align}
\begin{split}
&\exp(-F) = \Big{|}	\CZ^{S^3_b}_{U(1)_1+H}(b=1, m=0, \nu=0) \Big{|} = \sin \left( \frac{\pi}{8} \right) = (4+2\sqrt{2})^{-1/2}\;,
\\
&(S_{00} \textrm{ of  TFT}_\pm )= \Big{|}	\CZ^{S^3_b}_{U(1)_1+H}(b=1, m=0, \nu\rightarrow \pm 1) \Big{|} = \sin \left( \frac{3\pi }{8} \right)= (4-2\sqrt{2})^{-1/2}\;.
\label{F and S00 of U(1)_1+H}
\end{split}
\end{align}
The partition functions in the degenerate limits $\nu \rightarrow \pm 1$ are actually independent on the squashing parameter $b$, as we will check it perturbatively in \eqref{Perturbative invariants for U(1)+H},  and equality in the 2nd line holds  for arbitrary $b\in \mathbb{R}$.

\paragraph{Perturbative invariants $\CS_n^\alpha$} The integrand in  \eqref{S3b for U(1)+H} at $k=1$, after a  shift $Z\to Z+\frac{1}{2}(\p i+\frac{\hbar}2)$ of dummy integral variable,  can be expanded as
\begin{align}
	\begin{split}
		&\log \CI_{\hbar}(Z, m,\nu)|_{Z\rightarrow Z+\frac{1}{2}(\p i+\frac{\hbar}2) }\xrightarrow{\quad \hbar\rightarrow 0 \quad } \frac{1}\hbar \CW_0 (Z, m, \nu) +\CW_1 (Z, m, \nu)  + \ldots \quad \textrm{with }
		\\
		& \CW_0 (m=0) = \textrm{Li}_2 (e^{-Z+i \pi}) +\textrm{Li}_2 (e^{Z})+Z^2 +\pi i (1-\nu)Z +c_0 \pi^2 \;,
		\\
		& \CW_1 (m=0) = \frac{1}{2} \left(-i \pi  \nu -\nu  Z+Z-\log \left(1-e^Z\right)\right) +i  c_1 \pi \;,
	\end{split}
\end{align}
where $c_0$ and $c_1$ are $Z$-independent rational numbers.  There are two Bethe-vacua \eqref{Bethe-vacua} determined by a following algebraic equation
\begin{align}
	\textrm{Bethe-vacua of ($U(1)_1+H$) at $m=0$} \;:\; \Big{\{}z\;:\;  \frac{(-1)^\nu z (z+1)}{z-1}  =1 \Big{\}}\;.
\end{align}
In the degenerate limits, $m=0$ and $\nu \rightarrow \pm 1$, the two Bethe-vacua approach following values 
\begin{align}
	z_{\alpha =0} \rightarrow \left(\sqrt{2}-1\right), \quad z_{\alpha =1}  \rightarrow  \left(-\sqrt{2}-1\right)\;. 
\end{align}
Perturbative invariants  \eqref{perturbative invariants} of two saddle points associated to the two Bethe-vacua in the degenerate limits are
	\begin{align}
		\begin{split}
			& \CS^{\alpha =0 }_0 \rightarrow - \frac{11 }{12}  \pi ^2 \;, \quad \CS_1^{\alpha =0} \rightarrow -\frac{\pi i }{8} -\frac{1}2 \log \left(4-2 \sqrt{2} \right)\;, 
			\\
			&\CS^{\alpha =1 }_0 \rightarrow   \frac{1 }{12}  \pi ^2  \;, \quad \CS_1^{\alpha =1} \rightarrow -\frac{\pi i }{8} -\frac{1}2 \log \left(4+2 \sqrt{2} \right)\;, 
			\\
			& \textrm{Im}[\CS^\alpha_{n=2}] , \;\CS^\alpha_{n\geq 3}\rightarrow 0\;. \label{Perturbative invariants for U(1)+H}
		\end{split}
	\end{align}
That this is compatible with the expected properties in \eqref{perturbative invariants in the degenerate lmits} is  a highly non-trivial evidence for emergence of topological theory in the degenerate limits. 

 \paragraph{Fibering and Handle gluing } Using the formulae in   \eqref{Fibering} and \eqref{Handle gluing} combined with the above computation of $\CS^\alpha_n$,  we have
\beq
& \bigg{\{} \CF_{\alpha} (m=0, \nu \rightarrow  \pm 1, s=-1) \bigg{ \}}_{\alpha=0,1} \longrightarrow \;\left\{\exp \left(\frac{11 i \pi  }{24} \right), \exp \left(-\frac{ i \pi }{24} \right) \right\}\;, \nonumber
\\
& \bigg{\{} \CH_{\alpha} (m=0, \nu \rightarrow  \pm 1, s=-1) \bigg{ \}}_{\alpha=0,1} \longrightarrow \left\{ \left(4-2\sqrt{2}\right),\left(4+2\sqrt{2}\right)\right\}.
\eeq
Since the $1/\sqrt{\CH_{\a=0}}$ is equal to  $|\CZ^{S^3_b}|$ at $\nu=\pm 1 $ in \eqref{F and S00 of U(1)_1+H},  the $z_{\alpha=0}$ is indeed the Bethe-vacuum corresponding to the trivial object according to the criterion in \eqref{Consistency for trivial vacuum}. 

\paragraph{Supersymmetric loop operators} The consistency condition in \eqref{Consistency for Bethe-to-loop} is met  when we choose the (Bethe vacua)-to-(loop operators) map as follows:
\begin{align}
	&\CO_{\alpha=0} = (\textrm{identity operator})\;, \quad  \CO_{\alpha=1} =\CO_{(p,q)=(1,0)} \;.
\end{align}
Then, using the dictionary in Table \ref{Table : Dictionaries},
\begin{align}
	\begin{split}
		&W_{\beta=0,1} (0)=1 \;, \quad W_{\beta=0} (1) = z_{0} = \sqrt{2}-1\;,  \quad  W_{\beta=1} (1) = z_{1} = -\sqrt{2}-1\;.
	\end{split}
\end{align}
From $\CW_{\beta}(\alpha)$, one can compute the S-matrix using the formula $S_{\a \b }= S_{00} W_\beta (\a) W_{0}(\beta)$ and the result is
\begin{align}
S = \begin{pmatrix}
\sin  \frac{3\pi}8& \sin \frac{\pi}8 \\
 \sin \frac{\pi}8   & - \sin \frac{3\pi}8  \\ 
\end{pmatrix}\;.
\end{align}
Since the TQFTs, TFT$_\pm [U(1)_1+H]$, are spin TQFTs, only the modular $T^2$ matrix is well-defined and according to the dictionary in Table \ref{Table : Dictionaries}
\begin{align}
T^2 = \begin{pmatrix}
	1& 0 \\
	0  & -1 \\ 
\end{pmatrix}\;.
\end{align}
The modular data ($S$ and $T^2$) of the spin non-unitary TQFT, TFT$_\pm [U(1)_1+H]$, is identical to that of  $\textrm{Gal}_{d} (SU(2)_{6})/\mathbb{Z}_2^f$ with $d = \frac{1}2 \sin\left( \frac{3\pi}8\right)$.

\subsection{\texorpdfstring{$SU(2)_{k}^{\frac{1}2 \oplus \frac{1}2}$  : $\CN=5$ theory}{SU(2){k}{1/2 plus 1/2}  : N=5 theory} }
The theory is defined  as
\begin{align}
\begin{split}
&\left( SU(2)_{k}^{\frac{1}2 \oplus \frac{1}2} \right)  \;\; 
\\
&:=\textrm{$SU(2)$ gauge theory coupled to  a half hypermultiplet and  a half twisted-hypermultiplet }
\\
& \textrm{ \quad\; in fundamental representations with Chern-Simons level $k$}\;. \label{SU(2)k1/2+1/2}
\end{split}
\end{align}
The theory is can be regarded as  a special case of $O(M)\times Sp(2N)$ type quiver theories (with $M=N=1$) which have $\CN=5$ supersymmetry \cite{Hosomichi:2008jb}.

\subsubsection{IR phases}

\paragraph{Superconformal index} The superconformal index \eqref{Def : superconformal index} of the  $\CN=5$ theory is 
\begin{align}
	\begin{split}
		&\CI^{\rm sci}_{SU(2)_k^{\frac{1}2\oplus \frac{1}2}} (q,\eta, \nu; s=1) 
		\\
		&=\sum_{\mathfrak{m}=0}^{\infty} \oint_{|a|=1} \frac{da}{2\pi i a}   \Delta (\mathfrak{m},a)  a^{2k \mathfrak{m}} q^{\frac{|\mathfrak{m}|}2} \textrm{P.E.}\big{[}f_{\rm single}(q, \eta,a; \nu,\mathfrak{m})\big{]}\;. \label{{index for SU(2)+H-1}}
	\end{split}
\end{align}
Here we define
\begin{align}
	\begin{split}
		&f_{\rm single} (q, \eta,a; \nu,\mathfrak{m}) := \frac{q^{\frac{1}4 + \frac{|\mathfrak{m}|}2}}{1+q^{\frac{1}2}} (a+a^{-1})\left((\eta q^{1/2 \nu})^{\frac{1}2}+(\eta q^{1/2 \nu})^{-\frac{1}2} \right)\;,
		\\
		& \Delta (\mathfrak{m},a)  :=   \frac{1}{\textrm{Sym}(\mathfrak{m})} q^{-|\mathfrak{m}|} (1-a^2 q^{|\mathfrak{m}|}) (1-a^{-2} q^{|\mathfrak{m}|})\;,
		\\
		& \textrm{with }  \textrm{Sym}(\mathfrak{m}) :=  \begin{cases}
			2\; \quad  \textrm{if $\mathfrak{m}=0$}\;,\\
		    1\;  \quad \textrm{if $\mathfrak{m}>0$}\;.  \label{{index for SU(2)+H-2}}
		\end{cases}
	\end{split}
\end{align}
Using the formula, one can compute the superconformal index and check that
\begin{align}
	\begin{split}
		&\CI^{\rm sci}_{SU(2)_k^{\frac{1}2 \oplus \frac{1}2}} (q,\eta,\nu=0;s=1) 
		\\
		&=  \begin{cases}
		    \infty\;  \quad \textrm{if $k=0$}\;
			\\
			1\; \quad \textrm{if $|k|=1$}\;
			\\
			1+q^{\frac{1}2} + \left(-1-\frac{1}\eta-\eta\right) q + \left(2+ \eta +\frac{1}\eta \right)q^{\frac{3}2}+\ldots\; \;\;\textrm{if $|k|\geq 2$}\;. 
		\end{cases}
	\end{split}
\end{align}
The higher order terms (represented by $\ldots$) depend on $k$ for $|k|\geq 2$. From the computation, one can determine the basic property of theory appearing at the IR. The triviality of the index for $|k|=1$ implies that the theory has a mass gap and the IR physics is described by an unitary TQFT. The divergence of the index is a signal of emergence of a free chiral  theory that decouples with the other part of the theory \cite{Yaakov:2013fza}. The non-triviality of the index implies that the theory flows to a superconformal field theory.  In summary, from the index computation we  conclude that
\begin{align}
	\begin{split}
		&SU(2)_k^{\frac{1}2 \oplus \frac{1}2} \xrightarrow{\rm  \quad  at \; IR \quad  }  \begin{cases}
		   \textrm{Contains decoupled free chirals}\;  \quad \textrm{if $|k|=0$}\;,
			\\
		    \textrm{Unitary TQFT}\;  \quad \textrm{if $|k|=1$}\;,
			\\
			\textrm{3D $\mathcal{N}=5$ SCFT}\; \;\;\textrm{if $|k|\geq 2$}\;.
		\end{cases}
	\end{split}
\end{align}
In the degenerate limits, $\nu\rightarrow \pm 1$ and $\eta \rightarrow 1$, the index becomes  (for $|k| \geq 2$)
\begin{align}
\begin{split}
&\CI^{\rm sci}_{SU(2)_k^{\frac{1}2 \oplus \frac{1}2}} (q,\eta,\nu = \pm 1;s=1)  = 1+(1-\eta^{\mp 1}) q^{1/2} -(\eta^{\mp 1}-1) q+(2-\eta^{\pm 1} - \eta^{\mp2})q^{3/2}+\ldots \;, 
\\
&\CI^{\rm sci}_{SU(2)_k^{\frac{1}2 \oplus \frac{1}2}} (q,\eta=1,\nu = \pm 1;s=1)  = 1\;.
\end{split}
\end{align}
It implies that  the theory is of rank 0 and non-unitary TQFTs, $\textrm{TFT}_\pm[SU(2)_k^{\frac{1}2 \oplus \frac{1}2}]$, emerges when $|k| \geq 2$ in the limits. In addition, we expect that they are spin  TQFTs according to \eqref{criterion of spin/non-spin}. The  $\textrm{TFT}_\pm [SU(2)_k^{\frac{1}2 \oplus \frac{1}2}]$ in the Table \ref{table:non-unitary/SCFT} is indeed a  spin TQFT since it is given by a $\mathbb{Z}_2^f$ quotient (fermionic anyon condensation) of a bosonic TQFT. 

\subsubsection{Non-unitary TQFTs in  degenerate limits}

\paragraph{Squashed three-sphere partition function} The partition function of the  $\CN=5$ theory  is
\begin{align}
\begin{split}
&\CZ^{S^3_b}_{SU(2)_k^{\frac{1}{2}\oplus \frac{1}{2}}} (b, m, \nu) =  \int \frac{\text{d}Z}{\sqrt{2\pi \hbar}} \CI_\hbar (Z,m,\nu)\; \textrm{ with }
\\ 
& \CI_\hbar (Z,m,\nu) = \frac{1}2
\big{(}2\sinh(Z)\big{)}\bigg{(}2\sinh\Big{(}\frac{2\pi i Z}{\hbar}\Big{)}\bigg{)}  \exp \left( \frac{k Z^2} \hbar \right)
\\
&\qquad \qquad \qquad \times \prod_ {\epsilon_1, \epsilon_2 \in   \{\pm 1\}}\Psi_{\hbar}\Big[ \epsilon_1 Z+ \epsilon_2 \frac{m+\nu ( i \pi +\frac{\hbar}2)}2+\frac{\pi i}{2}+\frac{\hbar}{4}\Big] \;.
\end{split}
\end{align}
At $b=1$ and $m=0$, the localization integral is simplified as
\begin{align}
	\CZ^{S^3_b}_{SU(2)_k^{\frac{1}{2}\oplus \frac{1}{2}}}(b=1, m=0, \nu)
	=
	\frac{e^{\frac{\pi i}{12}}}{2\pi }
	\int
	\text{d}Z
	\frac{\sinh^2(Z)}{\cosh(Z)+\cosh(i \pi \nu)}
	\exp \left(\frac{k Z^2}{2\pi i}\right)
	.
\label{Simplified Z_su(2)_k half+half}
\end{align}
At $\nu= 0, \pm 1$, the partition function is exactly computable and we obtain
\begin{align}
\begin{split}
&\exp(-F) = \Big{|}	\CZ^{S^3_b}_{SU(2)_k^{\frac{1}{2}\oplus \frac{1}{2}}}(b=1, m=0, \nu=0) \Big{|} = \sqrt{\frac{2}{|k|}} \sin \left( \frac{\pi}{4|k|} \right)\;,
\\
&(S_{00} \textrm{ of  TFT}_\pm )= \Big{|}	\CZ^{S^3_b}_{SU(2)_k^{\frac{1}{2}\oplus \frac{1}{2}}}(b=1, m=0, \nu\rightarrow \pm 1) \Big{|} = \sqrt{\frac{2}{|k|}} \sin \left( \frac{\pi (2|k|-1)}{4|k|} \right)\;. 
\label{F and S00 of SU(2)+1/2+1/2}
\end{split}
\end{align}
\paragraph{Bethe-vacua and Handle gluing operators  in the degenerate limits } Since there is a $\mathbb{Z}_2$ symmetry $(m,\nu) \leftrightarrow (-m,-\nu)$, we only consider the limit $\nu\rightarrow  1$. 
In the  limit $\nu\rightarrow 1$, the asymptotic expansion coefficients  $\CW_0$ and $\CW_1$ of the integrand are
\begin{align}
\begin{split}
&
\log \CI_\hbar (Z,m, \nu) \xrightarrow{\quad \hbar \rightarrow 0 \quad } \frac{1}\hbar \CW_0 (Z, m, \nu) + \CW_1 (Z,m,\nu)+ O (\hbar)\; \textrm{ with }
\\
&\mathcal{W}_0(Z,m ,\nu=1) =-\frac{\pi^2}{2} + \frac{\pi i m}2  + \frac{m^2}4 \pm  2\pi i Z + (k+1)Z^2 
+ \sum_{\epsilon_1, \epsilon_2 \in  \{\pm 1\} }\text{Li}_2(e^{- \epsilon_1 Z -\epsilon_2 \frac{m}2})\;,
\\
&\mathcal{W}_1(Z,m, \nu=1) = \frac{1}{2}
\Big(
\pi i + \frac{m}2 - \log(1-e^{\frac{m}2-Z}) - \log(1-e^{\frac{m}2+Z})
\Big)+ \log(\sinh(Z))\;.
\end{split}
\end{align}
The Bethe vacua equation at $\nu=1$ is
\begin{eqnarray}
\exp(\partial_Z \mathcal{W}_0(Z,m, \nu=1))\bigg|_{Z\rightarrow \log(z),m \rightarrow \log(\eta) }
=
\frac{(\sqrt{\eta}-z)(\sqrt{\eta} z+1)}{(\sqrt{\eta}+z)(\sqrt{\eta} z-1)}z^{2k}=1\;.
\label{SU2k1212vacua}
\end{eqnarray}
In the degenerate limit $\eta \rightarrow 1 $,  the equation simplifies as  $z^{2k}=-1$ and there are $|k|$ Bethe-vacua  after taking into  account  of the  Weyl $\mathbb{Z}_2$ quotient, $z\leftrightarrow 1/z$,
\begin{eqnarray}
\textrm{Bethe-vacua  : }z_{\alpha} = (-1)^{\frac{2k-2\alpha+1}{2|k|}} \;,\quad \alpha = 1,\cdots,|k|\;.
\label{su21212sol}
\end{eqnarray}
Now, the handle gluing in the degenerate limit, $m=0,\nu=1$, is given by
\begin{align}
\mathcal{H}(z) 
&=  \exp \left( -2\CS_1^\alpha  \right)|_{Z\rightarrow \log(z),\nu\rightarrow 1, m\rightarrow 0 } = 
 \frac{1}{4}e^{-2 \mathcal{W}_1}\partial_Z\partial_Z\mathcal{W}_0\Big|_{Z\rightarrow \log(z),\nu\rightarrow 1, m\rightarrow 0 } 
\nonumber
\\
&=\frac{2k z}{(z+1)^2}\;. \label{su2hdz}
\end{align}
The factor $1/4$ comes from $1/|\textrm{Weyl}(SU(2))|^2$, see \eqref{S0 and S1}.
By plugging the eq.(\ref{su21212sol}) in the eq.(\ref{su2hdz}), we have
\begin{eqnarray}
\left(\mathcal{H}_{\alpha} \textrm{ of }  {SU(2)_k^{\frac{1}{2}\oplus \frac{1}{2}}}  \right)
&=&
\Bigg(
\sqrt{\frac{2}{|k|}}\sin\left(\frac{\pi(2\alpha-1)}{4|k|}\right)
\Bigg)^{-2}, \quad \alpha =1, \ldots, |k|\;.
\end{eqnarray}
The set of $\{|S_{0\alpha}| = (\mathcal{H}_{\alpha})^{-1/2}\}$ is identical to  the  set $\{|S_{0\alpha}|\}$ of the  $SU(2)_{4k-2}/\mathbb{Z}_{2}^{f}$ theory. It implies that the non-unitary TQFT $\textrm{TFT}_{\pm} [SU(2)^{\frac{1}2 \oplus \frac{1}2}_k ]$ is a Galois conjugate of  $SU(2)_{4|k|-2}/\mathbb{Z}_{2}^{f}$ with $S_{00}$ in \eqref{F and S00 of SU(2)+1/2+1/2} . 

\subsection{\texorpdfstring{$T[SU(2)]_{k_1, k_2}$ and $T[SU(2)]_{k_1, k_2}/\mathbb{Z}_2$}{T[SU(2)](k1, k2) and T[SU(2)](k1, k2)/Z(2)}}

$T[SU(2)]$ is the 3D theory living on the  S duality domain wall in 4D $\mathcal{N}=4$ SYM  \cite{Gaiotto:2008ak}.  The  theory is the 3d $\mathcal{N}=4$ SQED with two fundamental hyper-multiplets. Let the four $\mathcal{N}=2$ chiral fields in the two $\mathcal{N}=4$ hyper-multiplets be  $q_1,q_2,q_3,q_4$ and the adjoint $\mathcal{N}=2$ chiral field in the $\mathcal{N}=4$ vector multiplet be $\phi_0$. The theory has  $SU(2)^H\times SU(2)^C$ flavor symmetry  at the IR as well as the $SO(4)$ R-symmetry.  The charge assignments for chiral fields under the Cartan subalgebra of the gauge and global symmetries are:

\begin{table}[ht]
	\centering
	\begin{tabular}{|c|c|c|c|c|c|}
		\hline
		& $q_1$ &  $q_2$  & $q_3$  & $q_4$ &  $\phi_0$ \\
		\hline
		$U(1)_{\text{gauge}}$ & $1$ & $1$ & $-1$ & $-1$ & $0$ \\
		$U(1)^H$   & $1$ & $-1$ & $1$ & $-1$ & $0$ \\
		$U(1)_{\text{axial}}$ & $\frac{1}2$ & $\frac{1}2$ & $\frac{1}2$ & $\frac{1}2$ & $-1$ \\
		$U(1)^{C}$   & $0$ & $0$ & $0$ & $0$ & $0$ \\
		\hline
	\end{tabular}
	\caption{Charge assignment in $T[SU(2)]$ theory. $U(1)^H,U(1)^C$ and $U(1)_{\rm axial}$ denote the Cartans of $SU(2)^H,SU(2)^C$ and axial $U(1) \subset SO(4)_R$  symmetry respectively. }
\end{table}
By gauging the two $SU(2)$s with non-zero Chern-Simons level $k_1$ and $k_2$, we obtain infinitely many rank 0 3D $\mathcal{N}=4$ SCFTs which will be denoted as
\begin{align}
\begin{split}
&T[SU(2)]_{k_1, k_2} :=\frac{T[SU(2)]}{SU(2)^H_{k_1}\times SU(2)^C_{k_2}}  
\\
&:=\left( \textrm{Gauging  $SU(2)^{H}\times SU(2)^C$ of $T[SU(2)]$ with Chern-Simons levels $k_1$ and $k_2$} \right)\;.
\end{split}
\end{align}
As argued in \cite{Gang:2018huc}, the gauging does not break the supersymmetry down to $\CN=3$ thanks to the nilpotent property of the moment map operators, $\vec{\mu}^H$ and $\vec{\mu}^C$, of the  two $SU(2)$s. 

The theory has $\mathbb{Z}^H_2\times \mathbb{Z}^C_2$ one-form symmetry originating from the center symmetry $\mathbb{Z}_2\times \mathbb{Z}_2$ of the  $SU(2)^H \times SU(2)^C$ gauge group. The discrete one-form symmetry has 't Hooft anomaly characterized  by the following bulk action\footnote{The anomaly can be interpreted as a dependence  of the partition function $\CZ^{\CM_3 = \partial X_4}$ on  the choice of a 4-manifold $X_4$ having $\CM_3$ as a boundary. The difference between two partition functions, $\CZ^{\CM_3 = \partial X_4}$ and  $\CZ^{\CM_3 = \partial Y_4}$, with two choices of 4-manifolds,  $X_4$ and $Y_4$,  is determined by the  bulk action $S_{\rm anom}$ as  $\frac{\CZ^{\CM_3 = \partial X_4}}{\CZ^{\CM_3 = \partial Y_4}} = e^{i S_{\rm anom}[\CM_4]}$. Here $\CM_4 = X_4 \bigcup \overline{Y_4}$ is a closed 4-manifold obtained by gluing $X_4$ and orientation reversed $Y_4$ along the common boundary $\CM_3$. } 
\begin{align}
S_{\rm anom} = \pi \int_{\CM_4} \left( k_1 \frac{\CP (w_2^H)}2 + k_2 \frac{\CP (w_2^H)}2  + w_2^H \cup w_2^C\right) \quad \textrm{(mod $2\pi$)} \;. \label{anomaly poly for T[SU(2)]-k1k2}
\end{align}
Here $w_2^H$ ($w_2^C$) is the 2nd Stiefel-Whitney class, valued in $H^{2}(\CM_4, \mathbb{Z}_2)$, of the $SO(3)_H = (SU(2)_H)/\mathbb{Z}_2$ and $SO(3)_C  =  SU(2)_C/\mathbb{Z}_2$ bundle respectively.  $\CP$ is the Pontryagin square operation,
\begin{align}
\CP\;:\; \; H^2 (\CM_4, \mathbb{Z}_2) \rightarrow H^4 (\CM_4, \mathbb{Z}_2)\;,
\end{align}
which satisfies $\CP(w_2) = w_2^2 \;(\textrm{mod }2)$. On spin manifold $\CM_4$, the $\frac{1}2 \CP(\omega) \in \mathbb{Z}$.
The first two terms in \eqref{anomaly poly for T[SU(2)]-k1k2} come from the Chern-Simons action of two $SU(2)$ gauge fields \cite{Benini:2017dus} while the last term is from the anomaly polynomial of $T[SU(2)]$ theory \cite{Gang:2018wek}. From the anomaly polynomial, one can confirm that the following $\mathbb{Z}_2$ one-form symmetry is anomaly free
\begin{align}
\textrm{Anomaly free one-form $\mathbb{Z}_2$ symmetry}\;: \begin{cases} 
	\mathbb{Z}_2^H\;, \quad k_1 \in 2\mathbb{Z}\; \textrm{ and } \; k_2 \in 2\mathbb{Z}+1 \:,
	\\
	\mathbb{Z}_2^C\;, \quad k_2 \in 2\mathbb{Z}\; \textrm{ and }\; k_1 \in 2\mathbb{Z}+1\:,
	\\
    \mathbb{Z}_2^{\rm diag} \subset \mathbb{Z}^C_2 \times \mathbb{Z}^H_2\;, \quad  \textrm{otherwise}\;.
	\end{cases}
\end{align}
The $\mathbb{Z}_2$ one-form symmetry can be gauged and we define
\begin{align}
	\begin{split}
		&T[SU(2)]_{k_1, k_2}/\mathbb{Z}_2
		\\
		&:=\left( \textrm{Theory  after gauging  the anomaly free one-form $\mathbb{Z}_2$ symmetry in $T[SU(2)]_{k_1, k_2}$} \right)\;. \label{one-form gauged T[SU(2)]-k1k2}
	\end{split}
\end{align}
%

\subsubsection{IR phases}
\paragraph{Superconformal index}   Index of  the theory $T[SU(2)]_{k_1, k_2}$ (or  $T[SU(2)]_{k_1, k_2}/\mathbb{Z}_2$)   is
\begin{align}
\begin{split}
&\CI^{\rm sci} (q, \eta, \nu;s=1) 
\\
&= \sum_{\mathfrak{m}_1, \mathfrak{m}_2} \oint_{|a_1|=1,|a_2|=1}\left( \prod_{i=1}^2 \frac{\Delta (\mathfrak{m}_i, a_i)  da_i}{2\pi i a_i} (a_i(-1)^{\mathfrak{m}_i})^{2 k_i \mathfrak{m}_i} \right) \CI^{\rm sci}_{T[SU(2)]} (a_1,a_2, \eta,\nu ; \mathfrak{m}_1, \mathfrak{m}_2)\;. \label{index for T[SU(2)]k1k2-1}
\end{split}
\end{align}
Here the generalized superconformal index for $T[SU(2)]$ theory is 
\begin{align}
\begin{split}
&\CI^{\rm sci}_{T[SU(2)]} (a_1,a_2, \eta,\nu ; \mathfrak{m}_1, \mathfrak{m}_2)  = \CI^{\rm sci}_{T[SU(2)]} (a_1,a_2, \eta,\nu=0 ; \mathfrak{m}_1, \mathfrak{m}_2)\big{|}_{\eta \rightarrow \eta  q^{\frac{\nu}2}}\;\textrm{with}
\\
&\CI^{\rm sci}_{T[SU(2)]} (a_1,a_2, \eta,\nu=0 ; \mathfrak{m}_1, \mathfrak{m}_2) 
\\
&= \sum_{\mathfrak{n}} \oint_{|u|=1} \frac{du}{2\pi i u}((-1)^{\mathfrak{m}_2} a_2)^{-2 \mathfrak{n}}  ((-1)^{\mathfrak{n}} u)^{-2\mathfrak{m}_2}(q^{\frac{1}2} \eta^{-1})^{\frac{1}2 ( |\mathfrak{m}_1+ \mathfrak{n}| +|\mathfrak{m}_1- \mathfrak{n}|)}\textrm{P.E.} [f_{\rm single}]\;,
\\
&\textrm{where}
\\
&f_{\rm single}(a_1, a_2, \eta, u ;\mathfrak{m}_1, \mathfrak{n}) = \frac{q^{\frac{1}4}  \sqrt{\eta}  -q^{\frac{3}4 }  \sqrt{\eta^{-1}}}{1-q}  q^{\frac{1}2|\mathfrak{m}_1 + \mathfrak{n}|}\left(a_1 u  +\frac{1}{a_1 u} \right)  
\\
& \qquad \qquad  \qquad \qquad \qquad \qquad +  \frac{q^{\frac{1}4}  \sqrt{\eta}  -q^{\frac{3}4 }  \sqrt{\eta^{-1}}}{1-q}  q^{\frac{1}2|\mathfrak{m}_1 - \mathfrak{n}|}\left(\frac{a_1} u  +\frac{u}{a_1} \right) 
+ \frac{q^{\frac{1}2}}{1-q} \left(\frac{1}{\eta} -\eta \right) \;. \label{index for T[SU(2)]k1k2-2}
\end{split}
\end{align}
From Dirac quantization conditions, following monopole fluxes are allowed
\begin{align}
\mathfrak{n} ,  \mathfrak{m}_1, \mathfrak{m}_2 \in \frac{1}2 \mathbb{Z}\; \textrm{ with } \mathfrak{n}\pm \mathfrak{m}_1  \in \mathbb{Z}\;.
\end{align}
In the above formula, however, we are only summing over following monopole fluxes
\begin{align}
\textrm{for  $T[SU(2)]_{k_1, k_2}$  theory : } \mathfrak{n}, \mathfrak{m}_1, \mathfrak{m}_2 \in \mathbb{Z}\;, \label{index for T[SU(2)]k1k2-3}
\end{align}
since we are  summing over  $SU(2)$ bundles, i.e.\ $SO(3)$  bundles with trivial $w_2$. 
The localization saddle point  \eqref{Localization saddle for SCI} with $SU(2)$ monopole flux $\mathfrak{m}$ has non-trivial $w_2$ if and only if
\begin{align} 
	\mathfrak{m} \in \mathbb{Z}+\frac{1}2\;.
\end{align}
For the theory  \eqref{one-form gauged T[SU(2)]-k1k2} after gauging the $\mathbb{Z}_2$  one-form symmetry, we also need to sum over  gauge bundle with non-trivial $w_2^{\mathbb{Z}_2}$. For the superconformal index  for the $T[SU(2)]_{k_1, k_2}/\mathbb{Z}_2$ theory \eqref{one-form gauged T[SU(2)]-k1k2}, the summation range of monopole fluxes are
\begin{align}
\textrm{for  $T[SU(2)]_{k_1, k_2}/\mathbb{Z}_2$ theory : } \begin{cases} 
	\mathfrak{n}+\mathfrak{m}_1, 2\mathfrak{m}_1, \mathfrak{m}_2 \in \mathbb{Z}\;, \quad k_1 \in 2\mathbb{Z}\; \textrm{ and } \; k_2 \in 2\mathbb{Z}+1\:,
	\\
     	\mathfrak{n}, \mathfrak{m}_1, 2 \mathfrak{m}_2 \in \mathbb{Z}  \;, \quad k_2 \in 2\mathbb{Z}\; \textrm{ and }\; k_1 \in 2\mathbb{Z}+1\:,
	\\
	\mathfrak{n}+\mathfrak{m}_1, 2\mathfrak{m}_1 , \mathfrak{m}_1- \mathfrak{m}_2 \in \mathbb{Z}\;, \quad  \textrm{otherwise}\;. \label{index for T[SU(2)]k1k2-4}
\end{cases}
\end{align}
From the formulae in \eqref{index for T[SU(2)]k1k2-1},\eqref{index for T[SU(2)]k1k2-2},\eqref{index for T[SU(2)]k1k2-3},\eqref{index for T[SU(2)]k1k2-4}, one can compute the superconformal indices and check followings
\begin{align}
\begin{split}
&\CI^{\rm sci}(q, \eta, \nu=0;s=1)\textrm{ of } T[SU(2)]_{k_1, k_2} \;(\textrm{or }  T[SU(2)]_{k_1, k_2}/\mathbb{Z}_2 )
\\
& =  \begin{cases}
	\textrm{Non-trivial power series in $q^{1/2}$}\; \;\;\textrm{if $|k_1 k_2 -1|>3$ and min$(|k_1|, |k_2|)>1$}\:,
	\\
	0\;   \quad \textrm{if }|k_1 k_2 -1|=1\:,
	\\
	1\: \quad \textrm{if }|k_1 k_2-1|=3  \textrm{ or } (\textrm{$|k_1 k_2 -1|>3$ and min$(|k_1|, |k_2|)=1$})\:,
	\\
	1\; (\textrm{or } 2) \quad \textrm{if }|k_1 k_2-1|=2 \:,
	\\
	\infty\: \quad \textrm{if }|k_1 k_2 -1|=0 \label{SCI results for T[SU(2)]k1k2}\:.
\end{cases} 
\end{split}
\end{align}
For the case when $|k_1 k_2-1|=1$, the index vanishes and it implies that SUSY is spontaneously broken. For the case when $|k_1 k_2-1|=2$, on the other hand, the index for $T[SU(2)]_{k_1 k_2}$ is just $1$ while the index for  $T[SU(2)]_{k_1 k_2}/\mathbb{Z}_2$ is surprisingly $2$. It implies that theory $T[SU(2)]_{k_1 k_2}$ has a mass gap and flows to a  topological theory and  the UV $\mathbb{Z}_2$ one-form symmetry decouples at IR, i.e.\ the $\mathbb{Z}_2$ does not act faithfully on any IR observables. It means that the IR TQFT actually does not have the  $\mathbb{Z}_2$ symmetry. The index for $T[SU(2)]_{k_1 k_2}/\mathbb{Z}_2$ becomes $2$ just because we  perform the gauging of the  decoupled (so absent) one-form symmetry by hand. At the level of $S^2\times S^1$ partition function, the one-form gauging procedure is 
\begin{align}
\CZ^{S^2\times S^1}_{\textrm{TFT}/\mathbb{Z}_2} =\sum_{[\beta_2 ]\in H^2(S^2\times S^1, \mathbb{Z}_2)} \CZ^{S^2\times S^1}_{\rm TFT} ([\beta_2])\;.
\end{align}
Here the $[\beta_2 ] \in H^2 (S^2\times S^1, \mathbb{Z}_2) = \mathbb{Z}_2$ is the background 2-form $\mathbb{Z}_2$ flat connections coupled to the one-form symmetry. Alternatively, the RHS can be written as
\begin{align}
\CZ^{S^2\times S^1}_{\rm TFT}  +\CZ^{S^2\times S^1 + \CO^{[S^1]}_{\alpha_{\mathbb{Z}_2}}}_{\rm TFT} \;,
\end{align}
where first term is the $S^2\times S^1$ partition function (with trivial $[\beta_2]$) and the 2nd term is the partition function with insertion of loop operator along the $[S^1]$, the generator of $H_1(S^2\times S^1, \mathbb{Z}_2) = H^2 (S^2\times S^1, \mathbb{Z}_2)$, with  the $\mathbb{Z}_2$ symmetry generating anyon $\alpha_{\mathbb{Z}_2}$. For general TFT with a faithful $\mathbb{Z}_2$ one-form symmetry, the first term is just $1$ while the 2nd term vanishes and the total partition function $\CZ^{S^2\times S^1}_{\textrm{TFT}/\mathbb{Z}_2}$ becomes just 1 as expected. When the $\mathbb{Z}_2$ one-form symmetry is decoupled at IR, however, the 2nd term is also 1 since the anyon $\alpha_{\mathbb{Z}_2}$ actually becomes the trivial  operator, $\CO_{\alpha=0}$, at IR. So the result, $\CZ^{S^2\times S^1}_{\textrm{TFT}/\mathbb{Z}_2} =2$, is an artifact due to the ``gauging'' of the ``absent'' $\mathbb{Z}_2$ symmetry.   As we will see below, one can actually confirm that the $\mathbb{Z}_2$ symmetry act trivially on Bethe-vacua of $T[SU(2)]_{k_1, k_2}$ theory for the case when $|k_1 k_2-1|=2$. In summary, from the index computation, we can conclude that
\begin{align}
\begin{split}
&T[SU(2)]_{k_1, k_2}
\\
&\xrightarrow{\rm   \;  at \; IR \;  }    \begin{cases}
	\textrm{Non-trivial $\CN=4$ SCFT }\:\; \;\;\textrm{if $|k_1 k_2 -1|>3$ and min$(|k_1|, |k_2|)>1$}\:,
	\\
	\textrm{SUSY broken}\:   \quad \textrm{if }|k_1 k_2 -1|=1\:,
	\\
	\textrm{Unitary TQFT}\: \quad \textrm{if }|k_1 k_2-1|=3  \textrm{ or } (\textrm{$|k_1 k_2 -1|>3$ and min$(|k_1|, |k_2|)=1$} \:,
	\\
	\textrm{Unitary TQFT with decoupled $\mathbb{Z}_2$ }\; \quad \textrm{if }|k_1 k_2-1|=2\:,
	\\
	\textrm{Decoupled free chirals}\; \quad \textrm{if }|k_1 k_2 -1|=0 \:.
\end{cases} 
\end{split}
\end{align}
In the degenerate limits ($\nu\rightarrow \pm 1$ and $\eta \rightarrow 1$), on the other hand,  the indices are   (when  $|k_1 k_2 -1|>3$ and min$(|k_1|, |k_2|)>1$)
\begin{align}
\begin{split}
&\textrm{for  $T[SU(2)]_{k_1, k_2}$}\textrm{ theory },   
\\
& \CI^{\rm sci} \left(q,\eta ,\nu = \pm 1;s=1 \right)  = (\textrm{non-trivial power series in $q$})\; 
\\
&\textrm{ and }  \CI^{\rm sci} \left(q,\eta=1 ,\nu = \pm 1;s=1 \right)   =1\;, 
\end{split}
\end{align}
while
\begin{align}
\begin{split}
&\textrm{for } T[SU(2)]_{k_1, k_2}/\mathbb{Z}_2 \textrm{ theory },
\\
& \CI^{\rm sci} \left(q,\eta ,\nu = \pm 1;s=1 \right)  =   \begin{cases}
	\textrm{non-trivial power series in $q^{1/2}$}\; \;\;\textrm{if $k_1 k_2 \in 4\mathbb{Z}+1$}\;.
	\\
	\textrm{non-trivial power series in $q$}\; \;\;\textrm{otherwise}\;.
\end{cases} 
\\
&\textrm{ and } \CI^{\rm sci} \left(q,\eta=1 ,\nu = \pm 1;s=1 \right)   =1\;.  \label{SCI for T[SU(2)]-k1k2}
\end{split}
\end{align}
The computation implies that  the  IR theories are  $\CN=4$  SCFTs  of rank 0 and non-unitary TQFTs
\begin{align}
 \textrm{TFT}_\pm \big{[}T[SU(2)_{k_1, k_2}] \big{]}  \textrm{ or } \textrm{TFT}_\pm \big{[}T[SU(2)_{k_1, k_2}] /\mathbb{Z}_2 \big{]}
\end{align}
  emerge  in the degenerate limits, $\eta \rightarrow 1$ and $\nu  \rightarrow \pm 1$. According to the criterion in  \eqref{criterion of spin/non-spin}, we further expect that   $\textrm{TFT}_\pm \big{[}T[SU(2)]_{k_1,k_2}/\mathbb{Z}_2 \big{]}$  is a  spin TQFT  when $k_1 k_2 \in 4\mathbb{Z}+1$.  The non-unitary TQFTs associated to  the $\CN=4$ SCFT before the $\mathbb{Z}_2$ one-form symmetry gauging are given in Table~\ref{table:non-unitary/SCFT} and the TQFTs after the gauging  are 
\begin{align}
\begin{split}
&\textrm{TFT}_\pm \big{[}T[SU(2)]_{k_1,k_2}/\mathbb{Z}_2 \big{]} = \frac{\textrm{TFT}_\pm \big{[}T[SU(2)]_{k_1, k_2} \big{]}}{\mathbb{Z}_2}
\\
& =  \frac{\textrm{Gal}_{d_\pm} (SU(2)_{|k_1 k_2-1|-2}) \otimes U(1)_2}{\mathbb{Z}_2}\;, \quad \textrm{with }  d_+ =  \zeta_{|k_1 k_2-1|-2}^{|k_2|}, \; d_-=\zeta_{|k_1 k_2-1|-2}^{|k_1|}\;.
\end{split}
\end{align}
Here $\mathbb{Z}_2$ is the anomaly free $\mathbb{Z}_2$ one-form symmetry in $\textrm{Gal}_{d_\pm} (SU(2)_{|k_1 k_2-1|-2}) \otimes U(1)_2$. The topological theory  has $\mathbb{Z}^{SU(2)}_2\times \mathbb{Z}^{U(1)}_2$ one-form symmetry and the following $\mathbb{Z}_2$ one-form symmetry is non-anomalous
\begin{align}
\textrm{Anomaly free one-form $\mathbb{Z}_2$ symmetry}\;: \begin{cases} 
\mathbb{Z}_2^{SU(2)}\; \quad \textrm{if }k_1  k_2 \in 2 \mathbb{Z}+1 \;,
\\
\mathbb{Z}_2^{\rm diag}\; \quad \textrm{otherwise}\:.
\end{cases}
\end{align}
When $k_1 k_2 \in 4\mathbb{Z}_2+1$, the one-form $\mathbb{Z}_2$ symmetry is fermionic and the theory after the $\mathbb{Z}_2$ quotient becomes a spin TQFT,
\begin{align}
\frac{\textrm{Gal}_{d_\pm} (SU(2)_{|k_1 k_2-1|-2}) \otimes U(1)_2}{\mathbb{Z}_2} \;\textrm{ is a spin TQFT when } k_1 k_2 \in 4\mathbb{Z}+1\;.
\end{align}
It confirms the  criterion  \eqref{criterion of spin/non-spin}   combined with the   superconformal index computation \eqref{SCI for T[SU(2)]-k1k2}.
%

\subsubsection{Non-unitary TQFTs in  degenerate limits}
\paragraph{Squashed three-sphere partition function} The  partition function $\mathcal{Z}_{(k_1,k_2)}^{S_b^3}(b,m,\nu)$ of the $T[SU(2)]_{k_1, k_2}$ theory is
\begin{align}
\begin{split}
	&\mathcal{Z}_{(k_1,k_2)}^{S_b^3}(b,m,\nu) = \int \frac{dX_1 dX_2 dZ}{(2\pi \hbar)^{3/2}} \CI_\hbar (X_1, X_2, Z, m, \nu)\;,
	\\
	&\textrm{ with }\CI_\hbar (X_1, X_2, Z, m, \nu) = \CI_{\hbar}^{\rm vec}(X_1, X_2) \times \CI_{\hbar}^{T[SU(2)]} (X_1, X_2, Z,m,\nu)\;.
	\label{Simultaneous gauging partition function}
\end{split}
\end{align}
Here $\CI_\hbar^{\rm vec}$ is the contribution from the vector multiplet for the  $SU(2)^H_{k_1} \times SU(2)^C_{k_2}$ gauging:
\begin{align}
\CI_\hbar^{\rm vec} (Z_1, Z_2) = \exp \left(\frac{k_1 X_1^2+k_2 X_2^2}\hbar \right) \prod_{i=1}^2 \frac{1}2 \bigg{(} 2 \sinh X_i\bigg{)}  \left( 2 \sinh \frac{2\pi i X_i}\hbar \right) \;.
\end{align}
$\CI^{T[SU(2)]}_\hbar$ is the contribution from the $T[SU(2)]$ theory whose squashed three-sphere partition function is 
\begin{align}
\begin{split}
&\mathcal{Z}_{T[\SU(2)]}^{S_b^3}(b,X_1,X_2,m,\nu)= \int\! \frac{dZ}{\sqrt{2\pi\hbar}}\,\, \CI_{\hbar}^{T[SU(2)]} (X_1, X_2, Z, m, \nu), \textrm{ where}
\\
&\CI_\hbar^{T[SU(2)]} (X_1, X_2, Z, m, \nu) = \prod_{\epsilon_{1,2}=\pm1}\Psi_{\hbar}\left(\epsilon_1 Z+\epsilon_2 X_1+\frac{m+\nu ( i \pi +\frac{\hbar}2)}2+\frac{\pi i}{2}+\frac{\hbar}{4}\right)
\\
& \qquad \quad \times\Psi_{\hbar}\left(-2\frac{m+\nu ( i \pi \!\! +\!\! \frac{\hbar}2)}2+\pi i+\frac{\hbar}{2}\right)
\exp\Bigl(-\frac{2ZX_2}{\hbar}-\frac{\left(\pi i+\frac{\hbar}{2}\right)}{\hbar}\frac{m+\nu ( i \pi \!\! + \!\! \frac{\hbar}2)}2\Bigr)\;.
\label{TSU(2) partition function}
\end{split}
\end{align}
One of the non-trivial consistency checks is the mirror property
\begin{eqnarray}
\mathcal{Z}_{T[SU(2)]}^{S_b^3}(b, X_1 , X_2 , m=0, \nu) = \mathcal{Z}_{T[SU(2)]}^{S_b^3}(b, X_2 , X_1 , m=0 , -\nu) \;,
\label{mirror property}
\end{eqnarray}
which we have checked numerically  for various values of $X_1$, $X_2$, and $\nu$. Particularly for $b=1,m=0,\nu=0$, we have \cite{Gulotta:2011si,Benvenuti:2011ga,Nishioka:2011dq} (see also Appendix \ref{App : T[SU(2)] nu=0})
\begin{align}
\begin{split}
\mathcal{Z}^{S_b^3}_{T[SU(2)]}(b=1,X_1,X_2,m=0,\nu=0)
=
\frac{e^{\frac{2\pi i}{3}}}{2}
\frac{\sin(\frac{X_1 X_2}{\pi})}{\sinh(X_1)\sinh(X_2)}\;.
\label{TSU2 analytic three-sphere partition function}
\end{split}
\end{align}
As in \eqref{Simplified Z_su(2)_k half+half}, this localization \eqref{Simultaneous gauging partition function} also simplified at $b=1$, $m=0$, and is exactly computable at $\nu= 0, \pm 1$ (see Appendix \ref{App : Simultaneous nu=pm1}).
\begin{align}
\begin{split}
\exp(-F) &= \Big{|}	\CZ^{S^3_b}_{(k_1,k_2)}(b=1, m=0, \nu=0) \Big{|} = \sqrt{\frac{1}{|k_1 k_2 - 1|}} \sin \left( \frac{\pi}{|k_1 k_2 - 1|} \right)\;,
\\
(S_{00} \textrm{ of  TFT}_\pm ) &= \Big{|}	\CZ^{S^3_b}_{(k_1,k_2)}(b=1, m=0, \nu\rightarrow \pm 1) \Big{|} 
\\
&= \bigg\{ \begin{array}{ll}
\sqrt{\frac{1}{|k_1 k_2 - 1|}} \sin \left( \frac{\pi |k_2|}{|k_1 k_2- 1|} \right), \quad \nu \rightarrow +1\;,
\\
\sqrt{\frac{1}{|k_1 k_2 - 1|}} \sin \left( \frac{\pi |k_1|}{|k_1 k_2- 1|} \right),  \quad \nu \rightarrow -1\;.
\end{array}
\label{F and S00 of Simultaneous gauging}
\end{split}
\end{align}
\paragraph{Bethe-vacua and Handle gluing operators  in the degenerate limits }  The asymptotic expansions, $\CW_0$ and $\CW_1$, of the localization integral are 
\begin{align}
\begin{split}
&\log \CI_\hbar  (X_1, X_2, Z, m, \nu) \xrightarrow{\quad \hbar \rightarrow 0 \quad }  \frac{1}\hbar \CW_0 (X_1, X_2, Z, m, \nu)+ \CW_1 (X_1, X_2, Z, m, \nu) + O (\hbar)\;, 
\\
&\textrm{where}
\\
& \CW_0  = (k_1+1) X_1^2 +k_2 X_2^2 \pm 2\pi i X_1 \pm 2\pi i X_2-2 X_2 Z + Z^2+\sum_{\epsilon_1, \epsilon_2  = \pm 1} \textrm{Li}_2 (e^{\epsilon_1 X_1 + \epsilon_2 Z- \frac{m+ i \pi \nu}2 - \frac{\pi i }2} )\;,
\\
&\CW_1 = \frac{i \pi (\nu^2-\nu+1)}2 +\frac{2m\nu -m}4 +  \frac{1}4\sum_{\epsilon_1, \epsilon_2  = \pm 1} (\nu-1) \log \left(1+ e^{\epsilon_1 X_1 +\epsilon_2 Z - \frac{m+i \pi \nu}2 - \frac{\pi i}2}\right) 
\\
& \quad\quad \; - \frac{\nu}2 \log \left(1+e^{m+i \pi \nu}\right) +\log ( \sinh X_1) +\log ( \sinh X_2) \;.
\label{Simultaneous W0 W1}
\end{split}
\end{align}
In the $\CW_0$ above, we have ignored terms which are independent on $X_1, X_2$ and $Z$.  By extremizing the twisted superpotential
\begin{align}
\exp (\partial_{X_1} \CW_0) =\exp (\partial_{X_2} \CW_0) = \exp (\partial_{Z} \CW_0)  =1\;,
\end{align}
we have following Bethe-vacua equations
\begin{align}
\begin{split}
&\frac{x_1^{2k_1} (w x_1 + i z)(wx_1 z+i )}{(wz+ i x_1)(w+ i x_1 z)} =\frac{x_2^{2k_2}}{z^2}=\frac{(w z+ i x_1)(wx_1 z+i)}{x_2^2 (w x_1+ iz)(w+i x_1 z)}= 1\;,
\\
&\textrm{where } x_1 = e^{X_1}, \; x_2 = e^{X_2}, \; z = e^Z, \; w=e^{\frac{m+i \pi \nu}2}\;.
\end{split}
\end{align}
At generic choice of $w$, there are $2\times \big{|}(|k_1 k_2-1|-1)\big{|}$ Bethe-vacua, 
\begin{align}
\{(x_1, x_2, z) = ((x^*_1)_{n, \pm }, (x^*_2)_{n,\pm}, z^*_{n,\pm})\}_{n=1}^{\big{|}|k_1 k_2-1|-1\big{|}}\;,
\end{align}
 after removing the unphysical solutions, which are invariant under a non-trivial subgroup of the Weyl $\mathbb{Z}_2\times \mathbb{Z}_2$  acting as $x_i \rightarrow 1/x_i$, and quotienting by the Weyl group. The one-form symmetry $\mathbb{Z}^H_2\times \mathbb{Z}^C_2$, on the other hand, acts on the Bethe-vacua in the following way
 \begin{align}
\mathbb{Z}_2^H \;:\; x_1 \rightarrow  \pm x_1 \;, \quad \mathbb{Z}_2^C \;:\; x_2 \rightarrow \pm x_2\;.
 \end{align}
When $|k_1 k_2-1|=2$, the  anomaly free $\mathbb{Z}_2 = \mathbb{Z}_2^{\rm diag} \subset \mathbb{Z}_2^H \times \mathbb{Z}_2^{C}$ act trivially (modulo Weyl $\mathbb{Z}_2 \times \mathbb{Z}_2$) on the Bethe-vacua, i.e.
\begin{align}
x_1^* , x_2^* \in \{i , -i \}\;.
\end{align}
This explains why we obtain $2$ (instead of $1$) in the index computation of $T[SU(2)]_{k_1, k_2}/\mathbb{Z}_2$ theory when $|k_1 k_2-1|=2$, see the paragraph below \eqref{SCI results for T[SU(2)]k1k2}.

The handle gluing operator is
\begin{align}
\mathcal{H} (x_1, x_2,z;m,\nu)= \frac{1}{16} \exp (-2\CW_1) \det_{i,j}  \left(\partial_{i} \partial_{j} \CW_0 \right)\;.
\end{align}
The set of the handle gluing operators evaluated at the Bethe-vacua in the degenerate limits are
\begin{align}
\bigg{\{}\CH\big{(}(x^*_1)_{n,\pm}, (x^*_2)_{n,\pm},(z^*)_{n,\pm} \big{)}\bigg{\}}_{n=1}^{\big{|}|k_1 k_2-1|-1 \big{|}} \xrightarrow{\quad (m,\nu)\rightarrow (0, \pm 1) \quad } \bigg{\{} \frac{|k_1 k_2-1|}{\sin^2 \left( \frac{n \pi}{|k_1 k_2-1|}\right)}^{\otimes 2} \bigg{\}}_{n=1}^{\big{|}|k_1 k_2-1|-1 \big{|}} \;.
\end{align}
%

\subsection{\texorpdfstring{${T[SU(2)]}/{SU(2)^{\rm diag}_k}$ and ${T[SU(2)]}/{``PSU(2)^{\rm diag}_k"} $}{{T[SU(2)]}/{SU(2){diag}{k}} and {T[SU(2)]}/{``PSU(2){diag}{k}"}}}
Let us consider
\begin{align}
\frac{T[SU(2)]}{SU(2)^{\rm diag}_k}  =\left( \textrm{Gauging diagonal $SU(2)^{\rm diag}$ of $T[SU(2)]$ with Chern-Simons level $k$} \right)\;.
\end{align} 
Thanks to the nilpotency of the moment map $\vec{\mu}^{\rm diag}$ for the $SU(2)^{\rm diag}$ flavor symmetry, the theory remains $\CN=4$ theory even after the gauging with non-zero $k$ \cite{Gang:2018huc,Garozzo:2019ejm}.  

The theory has a $\mathbb{Z}_2$ one-form symmetry which corresponds to the center group of the gauged $SU(2)^{\rm diag}$
symmetry. The 't Hooft anomaly polynomial for the one-form symmetry is 
\begin{align}
S_{\rm anom} = \pi \int_{\CM_4 } \left( k \frac{\CP (w_2)}2 \right)\quad \textrm{(mod $2\pi$)}\;. \label{anomaly poly for T[SU(2)]-k}
\end{align}
Thus, the one-form symmetry is anomalous for $k=({\rm odd})$ while non-anomalous for $k=({\rm even})$. For odd $k$, the theory can be tensored with a topological theory $U(1)_{2} = SU(2)_{1}$, which also  has anomalous $\mathbb{Z}_2$ one-form symmetry, and  the diagonal $\mathbb{Z}_2$ one-form symmetry becomes non-anomalous.  We define 
\begin{align}
\begin{split}
&\frac{T[SU(2)]}{``PSU(2)^{\rm diag}_k"} 
\\
&:=  \begin{cases}
\textrm{Gauging the $\mathbb{Z}_2$ one-form symmetry of $\frac{T[SU(2)]}{SU(2)^{\rm diag}_k}$ }, \quad {\rm even \;}k\;, \\
\textrm{Gauging the  (diagonal) $\mathbb{Z}_2$ one-form symmetry of $\left(\frac{T[SU(2)]}{SU(2)^{\rm diag}_k} \otimes U(1)_2\right)$ }, \quad {\rm odd \;}k \;.\\
\end{cases}
\end{split} \label{T[SU(2)]/PSU(2)}
\end{align}

\subsubsection{IR phases}
\paragraph{Superconformal index} The superconformal index of the $\frac{T[SU(2)]}{SU(2)^{\rm diag}_{k}}$  \big(or $\frac{T[SU(2)]}{``PSU(2)^{\rm diag}_k"} $\big) theory is
\begin{align}
\begin{split}
&\CI^{\rm sci} (q, \eta, \nu;s=1) 
\\
&= \sum_{\mathfrak{m}} \oint_{|a|=1} \frac{\Delta (\mathfrak{m}, a)  da}{2\pi i a} \, (a(-1)^{\mathfrak{m}})^{2 k \mathfrak{m}} \,\CI^{\rm sci}_{T[SU(2)]} (a,a, \eta,\nu ; \mathfrak{m}, \mathfrak{m})\;, \label{index for T[SU(2)]k}
\end{split}
\end{align}
where $\CI^{\rm sci}_{T[SU(2)]}$ is the index of $T[SU(2)]$ theory given in \eqref{index for T[SU(2)]k1k2-2}.
From the Dirac quantization conditions, the following monopole fluxes are allowed
\begin{align}
\mathfrak{n} ,  \mathfrak{m} \in \frac{\mathbb{Z}}2 \; \textrm{ with } \mathfrak{n}\pm \mathfrak{m}  \in \mathbb{Z}\;.
\end{align}
The summation range of monopole fluxes is
\begin{align}
\begin{split}
&\textrm{for  $\frac{T[SU(2)]}{SU(2)^{\rm diag}_{k}}$  \;\;   : \;\;   } \mathfrak{n}, \mathfrak{m} \in \mathbb{Z}\;,
\\
&\textrm{for  $\frac{T[SU(2)]}{``PSU(2)^{\rm diag}_k"} $  \;\;   : \;\;   } \mathfrak{n}, \mathfrak{m} \in \frac{\mathbb{Z}}2 \; \textrm{ with } \mathfrak{n}-\mathbf{m} \in \mathbb{Z}\;.
 \label{index for T[SU(2)]k-2}
 \end{split}
\end{align}
From superconformal index computation \cite{Garozzo:2019ejm,Beratto:2020qyk}, 
\begin{align}
\begin{split}
&\CI^{\rm sci} (q, \eta,\nu=0;s=1)  \textrm{ for $\frac{T[SU(2)]}{SU(2)^{\rm diag}_k} $ or $\frac{T[SU(2)]}{``PSU(2)^{\rm diag}_k"} $ }
\\
& = \begin{cases}
\textrm{1}\;, \quad  |k|<2\;,
\\
\infty\;, \quad  |k|=2\;,
\\
\textrm{non-trivial power series in $q^{1/2}$}\;,\quad   |k|> 2\;.\ \\
\end{cases}
\end{split}
\end{align}
we expect that
\begin{align}
\begin{split}
& \textrm{$\frac{T[SU(2)]}{SU(2)^{\rm diag}_k} $ or $\frac{T[SU(2)]}{``PSU(2)^{\rm diag}_k"} $ }
 \\
 &
 \xrightarrow{ \quad \textrm{at IR} \quad } \begin{cases}
	\textrm{Unitary topological field theory}\; \quad\textrm{if $|k|<2$}\;,\\
	\textrm{Decoupled free chirals}\; \quad\textrm{if $|k|= 2$}\;,\ \\
	\textrm{3D $\mathcal{N}=4$ SCFT}\;  \quad \textrm{if $|k|> 2$}\;.\ \\
\end{cases} \label{IR phases of diag gauging}
\end{split}
\end{align}

\paragraph{Superconformal indices in degenerate limits} In the degenerate limits ($\nu\rightarrow \pm 1$ and $\eta \rightarrow 1$), on the other hand,  the indices are   (when  $|k|>2$)
\begin{align}
	\begin{split}
		&\textrm{for  $\frac{T[SU(2)]}{SU(2)_k^{\rm diag}}$}\textrm{ theory },   
		\\
		& \CI^{\rm sci} \left(q,\eta ,\nu = \pm 1;s=1 \right)  = (\textrm{non-trivial power series in $q$})\; 
		\\
		&\textrm{ and }  \CI^{\rm sci} \left(q,\eta=1 ,\nu = \pm 1;s=1 \right)   =1\;, 
	\end{split}
\end{align}
while
\begin{align}
	\begin{split}
		&\textrm{for } \frac{T[SU(2)]}{``PSU(2)^{\rm diag}_k"}  \textrm{ theory },
		\\
		& \CI^{\rm sci} \left(q,\eta ,\nu = \pm 1;s=1 \right)  =   \begin{cases}
			\textrm{non-trivial power series in $q^{1/2}$}\; \;\;\textrm{if $k \in 4\mathbb{Z}$}\;,
			\\
			\textrm{non-trivial power series in $q$}\; \;\;\textrm{otherwise}\;,
		\end{cases} 
		\\
		&\textrm{ and } \CI^{\rm sci} \left(q,\eta=1 ,\nu = \pm 1;s=1 \right)   =1\;.  \label{SCI for T[SU(2)]-k}
	\end{split}
\end{align}
The computation implies that non-unitary TQFTs, $\textrm{TFT}_\pm [\frac{T[SU(2)]}{SU(2)_k^{\rm diag}}]$ and  $\textrm{TFT}_\pm [\frac{T[SU(2)]}{``PSU(2)^{\rm diag}_k"}  ]$, emerge in the degenerate limits and that $\textrm{TFT}_\pm [\frac{T[SU(2)]}{``PSU(2)^{\rm diag}_k"}  ]$ are spin TQFTs when $k \in 4\mathbb{Z}$. From Table \ref{table:non-unitary/SCFT}, one can see that  $\textrm{TFT}_\pm [\frac{T[SU(2)]}{``PSU(2)^{\rm diag}_{k}"}  ]$ at $|k|=4$ are indeed spin TQFTs.

\subsubsection{Non-unitary TQFTs in  degenerate limits}
\paragraph{Squashed three-sphere partition function} The squashed three-sphere partition function of the ${T[SU(2)]}/{SU(2)^{\rm diag}_k}$ theory is realized as
\begin{align}
\mathcal{Z}_{\text{diag}_k}^{S_b^3}(b,m,\nu)
\equiv
\frac{1}{2} \int  \frac{\text{d} X}{\sqrt{2\pi \hbar}}  \big{(}2\sinh(X)\big{)}\bigg{(}2\sinh\Big{(}\frac{2\pi i X}{\hbar}\Big{)}\bigg{)}
e^{\frac{k X^2}{\hbar}}
\mathcal{Z}^{S_b^3}_{T[SU(2)]}(b,X,X,m,\nu)\;,
\nonumber\\
\label{Diagona gauging partition function}
\end{align}
where $\mathcal{Z}^{S_b^3}_{T[SU(2)]}$ is given in \eqref{TSU(2) partition function}. As in \eqref{Simplified Z_su(2)_k half+half}, this localization \eqref{Diagona gauging partition function} also simplified at $b=1$, $m=0$, and is exactly computable at $\nu= 0, \pm 1$ (see Appendix \ref{App : Diagonal nu=pm1}).
\begin{align}
\begin{split}
&\exp(-F) = \Big{|}	\CZ^{S^3_b}_{\text{diag}_k}(b=1, m=0, \nu=0) \Big{|} =
\frac{1}{\sqrt{8(|k|-2)}}-\frac{1}{\sqrt{8(|k|+2)}}\;,
\\
&(S_{00} \textrm{ of  TFT}_\pm )= \Big{|}	\CZ^{S^3_b}_{\text{diag}_k}(b=1, m=0, \nu\rightarrow \pm 1) \Big{|} = 
\frac{1}{\sqrt{8(|k|-2)}}+\frac{1}{\sqrt{8(|k|+2)}}\;.
\label{F and S00 of Diagonal gauging}
\end{split}
\end{align}
\paragraph{Bethe-vacua and Handle gluing operators  in the degenerate limits }
Similar to \eqref{Simultaneous W0 W1}, the asymptotic expansions $\mathcal{W}_0$ and $\mathcal{W}_1$ of the localization integral are
\begin{align}
\begin{split}
&\mathcal{W}_0 = (k+1)X^2 \pm 2\pi i X - 2 X Z + Z^2 +\sum_{\epsilon_1,\epsilon_2=\pm 1} \text{Li}_2(e^{\epsilon_1 X + \epsilon_2 Z -\frac{m+i\pi\nu}{2}-\frac{\pi i}{2}}) \;,
\\
&\mathcal{W}_1 = \frac{\pi i (\nu^2-\nu+1)}{2}+\frac{2m\nu-m}{4}+\frac{1}{4}\sum_{\epsilon_1,\epsilon_2=\pm 1}(\nu-1)\log\big(e^{\epsilon_1 X + \epsilon_2 Z -\frac{m+i\pi\nu}{2}-\frac{\pi i}{2}}\big)
\\
&\qquad
-\frac{\nu}{2}\log(1+e^{m+i\pi\nu})+\log(\sinh(X))\;.
\end{split}
\end{align}
We have ignored terms which are independent on $X$ and $Z$ in the expression for $\mathcal{W}_0$ given above. 
By extremizing the twisted superpotential
\begin{align}
\exp(\partial_{X}\mathcal{W}_0) = \exp(\partial_{Z}\mathcal{W}_0) = 1\;,
\end{align}
we have the  Bethe-vacua equations
\begin{align}
\begin{split}
&\frac{x^{2k}(wx+iz)(i+wxz)}{z^2(x-iwz)(iw-xz)}=\frac{(x-iwz)(i+wxz)}{x^2(z-iwx)(w+ixz)}=1\;,
\\
&\text{where}\quad x=e^{X}, \, z=e^{Z} \textrm{ and } w=e^{\frac{m+\pi i\nu}{2}}\;.
\end{split}
\end{align}
At generic choice of $w$, there are $2\times(2|k|+2)$ Bethe-vacua,
\begin{align}
 {(x,z)=(x^*_{n,\pm},z^*_{n,\pm})}_{n=1}^{2|k|+2}\;,
\end{align}
after removing the unphysical solutions, which are invariant under $x_i\rightarrow1/x_i$, and quotienting by the Weyl group.
The handle gluing operator is
\begin{align}
\mathcal{H}(x,z;m,\nu) = \frac{1}{4}\exp(-2\mathcal{W}_1) \det_{i,j}(\partial_i\partial_j \mathcal{W}_0)\;.
\end{align}
The set of the handle gluing operators evaluated at the Bethe-vacua in the degenerate limits are
\begin{align}
\begin{split}
&\bigg{\{}\CH\big{(} x^*_{n,\pm}, z^*_{n,\pm} \big{)}\bigg{\}}_{n=1}^{2|k|+2 } 
\!\!\!\!\!\!\!\!\!\!
\xrightarrow{\quad (m,\nu)\rightarrow (0, \pm 1) \quad } 
\\
&\bigg{\{} 
\frac{1}{\sqrt{2A_k}}^{\!\!\!\!\!\!\otimes (|k|-3)} 
\!\!\!\!\!
,
\!
\frac{1}{\sqrt{2B_k}}^{\!\!\!\!\!\!\otimes (|k|+1)} 
\!\!\!\!
,\Big(\frac{1}{\sqrt{8A_k}}\!\!+\!\!\frac{1}{\sqrt{8B_k}}\Big)^{\otimes 2}
\!\!
,\Big(\frac{1}{\sqrt{8A_k}}\!\!-\!\!\frac{1}{\sqrt{8B_k}}\Big)^{\otimes 2}
\bigg{\}} \;,
\end{split}
\end{align}
where we define $A_k := |k|-2$, $B_k := |k|+2$.

For $|k|>2$, as concluded in \eqref{IR phases of diag gauging}, the theory $T[SU(2)]/SU(2)^{\rm diag}_k$ lands on 3D $\CN=4$ SCFT of rank 0 at the end of RG.  According to the dictionary in Table \ref{Table : Dictionaries}, the above set should be equal to the set of $\{S_{0\alpha}^2\}$ for a non-unitary TQFT for the case when $|k|>2$. 

\paragraph{$|k|=3$ case } The set of handle gluing (and $S_{00}$ in \eqref{F and S00 of Diagonal gauging}) is identical to the  set of $\{|S_{0\alpha}^{-2}|\}$  (and $S_{00}$) for the  (Lee-Yang)$\otimes $(Lee-Yang)$\otimes U(1)_2$. From the computation, we arrive the conclusion in Table \ref{table:non-unitary/SCFT}.

\paragraph{$|k|=4$ case } The set is identical to the  set of $\{|S_{0\alpha}^{-2}|\}$  for   $\left( SU(2)_{10} \times SU(2)_2 \right)/{\mathbb{Z}_2^{\rm diag}}$, see \eqref{S0a of SU(2)10*SU(2)3/Z2}. Combined with the computation of $S_{00}$ in \eqref{F and S00 of Diagonal gauging}, we arrive the conclusion in Table \ref{table:non-unitary/SCFT}.

\paragraph{$|k|=5$ case } The set of handle gluing is identical to the  set of $\{|S_{0\alpha}^{-2}|\}$  for   $(G_2)_3\times U(1)_{-2}$, where $G_2$ is a exceptional group with dimension 14.  Combined with the computation of $S_{00}$ in \eqref{F and S00 of Diagonal gauging}, we arrive the conclusion in Table \ref{table:non-unitary/SCFT}.

\paragraph{$|k|>6$ case} For the cases, we could not identify  $\textrm{TFT}_{\pm}\left[\frac{T[SU(2)]}{SU(2)^{\rm diag}_k}\right]$ with previously known non-unitary TQFTs in the literature. It would be an interesting future work to better understand this novel series of non-unitary TQFTs. 

\subsection{Dualities among rank 0 theories}
In this paper, we introduce a pair of non-unitary TQFTs, TFT$_\pm [\CT_{\rm rank 0}]$, as an invariant of 3D rank 0 $\CN=4$ SCFT $\CT_{\rm rank 0 }$. One natural question is 
\begin{align}
\begin{split}
&\textrm{Q : How powerful are TFT$_\pm [\CT_{\rm rank 0}]$s in distinguishing $\CT_{\rm rank 0 }$? }
	\\
&\quad \quad \textrm{i.e.\ are TFT$_\pm [\CT_{\rm rank 0}]$s different for different $\CT_{\rm rank 0}$s?}
\end{split}
\end{align}
In this subsection, we provide evidences that the answer is affirmative. This is by examination of the duality between two $\CT_{\rm rank \;0}$s whose associated non-unitary TQFTs are  identical. Having common non-unitary TQFTs in the degenerate limits, $m=0$ and $\nu=\pm 1$, automatically implies that equivalence of all supersymmetric partition function $\CZ^{\mathbb{B}}$ in the  limits. Here we will confirm that the equivalence still holds at general values of $m$ and $\nu$.  

\paragraph{Duality among $(U(1)_{|k|=1}+H)$,   $SU(2)_{|k|=2}^{\frac{1}2 \oplus \frac{1}2}$ and  $\frac{T[SU(2)]_{k_1=3 , k_2=3}}{\mathbb{Z}_2}$ } According to Table \ref{table:non-unitary/SCFT}, the  non-unitary TQFTs associated to the 3 theories are all identical to the following theory
\begin{align}
	\textrm{TFT}_\pm   = \textrm{Gal}_{d = \sin \left( \frac{3\pi}8\right)} (SU(2)_6)/\mathbb{Z}_2^f\;.
\end{align}
Using the explicit formulas in \eqref{index for U(1)+H},\eqref{{index for SU(2)+H-1}},\eqref{{index for SU(2)+H-2}},\eqref{index for T[SU(2)]k1k2-1},\eqref{index for T[SU(2)]k1k2-2} and \eqref{index for T[SU(2)]k1k2-4}, one can confirm that the superconformal indices for the 3 theories are all equal to  
\begin{align}
\begin{split}
&\CI^{\rm sci} (q,\eta, \nu=1;s=1)
\\ &=1+q^{\frac{1}2}-\left(1+\eta+\frac{1}\eta\right)q+\left(2+\eta+\frac{1}\eta\right)q^{\frac{3}{2}}-\left(2+\eta+\frac{1}\eta\right)q^{2} +\ldots \;.
\end{split}
\end{align}
One can  check that computations of various other supersymmetric partition functions  also support the duality.   
\paragraph{Duality between  $(\CT_{\rm min})^{\otimes 2} \otimes U(1)_2$  and $\frac{T[SU(2)]}{SU(2)^{\rm diag}_{|k|=3}}$ } One can check that \cite{Gang:2018huc,Garozzo:2019ejm}
\begin{align}
	\begin{split}
		&\left(\CI^{\rm sci}(q, \eta, \nu=0;s=1) \textrm{ of } \frac{T[SU(2)]}{SU(2)_{|k|=3}^{\rm diag}}   \textrm{ given in \eqref{index for T[SU(2)]k1k2-2},\eqref{index for T[SU(2)]k},\eqref{index for T[SU(2)]k-2}} \right) 
		\\
		&= \left(\CI^{\rm sci}(q, \eta, \nu=0;s=1) \textrm{ of } \frac{T[SU(2)]}{``PSU(2)^{\rm diag}_{|k|=3}"}   \textrm{ given in \eqref{index for T[SU(2)]k1k2-2},\eqref{index for T[SU(2)]k},\eqref{index for T[SU(2)]k-2}}  \right) 
		\\
		&= \bigg{(}\CI^{\rm sci}(q, \eta, \nu=0;s=1) \textrm{ of } \CT_{\rm min} \textrm{ given in \eqref{sci for Tmin}}\bigg{)}^2\;
		\\
		&=  1-2q+2 \left(\eta +\frac{1}\eta \right)q^{3/2}- 3 q^2+\left(2 +\eta^2 +\frac{1}{\eta^2}\right) q^3 - 4 \left(\eta +\frac{1}\eta \right)q^{7/2}+\ldots \;.
		\\
	\end{split}
\end{align}
From the index computation, it is tempting to identify both $ \frac{T[SU(2)]}{SU(2)_{|k|=3}^{\rm diag}} $ and $ \frac{T[SU(2)]}{``PSU(2)^{\rm diag}_{|k|=3}"}$ with $(\CT_{\rm min})^{\otimes 2}$. But this cannot be true since $ \frac{T[SU(2)]}{SU(2)_{|k|=3}^{\rm diag}} $  has the one-form $\mathbb{Z}_2$ symmetry while the other two theories do not. Further, the Witten indices of the three theories do not match:
\begin{align}
	\begin{split}
		&\left(\textrm{Witten index of $(\CT_{\rm min})^{\otimes 2}$} \right) = (\textrm{Witten index of $(\CT_{\rm min})$})^2 = 4\;,
		\\
		&\left(\textrm{Witten index of $ \frac{T[SU(2)]}{``PSU(2)^{\rm diag}_{|k|=3}"} $} \right) = 4\;, \; \textrm{but }\left(\textrm{Witten index of $\frac{T[SU(2)]}{SU(2)^{\rm diag}_{|k|=3}}$} \right) = 8\;. \nonumber
	\end{split}
\end{align}
From the computation of superconformal indices, Witten indices, and one-form $\mathbb{Z}_2$ symmetry  matching, we propose that
\begin{align}
	\frac{T[SU(2)]}{SU(2)^{\rm diag}_{|k|=3}} =( \CT_{\rm min})^{\otimes 2} \otimes U(1)_{ 2}\; \;\; \textrm{at IR}.
\end{align}
The additional  $U(1)_{2}$ theory does not contribute to the superconformal index since there is no non-trivial local operator in the topological sector. The additional topological sector, however, doubles the Witten index since it has two ground states on two-torus and provide the one-form $\mathbb{Z}_2$ symmetry. The proposal is also compatible with following fact
\begin{align}
	\textrm{TFT}_\pm \bigg{[}\frac{T[SU(2)]}{SU(2)^{\rm diag}_{|k|=3}} \bigg{]} =\left(\textrm{TFT}_{\pm}[\CT_{\rm min}] \right)^{\otimes 2} \otimes U(1)_{ 2}\;,
\end{align}
which is obvious from Table \ref{table:non-unitary/SCFT}.
Using the duality, we can also confirm that
\begin{align}
	\begin{split}
		&\frac{T[SU(2)]}{``PSU(2)^{\rm diag}_{|k|=3}"}  := \left(\frac{T[SU(2)]}{SU(2)^{\rm diag}_{|k|=3}}\otimes U(1)_2 \right)/\mathbb{Z}^{\rm diag}_2 = \left(( \CT_{\rm min})^{\otimes 2} \otimes U(1)_{ 2} \otimes U(1)_2 \right)/\mathbb{Z}^{\rm diag}_2 
		\\
		& = ( \CT_{\rm min})^{\otimes 2} \otimes \frac{U(1)_2 \otimes U(1)_2}{\mathbb{Z}^{\diag}_2} \simeq ( \CT_{\rm min})^{\otimes 2}  \;.
	\end{split}
\end{align}
The theory $(U(1)_2\otimes U(1)_2)/\mathbb{Z}_2^{\rm diag}$ is an almost-trivial theory, whose partition function on any closed 3-manifold is a pure phase factor. Throughout this paper, we have ignored the overall phase factor of the partition function and thus will also ignore such a decoupled  almost trivial theory. The duality above has natural interpretation in terms of the 3D-3D correspondence for once-punctured torus bundles \cite{Terashima:2011qi,Gang:2013sqa},  for which readers are referred to Appendix \ref{App : 3D-3D for mapping torus} for details.

\paragraph{Duality between $SU(2)_{|k|=3}^{\frac{1}2 \oplus \frac{1}2}$    and $\frac{T[SU(2)]}{"PSU(2)^{\rm diag}_{|k|=4}"}$   } The superconformal index is 
\begin{align}
	\begin{split}
		&\left(\CI^{\rm sci}(q, \eta, \nu=0;s=1) \textrm{ of } \frac{T[SU(2)]}{SU(2)_{|k|=4}^{\rm diag}} \right) 
		\\
		&=  1-q+\left(\eta +\frac{1}\eta \right)q^{3/2}-  q^2-2\left(\eta +\frac{1}{\eta}\right) q^{5/2} +2 \left(3+\eta^2 +\frac{1}{\eta^2} \right)q^{3}+\ldots \;,
		\\
		& \left(\CI^{\rm sci}(q, \eta, \nu=0;s=1) \textrm{ of } \frac{T[SU(2)]}{``PSU(2)^{\rm diag}_{|k|=4}"}  \right) 
		\\
		&= 1+q^{1/2} - \left(1+\eta+\frac{1}\eta \right) q +\left(2+\eta +\frac{1}\eta \right) q^{3/2}- q^2 - \left(\eta +\frac{1}\eta \right)\left(2+\eta  +\frac{1}\eta \right) q^{5/2}+\ldots\;.
	\end{split}
\end{align}
Two indices are different unlike in the $|k|=3$ case. Surprisingly, the index computation shows that the two theories actually have different amount of supersymmetries \cite{Evtikhiev:2017heo}:
\begin{align}
	\begin{split}
		&\frac{T[SU(2)]}{SU(2)_{|k|=4}^{\rm diag}} \textrm{ has $\CN=4$ SUSY}\;,
		\\
		& \frac{T[SU(2)]}{``PSU(2)^{\rm diag}_{|k|=4}"} \textrm{ has $\CN=5$ SUSY}\;.
	\end{split}
\end{align}
We can thus conclude that supersymmetry is enhanced under the $\mathbb{Z}_2$ one-form symmetry gauging in \eqref{T[SU(2)]/PSU(2)}. We further claim that the $\CN=5$ theory  is actually dual to the following theory with manifest $\CN=5$ supersymmetry:
\begin{align}
	\left( \frac{T[SU(2)]}{``PSU(2)^{\rm diag}_{|k|=4}"}  \right) = \left(SU(2)_{|k|=3}^{\frac{1}2 \oplus \frac{1}2} \textrm{ in \eqref{SU(2)k1/2+1/2}} \right)\; \textrm{ at IR}.
\end{align}
The proposal can  be checked using the superconformal index and various other supersymmetric partition functions. Neither theories in the duality has any one-form $\mathbb{Z}_2$ symmetry. The duality is also compatible with Table \ref{table:non-unitary/SCFT} since 
\begin{align}
	\begin{split}
		&\textrm{TFT}_{\pm} \bigg{[} \frac{T[SU(2)]}{``PSU(2)^{\rm diag}_{|k|=4}"}\bigg{]} = \textrm{TFT}_{\pm} \bigg{[} \frac{T[SU(2)]}{SU(2)^{\rm diag}_{|k|=4}}\bigg{]}/\mathbb{Z}_2
		\\
		& = \left( \frac{\textrm{Gal}_{\zeta_{10}^7} (SU(2)_{10})\otimes SU(2)_2}{\mathbb{Z}_2^{\rm diag}}\right)/\mathbb{Z}_2 =   \left( \frac{\textrm{Gal}_{\zeta_{10}^7} (SU(2)_{10})}{\mathbb{Z}_2}\right) \otimes \left( \frac{ SU(2)_2}{ \mathbb{Z}_2}\right)
		\\
		&  \simeq   \left( \frac{\textrm{Gal}_{\zeta_{10}^7} (SU(2)_{10})}{\mathbb{Z}_2}\right) = \textrm{TFT}_{\pm} \bigg{[} SU(2)_{|k|=3}^{\frac{1}2 \oplus \frac{1}2} \bigg{]}  \;.
	\end{split}
\end{align}
In the last line, we again ignore the almost trivial spin TQFT $\frac{ SU(2)_2}{ \mathbb{Z}_2}$. In the second line, we use the following  fact
\begin{align}
	\left( \frac{\textrm{TFT}_1 \otimes \textrm{TFT}_2}{\mathbb{Z}_2^{\rm diag}} \right)/\mathbb{Z}_2 = \left( \frac{\textrm{TFT}_1}{\mathbb{Z}_2}\right) \otimes \left( \frac{\textrm{TFT}_2}{\mathbb{Z}_2} \right) \;,
\end{align}
which holds for any two TQFTs, TFT$_{1}$ and TFT$_{2}$, which have non-anomalous $\mathbb{Z}_2$ one-form symmetries.  The theory  ${T[SU(2)]}/{``PSU(2)^{\rm diag}_{|k|=4}"} $ has yet another dual description with only manifest $\CN=2$ supersymmetry which is  expected from the geometrical aspects of the 3D-3D correspondence for a once-punctured torus bundle \cite{Terashima:2011qi,Gang:2013sqa}. The $\CN=2$ dual  is  presented in Appendix \ref{App : 3D-3D for mapping torus}.
%

\section{Discussion}

There are several interesting questions we want to address in future. 
\paragraph{Relation with Rozansky-Witten theory}  
One well-known method of constructing a topological  theory from a  3D $\CN=4$ SCFT is using topological twisting as studied by Rozansky and Witten in \cite{Rozansky:1996bq}. An $\CN=4$ SCFT  $\CT$ has $SU(2)_L \times SU(2)_R$ R-symmetry and we can consider a pair of  topological twisted  theories, $\textrm{RW}_+ [\CT]$ and $\textrm{RW}_- [\CT]$,  using  the $SU(2)_L$  or $SU(2)_R$ in the twisting respectively. It would be interesting to clarify the exact relation between the pair of topological twisted theories, $\textrm{RW}_\pm [\CT_{\rm rank\;0}]$, and our $\textrm{TFT}_\pm[\mathcal{T}_{\rm rank 0}]$ for rank 0 SCFT $\CT_{\rm rank \;0}$.\footnote{For a 3D $\CN=4$ SCFT $\CT$ of non-zero rank, the topological twisted theories $\textrm{RW}_\pm [\CT]$ are  not genuine TQFTs  since they  do not obey the standard axioms of TQFT. Nevertheless, interesting  3-manifold/knot invariants can be studied using the $\textrm{RW}_\pm[\CT]$  \cite{Gukov:2020lqm}. }
Two theories, $\textrm{RW} [\CT_{\rm rank \;0}]$ and $\textrm{TFT} [\CT_{\rm rank \; 0}]$,  have the same ground state degeneracy $\textrm{GSD}_g$ for all $g \geq 0 $  since the partitial topological twisting on $\Sigma_{g}\times S^1$ using the $U(1)\subset SU(2)_R$ (or $U(1)\subset SU(2)_L$ ) symmetry is actually equivalent to the full topological twisting on the 3-manifold using the $SU(2)_R$ ($SU(2)_L$) symmetry  \cite{Gukov:2016gkn, Bullimore:2018jlp}. 
From the comparison, we naturally conjecture that
\begin{align}
	\textrm{TFT}_\pm [\CT_{\rm rank \;0}] \textrm{ is equal to }  \textrm{RW}_\pm [\CT_{\rm rank \;0}] \;.\label{relation between TFT and  RW}
\end{align} 
If the conjecture is true, we need to explain how the BPS partition functions of rank-0 theories in the degenerate limits compute  the partition functions of topologically twisted theories. 

\paragraph{3D non-unitary TQFTs from 4D $\CN=2$ SCFTs} In \cite{Dedushenko:2018bpp}, the authors constructed 3D non-unitary TQFTs from some 4D $\CN=2$ Argyres-Douglas theories. The construction is somewhat  similar to our construction of TFT, but the precise relation is not clear. In our construction, semi-simple non-unitary TQFTs appear  in 3D SCFTs of rank 0,  while their examples after naive dimension reduction to 3D are not of rank 0. It would be interesting  to clarify for which classes of 4D $\CN=2$ SCFTs their construction works, and to see if we can apply their construction in classification of 4D $\CN=2$ SCFTs. 

\paragraph{Are all non-unitary TQFTs correspond to rank 0  $\CN=4$ SCFTs?  } In our paper, we assign a pair of non-unitary TQFTs to 3D rank 0 $\CN=4$ SCFTs. But it is not clear if all non-unitary TQFTs can be constructed in this way.

\acknowledgments{We thank Hee-Cheol Kim for helpful discussions and comments at various stages of the project.  The research of DG is supported by the NRF Grant 2019R1A2C2004880 and by the New Faculty Startup Fund from Seoul National University. DG also acknowledges support by the appointment to the JRG program at the APCTP through the Science and Technology Promotion Fund and Lottery Fund of the Korean Government, as well as support by the Korean Local Governments, Gyeongsangbuk-do Province, and Pohang City. KL is supported in part by KIAS Individual Grant  PG006904 and by the National Research Foundation of Korea Grant NRF-2017R1D1A1B06034369.
MS is supported in part by National Research Foundation of Korea grant NRF-2019R1A2C2004880 and in part by scholarship from Hyundai Motor Chung Mong-Koo Foundation. MS also thanks for hospitality of APCTP during the visits.
MY is supported in part by JSPS KAKENHI Grant-in-Aid
for Scientific Research (Nos.~19H00689, 19K03820, 20H05850, 20H05860).}

\appendix
\section{Some reviews} \label{App : some reviews}
In order to help better understand the correspondence in \eqref{rank 0 to TFT}, we briefly review basic relevant aspects of (2+1)D non-unitary topological field theory and supersymmetric partition functions of 3D superconformal field theories. 

\subsection{\texorpdfstring{Localization on 3D $\CN\geq 3$ gauge theories}{Localization on 3D N>=3 gauge theories}} \label{App : review on localization}

In this paper,  we consider  following 4 types of   supersymmetric partition functions of 3D $\CN =4$ SCFT $\CT$ of rank 0 theory
\begin{align}
\begin{split}
&\CI^{\rm sci}_{\CT}(q, \eta,  \nu;s) \;:\; \textrm{Superconformal index \cite{Kim:2009wb,Imamura:2011su}}\;,
\\
& \CZ^{S^3_b}_{\CT}(b, m,\nu) \;:\; \textrm{Squashed three-sphere partition function \cite{Jafferis:2010un,Hama:2011ea}}\;,
\\
& \CI^{\Sigma_g}_{\CT}( \eta,\nu ;s) \;:\; \textrm{(Topologically) twisted indices on $\Sigma_g$ \cite{Gukov:2015sna,Benini:2015noa,Benini:2016hjo,Closset:2016arn} }\;,
\\
& \CZ^{\CM_{g,p}}_{\CT}( m,\nu ;s) \;:\; \textrm{Twisted partition function on $\CM_{g,p}$ \cite{Gang:2019juz,Alday:2012au,Imamura:2012rq,Closset:2017zgf,Closset:2018ghr}}\;.
\end{split}
\end{align}
These 4 types of partition functions are not totally exclusive. 
\begin{align}
\begin{split}
&\CZ^{S^3_b} (b=1, m , \nu) = \CZ^{\CM_{g=0,p=1}}(m,\nu;s=1)\;,
\\
&\CI^{\Sigma_g}(\eta, \nu;s) \big{|}_{\eta = e^m} = \CZ^{\CM_{g,p=0}}(m, \nu;s)\;.
\end{split}
\end{align}
In the partition function,  $m$ (resp. $\eta$) is the real mass parameter (resp. fugacity) associated to the axial $U(1)$ symmetry while $\nu$ is the R-symmetry mixing parameter  \eqref{R-symmetry mixing}.
Rank 0 $\CN=4$ SCFT cannot have  any flavor symmetry commuting with $\CN=4$ supersymmetries and thus the BPS partition functions cannot be further refined. 

 3D $\CN=4$ SCFTs  can appear as  IR fixed points of 3D quantum field theories. Since there could be supersymmetry enhancement along the RG flow, we do not need to start from a UV theory with manifest $\CN=4$ symmetry. For the exact computation of supersymmetric partition using localization, however, the UV theory should have at least $\CN=2$ supersymmetry.  In this paper, we study several examples of $\CN=4$ rank 0 SCFTs  which appears as IR fixed points of $\CN\geq 3$ supersymmetric theories.
 In the below, we summarize  localization  formulae for the BPS partition functions introduced above for $\CN\geq 3$ gauge theories.   In localization computations, $\CN \geq 3$ gauge theories have several advantages over $\CN=2$ gauge theories.  The local Lagrangian density of an $\CN=3$ gauge theory is uniquely determined by the choice of  gauge group $G$, its Chern-Simons levels $\vec{k}$ and matter contents (hypermultiplets and twisted hypermultilplets in unitary representations of $G$). When the CS levels are all zero, i.e.\ $\vec{k} = \vec{0}$, the theory has $\CN=4$ supersymmetry.  
 For $\CN\geq 3$ gauge theories, the R-symmetry is non-abelian ($SO(3)$), and thus the  IR R-symmetry is uniquely fixed (i.e.\ is  not   mixed with other abelian flavor symmetries) and we do not need to perform the  F-maximization \cite{Jafferis:2010un}.  The localization for general $\CN\geq 2$ theories can be done in a similar way but with some more complications. 
 
\paragraph{Superconformal index}The superconformal index   for a 3D $\mathcal{N}=4$ SCFT is defined as
\begin{align}
\mathcal{I}^{\rm sci}(q,\eta ,\nu;s ):= \begin{cases}
\textrm{Tr}_{\mathcal{H}_{\rm rad}(S^2)} (-1)^{2j_3} q^{\frac{R_\nu}2 + j_3} \eta^{A}\; \; , \quad s=1\;,\\
\textrm{Tr}_{\mathcal{H}_{\rm rad}(S^2)} (-1)^{R_\nu} q^{\frac{R_\nu}2 + j_3} \eta^{A}\; \;, \quad s= -1\;. \label{Def : superconformal index}
\end{cases}  
\end{align}
Here the trace is taken over the radially quantized Hilbert-space  $\mathcal{H}_{\rm rad} (S^2)$   on $S^2$ whose elements are local operators. $j_3 \in \frac{\mathbb{Z}}2$ is the Lorentz spin, the Cartan of $SO(3)$ isometry on the $S^2$.  The parameter $q$ plays role as an $\Omega$-deformation parameter. Only  BPS operators satisfying following relation contribute to the index 
\begin{align}
\Delta = R+R' + j_3\;,
\end{align}
where $\Delta$ is the conformal dimension.  The index alternatively can be regarded as a partition function on $(S^2\times S^1) = \CM_{g=0,p=0}$ with a fixed metric, background electric fields coupled to $U(1)_{R_\nu}$ and axial $U(1)$ symmetry and spin-structure along the $S^1$.  The indices at different $\nu$ are simply related as follows
\begin{align}
\begin{split}
&\mathcal{I}^{\rm sci}(q,\eta ,\nu;s  =1) = \mathcal{I}^{\rm sci}(q,\eta,\nu=0;s  =1) \big{|}_{\eta \rightarrow \eta q^{\frac{\nu}2}}\;,
\\
&\mathcal{I}^{\rm sci}(q,\eta ,\nu;s  =-1) = \mathcal{I}^{\rm sci}(q,\eta ,\nu=0;s  =-1) \big{|}_{\eta \rightarrow \eta (-q^{\frac {1}2})^\nu} \;.
\end{split}
\end{align}
The two indices with different choices of the spin structure are related to each other in the following way
\begin{align}
\mathcal{I}^{\rm sci}\left(q,\eta ,\nu = \pm 1 ;s  =1\right)  = \mathcal{I}^{\rm sci}\left(q,\eta ,\nu = \pm 1 ;s = -1\right)\big{|}_{q^{\frac{1}2} \rightarrow -q^{\frac{1}2}} \;. \label{SCI at different s}
\end{align}
Using localization, the superconformal index at $\nu=0$ is given as
\begin{align}
\begin{split}
&\CI^{\rm sci} (q,\eta,\nu=0;s=1) 
\\
&= \sum_{\mathbf{m} } \oint_{|a_i|=1}  \left(  \prod_{i=1}^{ \textrm{rank}G }  \frac{da_i}{2\pi i a_i}  \right) \Delta_G (\mathbf{m} , \mathbf{a}; q)  q^{\epsilon_0 (\mathfrak{n})} \CI_0^{cs} (\mathbf{m},\mathbf{a}) \textrm{P.E.}[f_{\rm single} (q, \mathbf{a},\eta;\mathbf{m})]\;.
\end{split}
\end{align}
In the localization, the saddle points are parametrized by $\{\mathfrak{m}_i , a_i \}_{i=1}^{\textrm{rank}(G)}$,
\begin{align}
\begin{split}
&\frac{1}{2\pi}\int_{S^2} F = \mathbf{m},  \;  \exp \left( i  \int_{S^1}  A \right) =  \mathbf{a} \;\;\textrm{and }\sigma = \frac{\mathbf{m}}2 
\\
& \textrm{ with }\mathbf{m}:=\sum_i \mathfrak{m}_i \mathbf{h}^i \textrm{ and } \mathbf{a}:=\exp (\mathbf{A}) :=\exp \left(\sum_i (\log a_i) \mathbf{h}^i \right)\;. \label{Localization saddle for SCI}
\end{split}
\end{align}
Here  $\sigma$ is the adjoint real scalar in the $\CN=2$ vector multiplet and $\{ \mathbf{h}^i \}$ is a normalized basis of Cartan subalgebra of $G$.  For $G=U(N)$ or $SU(2)$ case, the basis is chosen as
\begin{align}
	\begin{split}
		&G=U(N), \quad \mathbf{h}^i := \textrm{diag} \{0,\ldots, \stackrel{i-\text{th}}{1} , \ldots ,0 \}\;,
		\\
		&G= SU(2), \quad \mathbf{h}=\mathbf{h}^{i=1}= \textrm{diag} \{1,-1 \}\;.
	\end{split}
\end{align}
  $\Delta_G$ is the contribution from $\CN=2$ vector multiplet
\begin{align}
\begin{split}
&\Delta_G (\mathbf{m}, \mathbf{a};q) := \frac{1}{\textrm{Sym}(\mathbf{m})} \prod_{\lambda \in \Lambda^+_{\rm adj}} q^{- \frac{|\lambda (\mathbf{m})|}2} \left(1-q^{\frac{1}2 |\lambda (\mathbf{m})|} e^{\lambda (\mathbf{A}) }\right) \left(1-q^{\frac{1}2 |\lambda (\mathbf{m})|} e^{-\lambda ( \mathbf{A}) }\right)\;.
\end{split}
\end{align}
The monopole flux $\mathbf{m}$ in \eqref{Localization saddle for SCI} breaks the gauge group $G$ to its subgroup $H (\mathbf{m})$
\begin{align}
H(\mathbf{m}) := \{h \in G\;:\; [h,\mathbf{m}]=0 \}\;,  
\end{align}
and $\textrm{Sym}(\mathbf{m})$ is the order of the Weyl group of the subgroup,
\begin{align}
\textrm{Sym}(\mathbf{m}) := | \textrm{Weyl}(H(\mathbf{m}))|\;.
\end{align}
$\Lambda_{\rm adj}^+$ is the set of positive roots of $G$.
\begin{align}
\begin{split}
&G= U(N), \quad \{ \lambda (\mathbf{m})\;:\; \lambda  \in \Lambda^+_{\rm adj} \} = \{ \mathfrak{m}_i - \mathfrak{m}_j \;:\; 0<i<j\leq N\} \;,
\\
&G= SU(2), \quad \{ \lambda (\mathbf{m})\;:\; \lambda  \in \Lambda^+_{\rm adj} \} = \{ \mathfrak{m}_1 - \mathfrak{m}_2 \} \;.
\end{split}
\end{align}
The single particle index is
\begin{align}
f_{\rm single}(q,\mathbf{a}, \eta;\mathbf{m}) = \sum_{\Phi}\sum_{\beta \in \rho_{R_\Phi}}\left( \frac{q^{\frac{1}2 \Delta_\Phi+\frac{1}2 |\beta(\mathbf{m})|}e^{\beta(\mathbf{A})} \eta^{q_{A}(\Phi)}}{1-q}  - \frac{q^{\frac{1}2 (2- \Delta_\Phi)+\frac{1}2 |\beta(\mathbf{m})|}e^{-\beta(\mathbf{A})}  \eta^{-q_{A}(\Phi)} }{1-q} \right) \;.
\end{align}
Here the the summation is over $\CN=2$ chiral multiplets $\Phi$ in the representation of $R_\Phi$ under the gauge group $G$. $\rho_R$ is the set of weights of the representation $R$. 
$q_A (\Phi)$ and $\Delta _\Phi$ are the  axial $U(1)$ symmetry and the conformal dimension of the chiral field $\Phi$ respectively. An $\CN=4$ hypermultiplet consists of two chiral multiplets with gauge charges $R$ and $\overline{R}$, $q_A = \frac{1}2$ and $\Delta = \frac{1}2$.  The adjoint chiral multiplet in $\CN=4$ vector multiplet has $q_A = -1$ and $\Delta = 1$. If one wants to introduce a Chern-Simons interaction in an $\CN=4$ gauge theory, it will break the $\CN=4$ supersymmetry  down to $\CN=3$ symmetry. In the case, the R-symmetry is broken to $SO(3)$ and thus we cannot introduce the fugacity $\eta$ for the $U(1)$ axial symmetry.  The Casimir energy $\epsilon_0$ is
\begin{align}
\epsilon_0  = \frac{1}2 \big{(} \partial_q f_{\rm single} \big{)}\big{|}_{q, \eta \rightarrow 1}  = \sum_\Phi \sum_{\beta \in \rho_{R_\Phi}} \frac{(1-\Delta_{\Phi}) |\beta (\mathbf{m})|}4 \;.
\end{align}
$\CI_0^{cs} (\mathbf{m}, \mathbf{a})$ is the contribution from the classical CS term 
\begin{align}
\begin{split}
&U(N)_k \;:\; \prod_{i=1}^N (u_i(-1)^{\mathfrak{m}_i})^{k \mathfrak{m}_i}\;,
\\
& SU(2)_k \;:\; (u(-1)^{\mathfrak{m}})^{2 k \mathfrak{m}}\;.
\end{split}
\end{align}
For a $U(N)$ dynamical gauge group, there is a $U(1)$ topological symmetry whose Noether current  is
\begin{align}
j_{\rm top}^{\mu} = -\epsilon^{\mu \nu \rho} \textrm{Tr} (F_{\nu \rho})\;.
\end{align}
The fugacity $a$ and its background monpole flux $\mathfrak{m}_a$ for the topological symmetry can be introduced by including the following term to $\CI_0^{cs}$
\begin{align}
\left(a(-1)^{\mathfrak{m}_a}\right)^{-\sum_{i=1}^N \mathfrak{m}_i} \left(\prod_{i=1}^N u_i (-1)^{\mathfrak{m}_i}\right)^{-\mathfrak{m}_a}\;. 
\end{align}
The monopole flux $\mathbf{m}$ should satisfy the following Dirac quantization conditions
\begin{align}
\begin{split}
&\lambda(\mathbf{m}) \in \mathbb{Z}, \;\; \forall \lambda \in \Lambda^+_{\rm adj} \textrm{ and}
\\ 
&\beta (\mathbf{m})\in \mathbb{Z},\;\; \forall \beta \in R_\Phi\;.
\end{split}
\end{align}
There could be additional constraints on the monopole fluxes depending on the global structure of the $\CN\geq 3 $ gauge theories as we have seen in \eqref{index for T[SU(2)]k1k2-3} and \eqref{index for T[SU(2)]k1k2-4}. 
In the localization summation, we  need to sum over monopole flux $\mathbf{m}$  modulo the redundant Weyl symmetry of $G$.
 
\paragraph{Squashed three-sphere partition function $\CZ^{S^3_b}(b, m,\nu)$} 
This  is a partition function on $S^3 = \CM_{g=0,p=1}$ with the following metric
\begin{align}
ds^2 (S^3_b) =  |dz|^2 +|dw|^2 \;, \quad (z,w)\in \mathbb{C}^2 \textrm{ are subject to }   b^{-2}|z|^2 +b^2 |w|^2 =1\;.
\end{align}
To preserve some supercharges,  a background field coupled to the $U(1)_{R_\nu}$ symmetry  is properly turned on. Using localization, the partition function can be given in the following integral form
\begin{align}
\CZ^{S^3_b}(b, m,\nu) = \int  \left( \prod_{i=1}^{\textrm{rank}(G)} \frac{dZ_i}{\sqrt{2\pi \hbar}}  \right)  \Delta_G (\mathbf{Z};\hbar)\CI_\hbar (\mathbf{Z},m,\nu )\;, \quad \hbar := 2\pi i b^2\;.
\end{align}
Here $\{Z_i\}_{i=1}^{\textrm{rank}(G)}$ parametrizes the Cartan subalgebra of $G$. 
\begin{align}
\mathbf{Z} = \sum_{i=1}^{\textrm{rank}(G)}Z_i \mathbf{h}^i  \in \textrm{(Cartan subalgebra of $G$)}\;,
\end{align}
and $\Delta_G(\mathbf{Z})$ is the contribution from the  $\CN=2$ vector multiplet associated  to the gauge  group $G$
\begin{align}
\Delta_G (\mathbf{Z};\hbar):=	\frac{1}{|\textrm{Weyl}(G)|}\prod_{\lambda \in \Lambda^+_{\rm adj} } \bigg{[}4 \sinh \left( \frac{1}2 \lambda \cdot \mathbf{Z}\right)  \sinh \left( \frac{\pi i}\hbar \lambda \cdot \mathbf{Z}\right)  \bigg{]}\;.
\end{align}
\\
$|\textrm{Weyl}(G)|$ is the order of the Weyl group of $G$.  

The integrand $\CI_{\hbar} $ is determined by gauge group, matter contents and Chern-Simons levels of the $\CN=3$ gauge theory as follows:
\\
$\bullet $ An $\CN=2$ chiral  multiplet  in a representation $R$ under $G$ with $U(1)$ axial  charge $q_A$ and conformal dimension $\Delta$ contributes
\begin{align}
\begin{split}
&\prod_\Phi \prod_{\beta \in \rho_{R_\Phi}  }\Psi_\hbar \left(\beta\cdot \mathbf{Z} +q_A \big{(}m+ (i\pi +\frac{\hbar}2)\nu \big{)}   + (i\pi +\frac{\hbar}2)\Delta \right)\;.
\end{split}
\end{align}
We define $\Psi_{\hbar}$ as
\begin{align}
	\Psi_{\hbar}(X)&:=\psi_{\hbar}(X)\exp{\left(\frac{X^2}{4\hbar}\right)}\;,
	\label{Bigpsi}
\end{align}
with $\psi_\hbar(x)$ being the non-compact quantum dilogarithm function. (We refer to  \ref{App : QDL} for details of the definition and basic properties of the function.) 
An $\CN=4$ hypermultiplet consists of two $\CN=2$ chiral multiplets with gauge charge $R$ and $\overline{R}$, $q_A = \frac{1}2$ and $\Delta = \frac{1}2$.  The adjoint $\CN=2$ chiral multiplet in a $\CN=4$ vector multiplet has $q_A = -1$ and $\Delta = 1$.
\\
$\bullet$ Chern-Simons term of gauge $G$ of level $k$ contributes the following term to the integrand 
\begin{align}
\exp \left(\frac{k}{2\hbar }  \textrm{Tr} (\mathbf{Z}^2)\right)\;.
\end{align}
The real mass $m$ (FI parameter) and the R-symmetry mixing parameter $\nu$ of the $U(1)$ topological symmetry for $G=U(N)$ are introduced by adding the following term to the integrand
\begin{align}
\exp \left( - \frac{ W \textrm{Tr}(\mathbf{Z})}{\hbar}\right)\bigg{|}_{W = m + (i \pi +\frac{\hbar}2) \nu}\;.
\end{align}

The partition function at $b=1$, which corresponds to round three-sphere, enjoys interesting properties. Firstly, its free-energy is maximized at the superconformal R-charge choice, i.e.
\begin{align}
F_{\nu=0} > F_{\nu \neq 0 }\;, \quad \textrm{where } F_{\nu} := -\log |\CZ^{S^3_b} (b=1, m=0, \nu)|\;. \label{F-maximization}
\end{align}
Secondly, the round sphere free-energy $F$ at conformal point 
\begin{align}
F = -\log |\CZ^{S^3_b} (b=1, m=0, \nu=0)|\;, \label{F}
\end{align}
always monotonically decreases under the RG flow. So the quantity $F$ can be regarded as a proper measure of degrees of freedom. 

\paragraph{Perturbative expansion of squashed three-sphere partition function integral} One can consider formal perturbative expansion of the localization integral in an asymptotic  limit $\hbar \rightarrow 0 $,
to obtain infinitely many 3D SCFT invariants. 
In the limit, the integrand $\CI_\hbar$ can be perturbatively expanded in the following form
\begin{align}
\log \CI_\hbar (\vec{Z}, m, \nu) \xrightarrow{\quad \hbar \rightarrow 0 \quad} \sum_{n=0}^\infty \hbar^{n-1} \CW_{n} (\vec{Z}, n, \nu)\;. 
\end{align}
The leading part $\CW_0$ corresponds to the twisted superpotential. By extremizing the twisted superpotential, we obtain Bethe-vacua
\begin{align}
\textrm{Bethe-vacua : }\frac{ \bigg{\{} \vec{z} \;:\;  \left( \exp (  \partial_{Z_i } \CW_0 )\big{|}_{\vec{Z} \rightarrow  \log  \vec{z} } \right) =1 \;, \; w\cdot \vec{z} \neq \vec{z}\; \;\; \forall\; \textrm{non-trivial }w \in  \textrm{Weyl}(G) \bigg{\}}_{i=1}^{\textrm{rank}(G)}}{ \textrm{Weyl}(G)} \;.\label{Bethe-vacua}
\end{align}
Here $\textrm{Weyl}(G)$ is the Weyl group of gauge group $G$. A Bethe-vacuum $\vec{z}_\alpha$ can be promoted to a saddle point $\vec{Z}_\alpha = \log \vec{z}_\a$ of the localization integral by properly shifting $\CW_0$ as follows,
\begin{align}
 \CW_{0}^{\vec{n}_\alpha} =  \CW_0 +2\pi i \sum_{i} n_\alpha^i Z_i , \;\; n_\alpha^i \in \mathbb{Z}\; \textrm{is chosen such that } \partial_{Z_i } \CW^{\vec{n}_\alpha}_0 |_{\vec{Z}\rightarrow \log \vec{z}_\alpha} =0\;.
\end{align}
Then we can consider formal perturbative expansion of the localization integral around the saddle point 
\begin{align}
\begin{split}
&|\textrm{Weyl}(G)| \times \int \prod_{i=1}^{\textrm{rank}(G)} \frac{d(\delta Z_i)}{\sqrt{2\pi \hbar}}  \exp \left(\frac{1}\hbar \CW_0^{\vec{n}_\alpha}(\vec{Z}^\a + \delta \vec{Z}, m, \nu)  + \sum_{n=1}^\infty  \hbar^{n-1}\CW_n (\vec{Z}+ \delta \vec{Z},, m, \nu)\right)
\\
& \xrightarrow{ \quad \hbar \rightarrow 0 \quad }  \exp \left( \sum_{n=0}^\infty \hbar^{n-1} \CS^\alpha_n (m, \nu) \right)  \;. \label{perturbative invariants}
\end{split}
\end{align}
The factor $|\textrm{Weyl}(G)|$ is multiplied since that many saddle points, which all give the same perturbative expansion,  collapse into a single Bethe-vacuum after the Weyl quotient. 
The perturbative expansion can be formally computed by performing Gaussian integrals \cite{Dimofte:2012qj,Gang:2019jut}. For example,  
\begin{align}
&\CS^\alpha_0 = \CW_0^{\vec{n}_\alpha} (\vec{Z}_\alpha)\;,  \quad \CS_1^\alpha = -\frac{1}2 \log \left( \det_{i,j} \frac{\partial^2 \CW_0}{\partial Z_i \partial  Z_j}  \right) \bigg{|}_{\vec{Z} = \vec{Z}_\alpha} + \CW_1 (\vec{Z}_\alpha) + \log |\textrm{Weyl}(G)|\;. \label{S0 and S1}
\end{align}
The proposal in \eqref{rank 0 to TFT} implies the following highly non-trivial constraints on the perturbative invariants for rank 0 SCFTs,
\begin{align}
\begin{split}
&\textrm{Im} [\CS^\alpha_0(m=0,\nu )] \xrightarrow{\quad \nu\rightarrow \pm 1 \quad }0\;, \quad \textrm{Im} [\CS^\alpha_2(m=0,\nu  )] \xrightarrow{\quad \nu\rightarrow \pm 1 \quad }0\;,
\\
&\CS^\alpha_{n\geq 3}(m=0,\nu  )  \xrightarrow{\quad \nu\rightarrow \pm 1 \quad }0\;. \label{perturbative invariants in the degenerate lmits} 
\end{split}
\end{align}
This follows from the fact that the squashed three-sphere partition function becomes $b$-independent in the degenerate limits, $m=0$ and $\nu \rightarrow \pm 1$, modulo local counter terms which affect an overall factor of the following form 
\begin{align}
	\exp  \bigg{(} \pi i q_1 (b^2+\frac{1}{b^2}) + i \pi  q_2) \bigg{)}\bigg{|}_{q_1, q_2 \in \mathbb{Q}} \;. \label{local counter term}
\end{align}
\paragraph{Twisted indices and twisted partition functions} The twisted index is defined as  
\begin{align}
\CI^{\Sigma_g} (\eta, \nu;s) =   \begin{cases}
\textrm{Tr}_{\mathcal{H}(\Sigma_g;\nu)} (-1)^{2j_3} \eta^{A} \;, \quad s=1\;,\\
\textrm{Tr}_{\mathcal{H}(\Sigma_g;\nu)} (-1)^{R_\nu} \eta^{A} \;\; \;, \quad s= -1\;.   
\end{cases}
\end{align}
Here $\mathcal{H}(\Sigma_g;\nu)$ is the Hilbert-space on $\Sigma_g$ with topological twisting using the $U(1)_{R_\nu}$ symmetry.  Unlike the radially quantized Hilbert-space $\CH_{\rm rad}(S^2)$, the Hilbert-space depends on the choice of the R-symmetry mixing parameter $\nu$. 
Due to the topological twisting, the index is  well-defined only when following Dirac quantization condition is satisfied
\begin{align}
R_{\nu} \times  (g-1) \in \mathbb{Z} \quad  \textrm{for all local operators}\;. \label{Dirac quantization for twisted indices}
\end{align}
Note that the condition is always satisfied in the degenerate limit $\nu=\pm 1$ since $R_{\nu=\pm 1}  \in \mathbb{Z}$ which  obvious from the fact that $R_{\nu=1} = 2 R \in \mathbb{Z}$ and $R_{\nu=-1} = 2 R' \in \mathbb{Z}$.  For $g=0$ case, the quantization condition is  satisfied for all $\nu$ and the index is independent on the continuous parameter $\nu$.  Generally, the twisted indices  can be written as follows
\begin{align}
\CI^{\Sigma_g} (\eta, \nu;s)  = \sum_{\vec{z}_\alpha\;:\; \textrm{Bethe-vacua}}  (\CH_\alpha (\eta, \nu;s))^{g-1}\;.
\end{align}
Here $\CH_\alpha$ is called the {\it handle gluing} operator at the $\alpha$-th Bethe-vacuum. For $s=-1$ case, the operator is simply given as
\begin{align}
\CH_\alpha (\eta, \nu; s=-1) = e^{i \varphi} \exp\left(-2 \CS_1^\alpha (m, \nu) \right) \big{|}_{m = \log \eta} \;. \label{Handle gluing}
\end{align}
Here $e^{i \varphi}$ is a $\alpha$-independent overall phase factor, affected by the local counter term \eqref{local counter term}, which can be  fixed by requiring $\CI^{\Sigma_{g}} \in \mathbb{Z}$ for all $g$ up to a sign. For rank 0 SCFT, the phase factor is uniquely determined by requiring $\CI^{\Sigma_{g=0}} =1$ in the degenerate limits, $\eta\rightarrow 1$ and $\nu \rightarrow \pm 1$.  Upon the proper choice of the phase factor, furthermore,  the handle gluing operators become all positive real number in the degenerate limits,
\begin{align}
\CH_\alpha (\eta =1, \nu\rightarrow \pm 1, s)>0\;, \quad \textrm{for all }\alpha \;.
\end{align}
This is compatible with the dictionary for the handle gluing operators in Table \ref{Table : Dictionaries}.
More generally, the twisted partition function is given in the following form
\begin{align}
\CZ^{\CM_{g,p}} (m, \nu, s) =  \sum_{\vec{z}_\alpha\;:\; \textrm{Bethe-vacua}}  (\CH_\alpha (\eta =e^{m}, \nu;s))^{g-1} (\CF_\alpha(m, \nu ; s))^p \;. \label{Zgp from F and H}
\end{align}
Here $\CF_\alpha$ is called {\it fibering  operator} at the $\alpha$-th Bethe-vacuum. For $s=-1$ case, the operator is simply given as
\begin{align}
\CF_\alpha (\eta, \nu; s=-1) = \exp\left( \frac{\CS_0^\alpha (m, \nu)}{2\pi i } \right) \;.  \label{Fibering}
\end{align}
\paragraph{Supersymmetric loop operator $\CO (\vec{z})$} In  the twisted partition functions computation, an inclusion of a supersymmetric loop operator along the fiber $S^1$ in $\CM_{g,p}$ corresponds to an inclusion of a (Weyl invariant) finite Laurent polynomial $\CO(\vec{z})$ in $\{z_i\}_{i=1}^{\textrm{rank}(G)}$ with integer coefficients:
\begin{align}
\CZ^{\CM_{g,p}+\CO} (m, \nu, s) =  \sum_{\vec{z}_\alpha\;:\; \textrm{Bethe-vacua}}  (\CH_\alpha (m, \nu;s))^{g-1} (\CF_\alpha(m, \nu,s))^p \CO(\vec{z}_\alpha) \;. \label{Supersymmetric loop in Zgp}
\end{align}
Here $\CZ^{\CM_{g,p}+\CO} $ is the twisted partition function on $\Sigma_{g,p}$ with insertion of loop operator $\CO$. For example, dyonic loop operator $\CO_{(p,q)}$ of charge $(\textrm{electric charge, magnetic charge}) = (p,q)$ in a $U(1)$  gauge theory is given as
\begin{align}
\CO_{(p,q)}(z) = z^{p}\left(1-\frac{1}z \right)^q\;.
\end{align}
%

\subsection{Modular data of 3D TQFT}  \label{App : review on 3D TQFT}

\paragraph{In Bosonic (i.e.\ non-spin) TQFT} One basic characteristic quantity of 3D bosonic (i.e.\ non-spin) topological field theories is so-called {\it modular data},  which consists of $S$ and $T$ matrices. Let us denote  components of the two matrices by 
\begin{align}
S_{\alpha \beta}, \; T_{\alpha \beta}\;:\; \alpha, \beta =0, \ldots N-1\;. 
\end{align}
To understand the physical meaning of the matrices, let us consider the Hilbert space $\CH(\mathbb{T}^2)$ on two-torus. In topological field theory, there is no local operator and the only physical observables are loop operators $\CO^\Gamma_{\alpha=0,\ldots , N-1}$. $\alpha$ labels types (gauge charge) of loop operators, sometimes called {\it anyons}, and the natural number $N$ is called the rank of the topological field theory.   $\Gamma$ is the one-dimensional trajectory where the operator is supported.  $\CO_{\alpha=0}$ is the trivial loop operator, i.e.\ identity operator,
\begin{align}
\CO_{\alpha =0}=1\;.
\end{align}
 One natural basis of the Hilbert-space $\CH(\mathbb{T}^2)$ is
\begin{align}
\textrm{Basis of $\CH(\mathbb{T}^2)$} \;:\; \big{\{}|\alpha \rangle := \CO_\alpha^B |0\rangle \big{\}}_{\alpha=0}^{N-1}  \;,\label{Basis of torus Hilbert-space}
\end{align}
where $B$ is a generator of $H_1 (\mathbb{T}^2, \mathbb{Z}) = \langle A, B\rangle$. A mapping class element  $\varphi \in  SL(2,\mathbb{Z})$ acts on the Hibert-space as a unitary operator $\hat{\varphi}$. The operators $\{\hat{\varphi}\}_{\varphi \in SL(2,\mathbb{Z})}$ form a unitary representation of $SL(2,\mathbb{Z})$. The $S$ and $T$ matrices are nothing but\footnote{In our convention, $T_{00}$ is fixed to be 1.  Conventionally,  $T_{\alpha\beta}$ is defined as $\exp \left(-\frac{ 2\pi i c_{ 2d}}{24}\right) \times T^{\rm ours}_{\alpha \beta}$ such that $T_{\alpha\beta}$ is just $\langle \alpha |\hat{\mathbb{T}}|\beta \rangle$ without the phase factor.} 
\begin{align}
S_{\alpha\beta} = \langle \alpha |\hat{\mathbb{S}}|\beta \rangle \;,  \quad T_{\alpha\beta} =\exp \left(\frac{ 2\pi i c_{ 2d}}{24}\right)\times \langle \alpha |\hat{\mathbb{T}}|\beta \rangle \;.
\end{align} 
$c_{2d}$ (mod 24) is the chiral central charge of boundary 2d chiral CFT. Here $\mathbb{S}$ and $\mathbb{T}$ are two canonical generators of $SL(2,\mathbb{Z})$
\begin{align}
\begin{split}
& \mathbb{S} = \begin{pmatrix}
0 & 1 \\
-1 & 0
\end{pmatrix} \;, \quad  \mathbb{T} = \begin{pmatrix}
1 & 0 \\
1 & 1
\end{pmatrix} \;.
\end{split}
\end{align}
The modular matrices contain a lot of information of the topological field theory. According to the Verlinde formula, the fusion coefficients $N_{\alpha \beta}^\gamma$ can be given as
\begin{align}
N_{\alpha \beta}^\gamma  = \sum_{\delta =0}^{N-1} \frac{S_{\delta \alpha} S_{\delta \beta} S_{\delta \gamma}^*}{S_{0 \delta}}\;.
\label{Fusion Coefficient formula}
\end{align}
The S-matrix determines how the basic operators $\CO_\alpha^A$  and  $\CO_\alpha^B$ act on the Hilbert-space
\begin{align}
\CO_\beta^A |\alpha \rangle = W_\beta (\alpha) |\alpha \rangle = \frac{S_{ \alpha \beta}}{S_{ \alpha 0 }} |\alpha\rangle\;, \quad \CO_\beta^B |\alpha \rangle = \sum_\gamma N_{\alpha \beta}^\gamma |\gamma\rangle\;. \label{action of loop operators}
\end{align}
T-matrix is a diagonal unitary matrix
\begin{align}
\begin{split}
T_{\alpha \beta } &= \delta_{\alpha \beta} e^{2\pi i h_\alpha}\;(h_{\alpha=0}=0) \;,
\\
h_\alpha  &= \textrm{topological spin of $\alpha$-th anyon}\; .
\end{split}
\end{align}
The topological spin is defined only modulo 1.   $S$ and $T$ matrices satisfies
\begin{align}
S^2 = C \; (C^2=1)\;,\quad (ST)^3 = \exp \left(\frac{2\pi i c_{2d}}{8} \right) \times C\;.
\end{align}
The matrix $C_{\alpha \beta} $ is called charge conjugation. $S_{0\alpha}$ are real and they have following path-integral interpretation
\begin{align}
\begin{split}
&S_{0\alpha} =  \CZ^{S^3+ \CO_\beta^{\Gamma = \textrm{(unknot)}}} 
\\
&:=(\textrm{Partition function on $S^3$ with a loop operator $\CO^\Gamma_\alpha$ along the $\Gamma$=(unknot)})\;. \label{S0a}
\end{split}
\end{align}
Especially, $S_{00}$ is the partition function on $S^3$.  In an unitary  topological field theory,  S matrix satisfies the following conditions
\begin{align}
\textrm{Unitarity : } |S_{00}|\leq |S_{0\alpha}| \;. \label{Unitarity in S}
\end{align}
Partition functions  on $\CM_{g,p}$ with insertion of a loop operator $\CO^{[S^1]}_{\beta}$ along the fiber $[S^1]$ can be written as follows
\begin{align}
\CZ^{\CM_{g,p}+\CO_\beta^{[S^1]}}_{\rm TFT} = \sum_{\alpha=0}^{N-1} (S_{0\alpha})^{2-2g} (T_{\alpha \alpha})^p W_\beta (\alpha)=  \sum_{\alpha=0}^{N-1} (S_{0\alpha})^{2-2g-1} (T_{\alpha \alpha})^p S_{\a \beta} \;. \label{Zgp from S and T}
\end{align}
The partition function at $p=0$ without insertion of loop operator, i.e.\ $\beta=0$, counts ground state degeneracy $\textrm{GSD}_g$ on a genus $g$ Riemann surface,
\begin{align}
\textrm{GSD}_g =\sum_{\alpha=0}^{N-1} (S_{0\alpha})^{2-2g} \;. \label{GSDg from S}
\end{align}
Since this counts actual numbers, the partition function at $p=0$ can be defined without any phase factor ambiguity. 
By contrast the partition function at non-zero $p$ depends on the framing choice of the 3-manifold $\CM_{g,p}$ as well as of the knot along the fiber $[S^1]$, and consequently there is no canonical framing choice.  The formula above is only valid for a certain choice of the framing. The framing change affects the partition function by an overall phase factor of  the form $\exp \left(\frac{2\pi i c_{2d}}{24}  \mathbb{Z} +2\pi  i h_\beta \mathbb{Z}\right)$. For example, when $\CM_{g=0,p=1}=S^3$
\begin{align}
\CZ^{\CM_{g=0,p=1}+\CO_\beta^{[S^1]}}_{\rm TFT} =  \sum_{\alpha=0}^{N-1} S_{0\alpha} T_{\alpha \alpha} S_{\a \beta} = (STS)_{0 \beta} = \exp \left(\frac{2\pi i c_{2d}}{8} -2\pi i h_\beta \right) \times S_{0\beta}\;, \nonumber
\end{align}
which is different from the $\CZ^{S^3+ \CO_\beta^{\Gamma = \textrm{(unknot)} = [S^1]}}$ in \eqref{S0a} by a phase factor. 

\paragraph{In fermioinc (i.e.\ spin) TQFT} For this case the Hilbert space $\CH(\mathbb{T}^2)$  depends on the choice of a spin-structure $H^1 (\mathbb{T}^2, \mathbb{Z}_2) = \mathbb{Z}_2 \times \mathbb{Z}_2$. Let us consider following NS-NS sector
\begin{align}
\mathcal{H}_{--} (\mathbb{T}^2) = (\textrm{Hilbert-space on $\mathbb{T}^2$ with anti-periodic boundary conditions along both $S^1$}) \nonumber
\end{align}
Similarly, one can consider four Hilbert-spaces $\CH_{\pm \pm}$ depending on the choice of the spin-structure.  On the  $\mathcal{H}_{--} (\mathbb{T}^2)$, only  a subgroup of $SL(2,\mathbb{Z})$ generated by $S$ and $T^2$ can act since $T$ maps a state in $\mathcal{H}_{--} (\mathbb{T}^2)$ into a state in $\CH_{-+}(\mathbb{T}^2)$. In other words, topological spins of anyons are defined only modulo $1/2$ in spin TQFT. 

\paragraph{Under  anyon condensation, $/\mathbb{Z}_2$ and $/\mathbb{Z}^f_2$} In topological field theory, $\mathbb{Z}_2$ one-form symmetry is generated by an anyon $\CO_{\alpha_{\mathbb{Z}_2}}$ satisfying the fusion rule $\CO_{\a_{\mathbb{Z}_2}} \times  \CO_{\a_{\mathbb{Z}_2}} =1$,
\begin{align}
\CO_{a_{\mathbb{Z}_2}} \;:\; \textrm{The anyon generating the one-form $\mathbb{Z}_2$ symmetry}\;.
\end{align}
The topological spin for the anyon $\a_{\mathbb{Z}_2}$  can take only following values \cite{Hsin:2018vcg}
\begin{align}
h_{\a_{\mathbb{Z}_2 }} \in  \Big{\{}0, \frac{1}2, \pm \frac{1}4 \Big{\}} \; (\textrm{mod 1})\;.
\end{align}
When $h_{\a_{\mathbb{Z}_2 }} =  \pm \frac{1}4$, the $\mathbb{Z}_2$ symmetry  is anomalous and cannot be gauged. When  $h_{\a_{\mathbb{Z}_2 }} =0$ (resp. $h_{\a_{\mathbb{Z}_2 }} =1/2$), on the other hand, $\mathbb{Z}_2$ is non-anomalous and is called a bosonic $\mathbb{Z}_2$ (resp. fermionic $\mathbb{Z}_2$) symmetry and is  sometimes  denoted as $\mathbb{Z}_2^b$ (resp. $\mathbb{Z}_2^f$). 
\begin{align}
\mathbb{Z}_2 \textrm{ is }	
\begin{cases}
	\textrm{anomalous}\; \quad \textrm{if } h_{\alpha_{\mathbb{Z}_2}} =  \pm \frac{1}4 \;(\textrm{mod }1)\;,
	\\
	\textrm{non-anomalous and bosonic}\; \quad \textrm{if } h_{\alpha_{\mathbb{Z}_2}} =   0 \;(\textrm{mod }1)\;,
	\\
	\textrm{non-anomalous and fermionic}\; \quad \textrm{if } h_{\alpha_{\mathbb{Z}_2}} =  \frac{1}2 \;(\textrm{mod }1) \;.\label{3 cases of Z2 one-form}
\end{cases}
\end{align}
Starting from a bosonic topological field theory ${\rm TFT}$ with bosonic or fermionic $\mathbb{Z}_2$ one-form symmetry, one can gauge the one-form symmetry to obtain another topological field theory, ${\rm TFT}/\mathbb{Z}_2$ or ${\rm TFT}/\mathbb{Z}^f_2$.    The gauging procedure is sometimes called the {\it anyon condensation}. The resulting  theory after gauging  is a non-spin TQFT for bosonic $\mathbb{Z}_2$ case while the resulting theory is spin TQFT for fermionic $\mathbb{Z}_2^f$ case.
\begin{align}
\begin{split}
&{\rm TFT}/\mathbb{Z}^b_2 \;:\; \textrm{Bosonic  (non-spin) TQFT}\;,
\\
&{\rm TFT}/\mathbb{Z}^f_2 \;:\; \textrm{Fermionic  (spin) TQFT}\;. \label{Z2 gauging : spin/non-spin}
\end{split}
\end{align}
 The anomalous symmetry  can be gauged only after tensoring with another TQFT, such as $U(1)_{\pm2}$,  with anomalous $\mathbb{Z}_2$ one-form symmetry. At the level of modular data, the  anyon condensation process can be summarized as follows. First consider the bosonic $\mathbb{Z}_2$ one-form symmetry gauging. After the gauging, the Hilbert-space on $\CH(\mathbb{T}^2)$ is spanned by following basis
\begin{align}
\begin{split}
&\CH^{\textrm{TFT}/\mathbb{Z}_2}(\mathbb{T}^2) =\CH^{\textrm{TFT}/\mathbb{Z}_2}_{\rm untwisted} (\mathbb{T}^2)  \oplus \CH^{\textrm{TFT}/\mathbb{Z}_2}_{\rm twisted} (\mathbb{T}^2)\;,
\\
& \CH^{\textrm{TFT}/\mathbb{Z}_2}_{\rm untwisted} (\mathbb{T}^2) = \textrm{Span} \big{\{}|[\alpha]\rangle:=\frac{1}{\sqrt{2}} (|\alpha\rangle +|\alpha_{\mathbb{Z}_2}\cdot \alpha\rangle) \;:\; \CO^A_{\a_{\mathbb{Z}_2}}|\alpha \rangle = |\alpha \rangle, \;\;  |\alpha_{\mathbb{Z}_2}\cdot \alpha\rangle \neq |\alpha\rangle \big{\}}\;,
\\
& \CH^{\textrm{TFT}/\mathbb{Z}_2}_{\rm twisted} (\mathbb{T}^2) = \textrm{Span} \big{\{}|\alpha;\pm \rangle \;:\; \CO^A_{\a_{\mathbb{Z}_2}}|\alpha \rangle = |\alpha \rangle, \;\;  |\alpha_{\mathbb{Z}_2}\cdot \alpha\rangle  = |\alpha\rangle \big{\}}\;.
\label{Hilbert space structure on T^2}
\end{split}
\end{align}
Here $|\a \rangle$ is the basis in \eqref{Basis of torus Hilbert-space} and see \eqref{action of loop operators} for the action of  $\CO_\alpha^A$ and  $\CO_\alpha^B$ on the basis.  On the basis  given in \eqref{Basis of torus Hilbert-space}, the $\CO_{\alpha_{\mathbb{Z}_2}}^A$ acts as  a diagonal matrix whose entries are all $+1$ or $-1$ while $\CO_{\alpha_{\mathbb{Z}_2}}^B$ acts as a permutation matrix whose square is identity. In the above, we define $|\alpha_{\mathbb{Z}_2} \cdot \alpha \rangle:=\CO^B_{\a_{\mathbb{Z}_2}}|\alpha \rangle $.  In the gauging procedure, we first discard basis elements which are odd (having eigenvalue $-1$) under the $\CO_{\alpha_{\mathbb{Z}_2}}^A$. Then, we quotient the reduced Hilbert-space by the action of $\CO_{\alpha_{\mathbb{Z}_2}}^B$. When a basis $|\alpha\rangle$ is invariant under both $\CO_{\alpha_{\mathbb{Z}_2}}^A$ and $\CO_{\alpha_{\mathbb{Z}_2}}^B$, the basis will be doubled to $\{|\alpha;\pm\rangle\}$. The modular data of the gauged theory is
\begin{align}
\begin{split}
&S^{{\rm TFT}/\mathbb{Z}_2}_{[\a][\beta]} = 2 S^{{\rm TFT}}_{\a \b}\;, \quad  S^{{\rm TFT}/\mathbb{Z}_2}_{[\alpha=0](\beta;\pm)} =  S^{{\rm TFT}}_{0\b}\;,   \quad 
\\
&h^{{\rm TFT}/\mathbb{Z}_2}_{[\alpha]} =  h^{\rm TFT}_\alpha\;,  \quad h^{{\rm TFT}/\mathbb{Z}_2}_{(\alpha;\pm)} =   h_\alpha^{\rm TFT}
\end{split}
\end{align}
For other S-matrix elements, $S_{(\alpha;\pm)(\beta;\pm)}$ and $S_{[\alpha](\beta;\pm)}$, of $\textrm{TFT}/\mathbb{Z}_2$,  we need to know more information on the mother $\textrm{TFT}$ beyond modular data \cite{Delmastro:2021xox}. 

In the fermionic $\mathbb{Z}^f_2$ gauging, the Hilbert-space $\CH_{--}(\mathbb{T}^2)$ of the resulting spin TQFT is 
\begin{align}
&\CH^{\textrm{TFT}/\mathbb{Z}^f_2}_{--}(\mathbb{T}^2) = \textrm{Span} \big{\{}|[\alpha]\rangle:=\frac{1}{\sqrt{2}} (|\alpha\rangle +|\alpha_{\mathbb{Z}_2}\cdot \alpha\rangle) \;:\; \mathcal{O}^A_{\a_{\mathbb{Z}_2}}|\alpha \rangle = |\alpha \rangle \big{\}}\;.
\end{align} 
Unlike the bosonic $\mathbb{Z}_2$ gauging, there is no twisted sector in the $\mathbb{Z}_2^f$ gauging since $|\alpha_{\mathbb{Z}_2} \cdot \alpha \rangle \neq |\alpha \rangle$. The modular data of the gauged theory is
\begin{align}
\begin{split}
&S^{{\rm TFT}/\mathbb{Z}^f_2}_{[\a][\beta]} = 2 S^{{\rm TFT}}_{\a \b}\;,
\\
&h^{{\rm TFT}/\mathbb{Z}^f_2}_{[\alpha]} =  h^{\rm TFT}_\alpha\; (\textrm{mod 1/2})\;.
\end{split}
\end{align}
Note that the topological spin of $|[\alpha]\rangle$ is only defined modulo $1/2$ (instead of $1$) after the quotient since  $h_{\alpha}^{\rm TFT} - h_{\a_{\mathbb{Z}_2}\cdot \alpha}^{\rm TFT} =\pm 1/2$ for fermionic $\mathbb{Z}_2$. This is also compatible with the fact that  anyon spins are defined only modulo 1/2 in spin-TQFT.

\paragraph{Under Galois conjugation } For a given unitary TQFT satisfying \eqref{Unitarity in S}, there could be  non-unitary TQFTs  violating   \eqref{Unitarity in S} called Galois conjugates. These non-unitary theories have several properties in common with the unitary TQFT.  Galois conjugate pair has the same ground state degeneracy, $\textrm{GSD}_g= \mathcal{Z}^{\CM_{g,p=0}}$, on any Riemann surface $\Sigma_g$. According to the formula in \eqref{Zgp from S and T}, it implies that 
\begin{align}
\textrm{Galois conjugate pair has the same set of } \{S_{0\alpha}^2\}_{\alpha=0}^{N-1}\;.
\end{align}
But the pair has different $S_{00} =| \CZ^{S_3}|$ and the unitarity condition in \eqref{Unitarity in S} says that
\begin{align}
S_{00}(\textrm{non-unitary Galois conjugate})>S_{00}(\textrm{unitary  TQFT})\;.
\end{align}
From the computation of $\CZ^{\CM_{g,p=0}}$ one can determines the set $\{S_{0\alpha}^2\}_{\a=0}^{N-1}$,  
while from the $|\CZ^{S^3}|$ one can  determine $S_{00}$. So, from the two  computations, one can determine whether the TQFT is  unitary satisfying \eqref{Unitarity in S} or not. 
\paragraph{Example : $U(1)_k$ theory} The action for the theory is given as
\begin{align} 
S = \frac{k}{4\pi} \int_{\CM_3} A\wedge dA = \frac{k}{4\pi} \int_{X_4\;:\; \partial X_4 = \CM_3} F\wedge F\;.
\end{align}
The action depends on the choice of a 4-manifold $X_4$ whose boundary is $\CM_3$. Two different choices of the 4-manifolds, say $X_4$ and $Y_4$, give the following difference in the action
\begin{align}
\Delta S = \frac{k}{4\pi} \left(\int_{X_4 } F\wedge F - \int_{Y_4} F\wedge F\right) = \frac{k}{4\pi} \int_{\CM_4 := X_4 \cup \overline{Y_4}} F\wedge F\;.
\end{align}
Here $\CM_4$ is a closed orientable 4-manifold obtained by gluing $X_4$ and $\overline{Y}_4$, an orientation reversal of $Y_4$, along the common boundary $\CM_3$. Since 
\begin{align}
\;\int_{\CM_4} F\wedge F \in 4\pi^2 \mathbb{Z}\quad (\textrm{for any closed orientiable $\CM_4$})\;,
\end{align}
the action is well-defined modulo $\pi k$ and  thus
\begin{align}
	&\textrm{$e^{iS}$ depends only on $\CM_3$ (but not on $X_4$)   when $k\in 2\mathbb{Z}$}\;.
\end{align}
On the other hand, if we restrict the case when $\CM_4$ is a spin 4-manifold
\begin{align}
	\;\int_{\CM_4} F\wedge F \in 8\pi^2 \mathbb{Z}\quad (\textrm{for any closed spin  $\CM_4$})\;.
\end{align}
It means that we choose  a particular spin choice on $\CM_3$ and the  4-manifold $X_4$  is chosen such that it has a spin structure which is compatible the spin structure of the boundary $\CM_3$. Then the $\CM_4 = X_4 \cup \overline{Y}_4$ for two possible such extensions of $\CM_3$ has a spin structure.  Thus, 
\begin{align}
	&\textrm{$e^{iS}$ depends  only on $\CM_3$ and its spin-structure  (but not on $X_4$)  when $k\in 2\mathbb{Z}+1$}\;.
\end{align}
Actually,  the $U(1)_k$ theory is a spin or non-spin  TQFT depending on evenness/oddness of $k$
\begin{align}
\textrm{$U(1)_k$ is } 
\begin{cases}
\textrm{non-spin (bosonic) TQFT}  \quad \textrm{ if $k\in 2\mathbb{Z}$}\;,
\\
\textrm{spin (fermionic) TQFT}  \quad \textrm{ if $k\in 2\mathbb{Z}+1$}\;.
\end{cases}
\end{align}
Modular data ($S$ and $T$ matrices) of $U(1)_{k}$ with even $k$ is
\begin{align}
\begin{split}
&k \in 2\mathbb{Z}_{>0}\;:\; S_{\a\b}=\frac{1}{\sqrt{k}}e^{\frac{2\pi i \a\b}{k}} ,\;\;
T_{\a\b}=\delta_{\a\b}e^{2\pi i h_{\a}} \textrm{ with } h_{\a}:=\frac{\a^2}{2k} \;(\textrm{mod }1) \;, 
\\
&\textrm{where $\a,\b = 0,1,\cdots,k-1$\;. }
\label{U(1)_k MTC}
\end{split}
\end{align}
On the other hand, modular data ($S$ and $T^2$ matrices) of $U(1)_{k}$ with odd $k$ is
\begin{align}
	\begin{split}
		&k \in 2\mathbb{Z}_{>0}-1\;:\; S_{\a\b}=\frac{1}{\sqrt{k}}e^{\frac{2\pi i \a\b}{k}} ,\;\;
		(T^2)_{\a\b}=\delta_{\a\b}e^{4\pi i h_{\a}} \textrm{ with } h_{\a}:=\frac{\a^2}{2k} \;(\textrm{mod }\frac{1}2) \;,   
		\\
		&\textrm{where $\a,\b = 0,1,\cdots,k-1$\;.}
		\label{U(1)_k MTC-2}
	\end{split}
\end{align}
The loop operator $\CO_\alpha^\Gamma$ corresponds to the Wilson loop of $U(1)$ gauge charge $\alpha$, i.e.
\begin{align}
\CO^\Gamma_\alpha = \exp \left( i \alpha \oint_\Gamma A \right)\;.
\end{align}
The fusion coefficients of $U(1)_{k}$ can be computed from \eqref{Fusion Coefficient formula} and \eqref{U(1)_k MTC} as
\begin{align}
	N^{\gamma}_{\a\b} = \delta^{\gamma}_{\a+\b \, (\text{mod } k)}\;, \;\textrm{i.e.} \quad \CO_\alpha \times \CO_{\beta } = \CO_{\alpha+\beta (\textrm{mod }k)}\;.
\end{align}
The $U(1)_k$ theory has one-form $\mathbb{Z}_k$ symmetry generated by 
\begin{align}
\CO_{\alpha_{\mathbb{Z}_k}} = \CO_{\alpha=1}\;, \textrm{ which satisfies }  (\CO_{\alpha_{\mathbb{Z}_k}})^k=1\;.
\end{align}
For even $k$, the theory has $\mathbb{Z}_2 \subset \mathbb{Z}_k$ one-form symmetry generated by $\CO_{\alpha_{\mathbb{Z}_2}} = \CO_{\alpha = \frac{k}2}$. The topological spin of the symmetry generating anyon is
\begin{align}
h_{\alpha_{\mathbb{Z}_2}} = \frac{k}{8} \;(\textrm{mod }1) = \begin{cases} \frac{1}4 \; (\textrm{mod } 1)\; \quad \textrm{ if }k \in 4\mathbb{Z}+2\;,
	\\
	0 \; (\textrm{mod } 1)\; \quad \textrm{ if } k \in 8 \mathbb{Z}\;,
	\\
	\frac{1}2 \; (\textrm{mod } 1)\;\quad \textrm{ if } k \in 8 \mathbb{Z}+4\;.
	\end{cases}
\end{align}
Thus, according to \ref{3 cases of Z2 one-form}
\begin{align}
\mathbb{Z}_2 \textrm{ in $U(1)_{k \in 2\mathbb{Z}}$ is } \begin{cases} \textrm{anomalous}\quad \textrm{if }k \in 4\mathbb{Z}+2\;,
	\\
	\textrm{non-anomalous and bosonic} \quad  \textrm{if } k \in 8 \mathbb{Z}\;,
	\\
	\textrm{non-anomalous and fermionic}\quad  \textrm{if } k \in 8 \mathbb{Z}+4\;.
	\end{cases}
\end{align}
\paragraph{Example : Modular data of $SU(2)_k$} The action of the topological field theory is 
\begin{align}
S = \frac{k}{4\pi} \int_{\CM_3} \textrm{Tr} \left(A\wedge dA +\frac{2 i }3 A\wedge A \wedge A \right) = \frac{k}{4\pi} \int_{X_4 \;:\; \partial X_4 = \CM_3} \textrm{Tr} \left(F\wedge F\right)\;.
\end{align}
The topological theory is a  non-spin TQFT with a $\mathbb{Z}_2$ one-form symmetry. 
Modular data $(S,T)$ of $SU(2)_{k>0}$ TFT are ($\alpha, \beta = 0,1, \ldots k$)
\begin{align}
S_{\a\b}=\sqrt{\frac{2}{k+2}}\sin\Big({\frac{\pi (\a+1) (\b+1)}{k+2}}\Big)
\;,\quad
T_{\a\b}=\delta_{\a\b}e^{2\pi i h_{\a}}
\;  \textrm{ with } h_{\a}=\frac{\a(\a+2)}{4(k+2)} \;(\textrm{mod }1)\;.
\label{SU(2)_k MTC}
\end{align}
The loop operator $\CO_\alpha$ corresponds to the Wilson loop operator in the representation $\textrm{Sym}^{\otimes \alpha} \Box$, $\alpha$-th symmetric product of the fundamental representation, i.e.
\begin{align}
\CO_\alpha^\Gamma = \textrm{Tr}_{R = \textrm{Sym}^{\otimes \alpha}\Box} \left( P \exp \left(i \oint_\Gamma A \right) \right)\;.
\end{align}
The $\mathbb{Z}_2$ one-form symmetry is generated  by 
\begin{align}
\CO_{\alpha_{\mathbb{Z}_2}} = \CO_{\alpha = k } \textrm{  with } h_{\alpha_{\mathbb{Z}_2}} = \frac{k}4\;.
\end{align}
According to \eqref{3 cases of Z2 one-form}
\begin{align}
\textrm{The $\mathbb{Z}_2$ is } 
\begin{cases}
\textrm{anomalous}\quad \textrm{if }k\in 2\mathbb{Z}+1	\;,
	\\
\textrm{non-anomalous and bosonic} \quad \textrm{if }k\in 4\mathbb{Z}	\;,
	\\
\textrm{non-anomalous and fermionic} \quad \textrm{if }k\in 4\mathbb{Z}+2	\;.
\end{cases}
\end{align}
\paragraph{Example : $U(1)_{4k}/\mathbb{Z}_2 = U(1)_k$} The non-anomalous $\mathbb{Z}_2$ symmetry is generated by $\CO_{\alpha=2k}$ which is bosonic (resp. fermionic) for  even $k$ (resp. odd $k$). 
After the one-form $\mathbb{Z}_2$ gauging, the Hilbert-space on the two-torus is 
\begin{align}
\begin{split}
&k \in 2\mathbb{Z}_{>0} \;:\;\CH^{U(1)_{4k}/\mathbb{Z}_2}(\mathbb{T}^2) = \textrm{Span} \bigg{\{} |[\alpha]\rangle := \frac{1}{\sqrt{2}} (|2\alpha \rangle + |2\alpha+2k \rangle  )\;:\; \alpha =0, 1,\ldots,  k-1 \bigg{\}}\;,
\\
&k \in 2\mathbb{Z}_{>0}-1 \;:\;\CH^{U(1)_{4k}/\mathbb{Z}_2}_{--}(\mathbb{T}^2) = \textrm{Span} \bigg{\{} |[\alpha]\rangle := \frac{1}{\sqrt{2}} (|2\alpha \rangle + |2\alpha+2k \rangle  )\;:\; \alpha =0, 1,\ldots,  k-1 \bigg{\}}\;.
\end{split}
\end{align}
The modular data of the $U(1)_{4k}/\mathbb{Z}_2$ theory is ($\a,\b = 0,1,\cdots,k-1$)
\begin{align}
&S_{[\a][\b]}^{U(1)_{4k}/\mathbb{Z}_2} = \frac{2}{\sqrt{4k}} e^{\frac{2\pi i (2\a)(2\b)}{4k}} = S_{\a\b}^{U(1)_k}\;, \quad h_{[\a]}^{U(1)_{4k}/\mathbb{Z}_2}  
= 
\begin{cases}
 \frac{(2\a)^2}{8k}  \;(\text{mod}\,1 )\; \;\; \textrm{if } k \in 2 \mathbb{Z}_{>0}\;,
 \\
  \frac{(2\a)^2}{8k} \;(\text{mod}\,\frac{1}2 ) \; \;\; \textrm{if } k \in 2 \mathbb{Z}_{>0}-1\;.
\end{cases}
\end{align}
It implies that  $U(1)_{4k}/\mathbb{Z}_2$ is actually the $U(1)_k$ theory. From the $\mathbb{Z}_2$ gauging, one can also confirm that the $U(1)_k$ is a non-spin (resp.\ spin) TQFT for even $k$ (resp.\ odd $k$) since the $\mathbb{Z}_2$ theory is bosonic (resp.\ fermionic).
\paragraph{Example : $SU(2)_{2k}/\mathbb{Z}_2 $} 
For odd $k$, the $\mathbb{Z}_2$ one-form symmetry is fermionic and we have a spin topological theory after the $\mathbb{Z}_2$ gauging. The Hilbert-space on the two-torus in the NS-NS sector  is
\begin{align}
k \in 2 \mathbb{Z}_{\geq 0}+1\;:\;\CH^{SU(2)_{2k}/\mathbb{Z}_2}_{--} (\mathbb{T}^2) = \textrm{Span} \bigg{\{} |[\a]\rangle := \frac{1}{\sqrt{2}}\left(|2\alpha \rangle + |2k-\a \rangle \right) \;:\; \alpha =0, \ldots, \frac{(k-1)}2 \bigg{\}}\;.
\end{align}
%
%
On the basis, the modular $S, T^2$ matrices are 
\begin{align}
\begin{split}
&S_{[\a][\b]}^{SU(2)_{2k}/\mathbb{Z}_2} = 2\sqrt{\frac{2}{2k+2}}\sin\Big(\frac{\pi(2\a+1)(2\b+1)}{2k+2}\Big)\;,
\\
&(T^2)^{SU(2)_{2k}/\mathbb{Z}_2}_{[\alpha] [\beta]} = \delta_{\alpha \beta } e^{2\pi i h_{[\a]}}\; \textrm{ with } h_{[\a]} = \frac{\a(\a+1)}{2(k+1)} \; (\text{mod}\, \frac{1}2) \;.
\end{split}
\end{align}
%
 For even $k$, on the other hand, the $\mathbb{Z}_2$ one-form symmetry is bosonic and the resulting theory after the gauging is a bosonic topological field theory.  The Hilbert space on the two-torus is
\begin{align}
\begin{split}
&k \in 2 \mathbb{Z}_{> 0}\;:\;\CH^{SU(2)_{2k}/\mathbb{Z}_2} (\mathbb{T}^2) = \CH^{SU(2)_{2k}/\mathbb{Z}_2}_{\rm untwisted}(\mathbb{T}^2) \oplus  \CH^{SU(2)_{2k}/\mathbb{Z}_2}_{\rm twisted}(\mathbb{T}^2)\; \textrm{ where}
\\
& \CH^{SU(2)_{2k}/\mathbb{Z}_2}_{\rm untwisted}= \textrm{Span} \bigg{\{} |[\a]\rangle := \frac{1}{\sqrt{2}}\left(|2\alpha \rangle + |2k-\a \rangle \right) \;:\; \alpha =0, \ldots, \frac{(k-2)}2 \bigg{\}}\;,
\\
&\CH^{SU(2)_{2k}/\mathbb{Z}_2}_{\rm twisted} = \textrm{Span} \big{\{} |k ;+ \rangle, \;  |k;- \rangle  \big{\}}\;.
\end{split}
\end{align}
The modular data of the topological theory is 
\begin{align}
\begin{split}
	&S_{[\a][\b]}^{SU(2)_{2k}/\mathbb{Z}_2} = 2\sqrt{\frac{2}{2k+2}}\sin\Big(\frac{\pi(2\a+1)(2\b+1)}{2k+2}\Big)\;, \;\; S^{SU(2)_{2k}/\mathbb{Z}_2}_{[0](k;\pm)} = \sqrt{\frac{2}{2k+2}} \;,
	\\
	&\textrm{and}
	\\
	&h_{[\a]} = \frac{\a(\a+1)}{2(k+1)} \; (\text{mod}\, 1)\;, \;\;
h_{(k;\pm)} = \frac{k(k+2)}{8(k+1)}\;(\text{mod}\; 1)\;.
\end{split}
\end{align}

\paragraph{Example : $\frac{SU(2)_{10}\times SU(2)_2}{\mathbb{Z}_2^{\rm diag}}$} The Hilbert-space of $SU(2)_{10}\times SU(2)_2$ theory on the two torus is 
\begin{align}
\CH^{SU(2)_{10}\times SU(2)_2}(\mathbb{T}^2) =  \textrm{Span} \big{\{} |\alpha_1, \alpha_2\rangle \; :\; 0\leq \alpha_1 \leq 10, \;\;  0 \leq \alpha_2 \leq 2 \big{\}}\;.
\end{align}
Modular data is
\begin{align}
S^{SU(2)_{10}\times SU(2)_2}_{(\alpha_1, \a_2), (\b_1, \b_2)} =  S^{SU(2)_{10}}_{\alpha_1 \beta_1} \times S^{SU(2)_{2}}_{\alpha_2 \beta_2}, \quad T^{SU(2)_{10}\times SU(2)_2}_{(\alpha_1, \a_2), (\b_1, \b_2)} =  T^{SU(2)_{10}}_{\alpha_1 \beta_1} \times T^{SU(2)_{2}}_{\alpha_2 \beta_2}\;.
\end{align}
The theory has $\mathbb{Z}^{(1)}_2 \times \mathbb{Z}^{(2)}_2$ one-form symmetry generated by 
\begin{align}
\CO_{\alpha_{\mathbb{Z}_2^{(1)}}} = \CO_{(\alpha_1=10,\alpha_2=0)}\;, \quad \CO_{\alpha_{\mathbb{Z}_2^{(2)}}} = \CO_{(\alpha_1=0,\alpha_2=2)}\;.
\end{align}
Both $\mathbb{Z}_2^{(1)}$ and $\mathbb{Z}_2^{(2)}$ are fermionic. The diagonal $\mathbb{Z}_2^{\rm diag}$ one-form symmetry is generated by 
\begin{align}
\CO_{\alpha_{\mathbb{Z}_2^{\rm diag}}} = \CO_{(\alpha_1, \alpha_2)=(10,2)}
\end{align}
and is bosonic. After the $\mathbb{Z}_2^{\rm diag}$ gauging, the Hilbert-space on the two torus is 
\begin{align}
\begin{split}
&\CH^{(SU(2)_{10}\times SU(2)_2)/\mathbb{Z}_2^{\rm diag}} (\mathbb{T}^2) = \CH^{(SU(2)_{10}\times SU(2)_2)/\mathbb{Z}_2^{\rm diag} }_{\rm untwisted}(\mathbb{T}^2) \oplus  \CH^{(SU(2)_{10}\times SU(2)_2)/\mathbb{Z}_2^{\rm diag}}_{\rm twisted}(\mathbb{T}^2),
\\
& \CH^{(SU(2)_{10}\times SU(2)_2)/\mathbb{Z}_2^{\rm diag} }_{\rm untwisted}(\mathbb{T}^2)  = \textrm{Span} \bigg{\{} |[\alpha] \rangle :=\frac{1}{\sqrt{2}} \left(|2\alpha,0 \rangle +  |10-2\alpha,2   \rangle  \right) \;:\; \alpha=0, \ldots, 5 \bigg{\}}
\\
& \qquad \qquad \qquad \qquad \qquad \qquad \oplus  \textrm{Span} \bigg{\{} |[\tilde{\alpha}]\rangle :=\frac{1}{\sqrt{2}} \left(|2\tilde{\alpha}+1,1 \rangle +  |9-2\tilde{\alpha},1   \rangle  \right) \;:\; \tilde{\alpha} =0,\ldots, 1  \bigg{\}}, 
\\
& \CH^{(SU(2)_{10}\times SU(2)_2)/\mathbb{Z}_2^{\rm diag} }_{\rm twisted}(\mathbb{T}^2)  = \textrm{Span} \bigg{\{} |5,1;+\rangle,\; |5,1;-\rangle \bigg{\}}\;.
\end{split}
\end{align}
There are 10 simple objects and their $\{S_{0\alpha}\}$ are ($|0\rangle = |[\alpha=0]\rangle$)
\begin{align}
\begin{split}
&S_{0[\alpha]} = \sqrt{\frac{1}6} \sin \left( \frac{\pi (2\alpha+1)}{12}\right)\;, \quad  S_{0[\tilde{\alpha}]} = \sqrt{\frac{1}3} \sin \left( \frac{\pi (2\tilde{\alpha}+2)}{12}\right)\;,   
\\
&S_{0,(5,1;+)} = S_{0,(5,1;-)} = \frac{1}{2\sqrt{3}}\;. \label{S0a of SU(2)10*SU(2)3/Z2}
\end{split}
\end{align}
\paragraph{Example : Lee-Yang TQFT as a Galois conjugation of Fibonacci TQFT} The Fibonacci topological field is 
\begin{align}
\textrm{Fibonacci TQFT : } \frac{SU(2)_3 \otimes U(1)_2}{\mathbb{Z}_2}\;\textrm{ or equivalently } (G_2)_1 \;.
\end{align}
The modular data of the bosonic topological field theory is
\begin{align}
 S =  \begin{pmatrix}
\sqrt{\frac{1}{10} \left(5-\sqrt{5}\right)}   &	\sqrt{\frac{1}{10} \left(\sqrt{5}+5\right)}\\
	\sqrt{\frac{1}{10} \left(\sqrt{5}+5\right)} & -\sqrt{\frac{1}{10} \left(5-\sqrt{5}\right)}  
\end{pmatrix}\;, \quad  T =  \begin{pmatrix}
	1 &  0 \\
	0  & \exp (\frac{ 4\pi i }5) \;.
\end{pmatrix}
\end{align}
The non-unitary Lee-Yang TQFT, whose modular data is given in \eqref{S,T of Lee-Yang}, is a Galois conjugate of the Fibonacci TQFT. 
%
%

\section{\texorpdfstring{Dual description for $\frac{T[SU(2)]}{``PSU(2)^{\rm diag}_k"}$}{Dual description for {T[SU(2)]}/{``PSU(2)(diag)(k)"}}} \label{App : 3D-3D for mapping torus}

In term of the 3D-3D correspondence, the theory $\frac{T[SU(2)]}{``PSU(2)^{\rm diag}_k"}$ corresponds to a 3-manifold called the once-punctured torus bundle with monodromy matrix $\varphi = \mathbb{S}\mathbb{T}^k$ \cite{Terashima:2011qi,Gang:2013sqa,Gang:2015wya}
\begin{align}
\mathcal{T}[(\Sigma_{1,1}\times S^1)_{\varphi = \mathbb{ST}^k} ; A =S^1_{\textrm{punct}})] = 	\frac{T[SU(2)]}{``PSU(2)^{\rm diag}_k"}\;.  \label{Duality for diag gauging}
\end{align}
The once-punctured torus bundle $(\Sigma_{1,1}\times S^1)_\varphi $ with $\varphi \in SL(2, \mathbb{Z})$ is defined as
\begin{align}
\begin{split}
&(\Sigma_{1,1}\times S^1)_\varphi  = (\Sigma_{1,1} \times [0,1])/\sim\;,  \quad \textrm{where}
\\
& \left(x, 0 \right) \sim \left(\varphi(x), 1\right)\;.
\end{split}
\end{align}
Here $\Sigma_{g=1,h=1}$ is the once-punctured torus and $\varphi \in SL(2,\mathbb{Z})$ is an element of mapping class group of the Riemann surface. The mapping torus  actually depends only on the conjugacy class of  $\varphi$  in $SL(2,\mathbb{Z})$, i.e.
\begin{align}
(\Sigma_{1,1}\times S^1)_{\varphi_1} = (\Sigma_{1,1}\times S^1)_{\varphi_2} \;\textrm{ if and only if } \; \varphi_1 \sim \varphi_2\;.
\end{align}
Here $\varphi_1\sim \varphi_2$ means that $\varphi_1$ are $\varphi_2$ are related to each other by conjugation in $SL(2,\mathbb{Z})$.
The mapping torus has a torus boundary. Generally, for 3-manifold $N$ with a torus boundary, we need to choose primitive boundary 1-cycle $A \in H_1 (\partial N , \mathbb{Z})$ to specify its associated 3D gauge theory $\mathcal{T}[N; A]$ \cite{Dimofte:2011ju,Gang:2018wek}. In the once-puncture torus bundle, there is a natural  choice of the boundary 1-cycle $A = S^1_{\rm punct}$, which is the cycle encircling  the  puncture  in $\Sigma_{1,1}$.

In the view point of 3D-3D correspondence, the conformal window in \eqref{IR phases of diag gauging} can be geometrically understood from the following topological fact:
\begin{align}
(\Sigma_{1,1}\times S^1)_{\varphi = \mathbb{ST}^k} \textrm{ is } \begin{cases}
\textrm{non-hyperbolic}\quad\textrm{if $|k|< 2$} \;,\\
\textrm{hyperbolic} \quad  \textrm{if $|k|> 2$}\;. \\
\end{cases}
\end{align}

We will focus on the case when the mapping torus $(\Sigma_{1,1}\times S^1)_\varphi$ is hyperbolic. For the case, the conjugacy class of  $\varphi $ can be decomposed into positive powers of $\mathbb{L}$ and $\mathbb{R}$ (up to sign)
\begin{align}
\begin{split}
&\varphi  = \pm g   \left(\mathbb{L}^{n_1} \mathbb{R}^{n_2} \mathbb{L}^{n_3}\ldots \mathbb{L}^{n_L} (\textrm{or } \mathbb{R}^{n_L})  \right) g^{-1}\; \quad n_i \in \mathbb{Z}_{>0}\;,
\\
& \mathbb{L} = \begin{pmatrix}
1 & 1 \\
0 & 1
\end{pmatrix} \;, \quad  \mathbb{R} = \begin{pmatrix}
1 & 0 \\
1 & 1
\end{pmatrix} \;.
\end{split}
\end{align}

The mapping torus $(\Sigma_{1,1}\times S^1)_\varphi$ has an alternative topological description based on an ideal triangulation. Using an ideal triangulation of $(\Sigma_{1,1}\times S^1)_\varphi$, one can give an alternative description for $\mathcal{T}[(\Sigma_{1,1}\times S^1)_\varphi ;A =S^1_{\rm punct}] $ following the algorithm proposed in \cite{Dimofte:2011ju}.  
Interestingly, the 3D gauge theory based on an ideal triangulation has only manifest $\mathcal{N}=2$ supersymmetry. We expect that the $\mathcal{N}=2$ gauge theories have enhanced $\mathcal{N}=4$ supersymmetry at IR. 

\paragraph{$k=3\; (\varphi= \mathbb{LR} \sim \mathbb{ST}^3)$ case : $(\mathcal{T}_{\rm min})^{\otimes 2}$}  The corresponding mapping-torus  can be decomposed into two ideal tetrahedrons \cite{Gang:2013sqa}. The corresponding 3D $\CN=2$ theory is \cite{Gang:2018huc}
\begin{align}
\mathcal{T}[(\Sigma_{1,1}\times S^1)_{\varphi =\mathbb{LR}} ;A=S^1_{\rm punct}] =   \left(U(1)_{3/2} +\Phi\right) \otimes  \left(U(1)_{-3/2} +\Phi\right)\;.
\end{align}
The theory is nothing but $(\CT_{\rm min})^{\otimes 2}$ using the duality between $(U(1)_{3/2}+\Phi)$ and $(U(1)_{-3/2}+\Phi)$. 
\paragraph{$k=4\; (\varphi =\mathbb{LLR} \sim \mathbb{ST}^4$) case : $\mathcal{N}=2 \rightarrow \mathcal{N}=5$ }The corresponding mapping torus can be decomposed into three ideal tetrahedrons \cite{Gang:2013sqa}. According to the algorithm in \cite{Dimofte:2011ju}, the corresponding 3D $\CN=2$ field theory is 
\begin{align}
\begin{split}
&\mathcal{T}[(\Sigma_{1,1}\times S^1)_{\varphi = \mathbb{LLR}} ;A =S^1_{\rm punct}] 
\\
&=  \big{(}\textrm{3D $\CN=2$ $U(1)\times U(1)$ gauge theory with mixed CS level $K$ coupled to } 
\\
&\quad \quad\textrm{3 chiral multiplets ($\Phi_1, \Phi_2, \Phi_3$)  of charge  $\mathbf{Q}$}
\\
& \quad \quad \textrm{with  superpotential }W = (\Phi_1 \Phi_2 \Phi_3)^2 + V_{\mathbf{m}=(1,-1)} \big{)}\;.
\end{split}
\end{align}
The mixed CS level $K$ for $U(1) \times U(1)$ gauge group is
\begin{align}
K = \begin{pmatrix} 
-1 & -1/2 \\
-1/2 & -1 
\end{pmatrix}
\;.
\end{align}
Gauge charges $\mathbf{Q}$ for 3 chirals are assigned  as follows
\begin {table}[ht]
\begin{center}
	\begin{tabular}{| c| c | c|}
		\hline
		& $U(1)$ & $U(1)$
		\\
		\hline
		$\Phi_1$ & $1$ & $0$
		\\ 
		\hline
		$\Phi_2$ & $-1$ & $1$
		\\ 
		\hline
		$\Phi_3$ & $0$ & $-1$
		\\
		\hline
	\end{tabular} 
\end{center}
\end{table}
\\
$V_{\mathbf{m}=(\mathfrak{m}_1, \mathfrak{m}_2)}$  denotes the 1/2 BPS bare monopole operator with monopole fluxes $(\mathfrak{m}_1, \mathfrak{m}_2)$ coupled to the two gauge $U(1)$s. The bare monopole operator is a gauge-invariant 1/2 BPS chiral primary  when $\mathfrak{m}_1 +\mathfrak{m}_2 =0$.  
 The superpotential breaks the $U(1)^3$ flavor symmetry  to $U(1)$. From the F-maximization, the IR superconformal R-charge ($\nu=0$) is determined as
\begin{align}
R_{\nu=0 }(\Phi_1) = R_{\nu=0} (\Phi_3) = 1, \; R_{\nu=0}(\Phi_2)=-1\;.
\end{align}
The superconformal index at the IR conformal fixed point is 
\begin{align}
\begin{split}
&\CI^{\rm sci}_{\CT[(\Sigma_{1,1}\times S^1)_{\varphi = \mathbb{LLR}},A =S^1_{\rm punct}]}(q,\eta,\nu=0;s=1) 
\\
&= 1+ q^{1/2} - \left(\eta +\frac{1}\eta  +1\right)q-(2+ \eta + \eta^{-1})q^{3/2}+\ldots \;.
\end{split}
\end{align}
Surprisingly, the index show $\CN=5$ supersymmetry instead of $\CN=4$ \cite{Evtikhiev:2017heo}. Actually, from the superconformal index computation, one can confirm that the theory is  dual to the following $\CN=5$ gauge theory
\begin{align}
\begin{split}
&\CT[(\Sigma_{1,1}\times S^1)_{\varphi =\mathbb{ LLR}};A = S^1_{\rm punct}] =   \left(SU(2)^{\frac{1}2\oplus \frac{1}2}_{|k|=3} \textrm{in \eqref{SU(2)k1/2+1/2}}\right)\;.
\end{split}
\end{align}
%

\section{Contour integrals}  \label{App : Contour}
%
We explicitly evaluate the contour integrals that appear in this paper.
%
\subsection{\texorpdfstring{$\mathcal{Z}^{S_b^3}_{T[SU(2)]}(b=1,X_1, X_2 , m=0 , \nu)$}{Z(S3)(T[SU(2)](b=1,X1, X2 , m=0 , nu))}}\label{App : T[SU(2)] nu=0}
With the properties in Appendix \ref{App : QDL}, the partition function \eqref{TSU(2) partition function} for $b=1$, $m=0$, and $\nu=0$ is simplified as
\begin{align}
    \mathcal{Z}^{S_b^3}_{T[SU(2)]}(b=1,X_1, X_2 , m=0 , \nu=0) 
    = \frac{e^{\frac{2\pi i}{3}}}{4\pi} \int \text{d}Z \frac{e^{\frac{Z X_2}{\pi i}}}{\cosh (Z) + \cosh(X_1)} \;.
    \label{T[SU(2)] b=1 nu=0 partition function}
\end{align}
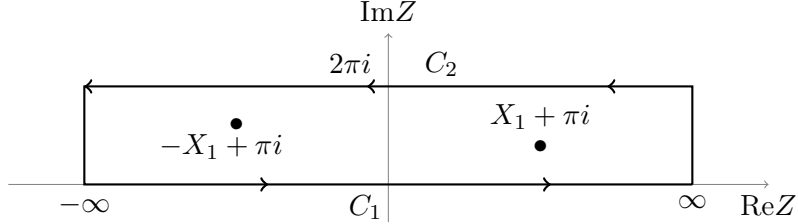
\begin{figure}[H]
    \centering
    \begin{tikzpicture}
[decoration={markings,
	mark=at position 0.13 with {\arrow[line width=1pt]{>}},
	mark=at position 0.3 with {\arrow[line width=1pt]{>}},
	mark=at position 0.5 with {\arrow[line width=1pt]{>}},
	mark=at position 0.7 with {\arrow[line width=1pt]{>}},
	mark=at position 0.9 with {\arrow[line width=1pt]{>}}
}
]
\draw[help lines,->] (-5,0) -- (5,0) coordinate (xaxis);
\draw[help lines,->] (0,-0.5) -- (0,2) coordinate (yaxis);

\path[draw,line width=0.8pt,postaction=decorate] 
(4,0) node[below] {$\infty$} -- (4,1.3) --  (-4,1.3) -- (-4,0) node[below] {$-\infty$} -- (4,0);

\node at (-2,0.8) {$\bullet$};
\node at (-2.2, 0.5) {$-X_1 + \pi i$};
\node at (2,0.5) {$\bullet$};
\node at (2,0.9) {$X_1 + \pi i$};

\node[below] at (xaxis) {Re$Z$};
\node at (0,2.3) {Im$Z$};
\node at (-0.3,-0.3) {$C_1$};
\node at (0.7,1.6) {$C_2$};
\node at (-0.5,1.6) {$2 \pi i$};

\end{tikzpicture}
    \caption{A contour for the evaluation of \eqref{T[SU(2)] b=1 nu=0 partition function}. Assuming $|\text{Im}[X_1]| < \pi$, there are two simple poles at $Z = \pm X_1 + \pi i$ inside the path.}
    \label{T[SU(2)] nu=0 contour}
\end{figure}
The contour integral of the integrand in \eqref{T[SU(2)] b=1 nu=0 partition function} along the path in Fig. \ref{T[SU(2)] nu=0 contour} is
\begin{align}
\int_{-\infty}^{\infty} & \text{d}Z \frac{e^{\frac{i Z X_2}{\pi}}}{\cosh(Z) + \sinh(Z)} + \int_{\infty}^{-\infty} \text{d}Z \frac{e^{\frac{i Z X_2}{\pi}}e^{-2 X_2}}{\cosh(Z) + \sinh(Z)} = 2\pi i \Bigg( 
\frac{e^{-\frac{i X_1 X_2}{\pi}}e^{-X_2}}{\sinh(X_1)} - \frac{e^{\frac{i X_1 X_2}{\pi}}e^{-X_2}}{\sinh(X_1)}
\Bigg)
\nonumber\\
&\rightarrow
\int_{-\infty}^{\infty} \text{d}Z \frac{e^{\frac{Z X_2}{\pi i}}}{\cosh(Z) + \sinh(Z)}
=
\frac{4\pi \sin\Big( \frac{X_1 X_2}{\pi}\Big) }{\sinh(X_1)\Big(e^{X_2}-e^{-X_2}\Big)}
=
\frac{2\pi \sin\Big(\frac{X_1 X_2}{\pi}\Big) }{\sinh(X_1) \sinh(X_2) } \;.
\label{T[SU(2)] integration}
\end{align}
Here, the first and second term of the first line of \eqref{T[SU(2)] integration} are for the path $C_1$ and $C_2$ in the Fig. \ref{T[SU(2)] nu=0 contour} respectively. Restoring the factor $\frac{e^{\frac{2\pi i}{3}}}{4\pi}$, we have
\begin{align}
\mathcal{Z}^{S_b^3}_{T[SU(2)]}(b=1,X_1, X_2 , m=0 , \nu=0) = \frac{e^{\frac{2\pi i}{3}}}{2} \frac{ \sin\Big(\frac{X_1 X_2}{\pi}\Big) }{\sinh(X_1) \sinh(X_2) }
\,
,\text{ for } 
\,
\big|\text{Im}[X_1]\big| \leq \pi \;.
\end{align}
The computations for gauging $SU(2)^H$ or $SU(2)^C$ of this $T[SU(2)]$ theory at the conformal limit ($\nu=0$) are straightforward since they are always reduced to the simple Gaussian integration at the three-sphere partition function level.

For the degenerate limit, say $\nu = 1$, the three-sphere partition function $\mathcal{Z}^{S_b^3}_{T[SU(2)]}(b=1,m=0,\nu=1)$ diverges due to the factor $\psi_{\hbar=2\pi i}\Big(2\pi i(1-\nu)\Big)$ from the adjoint matter. To handle it, we expand the partition function divided by this divergence around $\nu=1$ as
\begin{align}
&\frac{\mathcal{Z}^{S_b^3}_{T[SU(2)]}(b=1,m=0,\nu)}{\psi_{\hbar=2\pi i}\Big( 2\pi i (1-\nu) \Big)}
=
\Bigg(\frac{e^{\frac{13\pi i}{12}}}{2\pi} \int \text{d}Z \quad e^{\frac{i Z X_2}{\pi}}\Bigg) 
\nonumber\\
&\qquad\qquad + \Bigg( \frac{e^{\frac{19\pi i}{12}}}{2} \int \text{d}Z \quad e^{\frac{i Z X_2}{\pi}}\Big(1-\frac{i}{\pi} \frac{Z\sinh(Z)-X_1\sinh(X_1)}{\cosh(Z)-\cosh(X_1)} \Big) \Bigg)(\nu-1)
+\mathcal{O}\big((\nu-1)^2 \big)\;.
\label{degenerate limit T[SU(2)] expansion}
\end{align}
The divergence of $\psi_{\hbar=2\pi i}\Big( 2\pi i (1-\nu) \Big)$ comes from the simple pole at $\nu=1$
\begin{align}
\psi_{\hbar=2\pi i}\Big( 2\pi i (1-\nu) \Big) = -\frac{e^{\frac{\pi i}{12}}}{2\pi i}\frac{1}{(\nu - 1)}  + \frac{e^{\frac{\pi i}{12}}(\pi - i)}{2\pi} + \mathcal{O}\big((\nu-1)^1\big)\;.
\label{diverging adjoint expansion}
\end{align}
The first term in \eqref{degenerate limit T[SU(2)] expansion} vanishes after $SU(2)_k$ gauging since
\begin{align}
\int \text{d}X \text{d}Z \,\, e^{\frac{i Z X}{\pi}} e^{\frac{k X^2}{2\pi i}} \sinh^2(X) = 0 \;.
\label{vanishing int}
\end{align}
This means that there is no divergence problem even for $\nu=1$ if we are considering, say, the $SU(2)_k$ gauged $T[SU(2)]$ theory, since \eqref{degenerate limit T[SU(2)] expansion} always starts with linear $(\nu-1)$ term which cancels the diverging simple pole in \eqref{diverging adjoint expansion} from the adjoint matter.
%
\subsection{\texorpdfstring{$\mathcal{Z}^{S_b^3}_{(k_1,k_2)}(b=1,m=0,\nu=\pm 1)$}{Z(S3b)(k1,k2)(b=1,m=0,nu=+-1)}}
\label{App : Simultaneous nu=pm1}
With the help of Appendix \ref{App : QDL}, \eqref{diverging adjoint expansion}, and \eqref{vanishing int}, the partition function \eqref{Simultaneous gauging partition function} for $b=1$, $\nu=1$ is simplified as (with Gaussian integral of $X_2$)
\begin{align}
\mathcal{Z}^{S_b^3}_{(k_1,k_2)}(b=1,m=0,\nu=1)
&=
\frac{e^{\frac{11\pi i}{12}}}{16\pi^3}\sqrt{\frac{2}{k_2}}
\int \text{d}Z \text{d}X_1 \sinh^2(X_1)
\bigg( \frac{Z\sinh(Z)-X_1\sinh(X_1)}{\cosh(Z)-\cosh(X_1)} \bigg)
\nonumber\\
&\qquad \times
\Big( e^{-\frac{2\pi i}{k_2}} e^{\frac{2Z}{k_2}} + e^{-\frac{2\pi i}{k_2}} e^{-\frac{2Z}{k_2}} -2 \Big)
e^{-\frac{Z^2}{2\pi i k_2}} e^{\frac{k_1 X_1^2}{2\pi i}}\;.
\end{align}
Changing the variables as $X_1\rightarrow A+B$, $Z\rightarrow A-B$, we have
\begin{align}
\mathcal{Z}^{S_b^3}_{(k_1,k_2)} (b=1,m=0,\nu=1)
&=
\frac{e^{\frac{11\pi i}{12}}}{2\pi^3}\sqrt{\frac{2}{k_2}}
\int \text{d}A \text{d}B
\sinh^2(A+B)
\frac{A\cosh(A)}{\sinh(A)}
\nonumber\\
&\times
e^{\frac{(k_1 k_2 -1)}{2\pi i k_2} A^2}
e^{\frac{(k_1 k_2 -1)}{2\pi i k_2} B^2}
e^{\frac{(k_1 k_2 +1)}{\pi i k_2} A B}
\Big(
e^{\frac{2A-2B-2\pi i}{k_2}}
-1
\Big)\;.
\label{Simultaneous AB integral}
\end{align}
where several even terms in the integrand under $A\rightarrow -A$, $B\rightarrow -B$ have been stacked up. With Gaussian integral of $B$, \eqref{Simultaneous AB integral} is further evaluated as
\begin{align}
\mathcal{Z}^{S_b^3}_{(k_1,k_2)} &(b=1,m=0,\nu=1)
=
\frac{e^{\frac{2\pi i}{3}}}{4\pi^2}\frac{1}{\sqrt{k_1 k_2 -1}}
\int \text{d}A
\frac{A \cosh(A)}{\sinh(A)}
\nonumber\\
&\times
\Bigg[
e^{-\frac{2(A+\pi i)^2}{\pi i k_2}}
\Big(
e^{-\frac{2(A+\pi i + k_2 \pi i)^2}{\pi i k_2 (k_1 k_2 -1)}}
+
e^{-\frac{2(A+\pi i - k_2 \pi i)^2}{\pi i k_2 (k_1 k_2 -1)}}
-
2 e^{-\frac{2(A+\pi i)^2}{\pi i k_2 (k_1 k_2 -1)}}
\Big)
\nonumber\\
&\qquad\qquad
-e^{-\frac{2 A^2}{\pi i k_2}}
\Big(
e^{-\frac{2(A + k_2 \pi i)^2}{\pi i k_2 (k_1 k_2 -1)}}
+
e^{-\frac{2(A-k_2 \pi i)^2}{\pi i k_2 (k_1 k_2 -1)}}
-
2 e^{-\frac{2A^2}{\pi i k_2(k_1 k_2 -1)}}
\Big)
\Bigg]\;.
\label{Simultaneous A integral}
\end{align}
The third and sixth terms of \eqref{Simultaneous A integral} can be evaluated from the following contour integrals:
\begin{figure}[H]
    \centering
    \begin{tikzpicture}
[decoration={markings,
	mark=at position 0.13 with {\arrow[line width=1pt]{>}},
	mark=at position 0.3 with {\arrow[line width=1pt]{>}},
	mark=at position 0.5 with {\arrow[line width=1pt]{>}},
	mark=at position 0.7 with {\arrow[line width=1pt]{>}},
	mark=at position 0.9 with {\arrow[line width=1pt]{>}}
}
]
\draw[help lines,->] (-5,0) -- (5,0) coordinate (xaxis);
\draw[help lines,->] (0,-0.5) -- (0,1.5) coordinate (yaxis);

\path[draw,line width=0.8pt,postaction=decorate] 
(4,0) -- (4,1) -- (0.3,1)  arc (0:-180:0.3) -- (-4,1) -- (-4,0)  -- (4,0);

\node at (0,1) {$\bullet$};

\node[below] at (xaxis) {Re$A$};
\node at (0,1.8) {Im$A$};
\node at (2,1.3) {$l$};
\node at (0.3,1.3) {$\pi i$};
\node at (4,-0.15) {$\infty$};
\node at (-4,-0.15) {$-\infty$};
\end{tikzpicture}
    \caption{A path $l$ for \eqref{Simultaneous contour 1}. There is a simple pole at $A = \pi i$.}
    \label{Diagram for Simultaneous contour 1}
\end{figure}
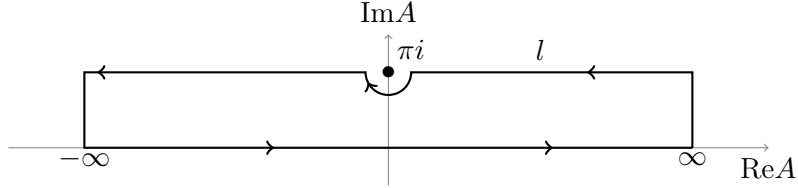
\begin{align}
&\oint_l \text{d}A \frac{A \cosh(A)}{\sinh(A)} e^{-\frac{2A^2}{\pi i k_2}} e^{-\frac{2A^2}{\pi i k_2(k_1 k_2 -1)}} = 0
\nonumber\\
&\rightarrow
\int \text{d}A \frac{A \cosh(A)}{\sinh(A)}
\Big(
e^{-\frac{2A^2}{\pi i k_2}} e^{-\frac{2A^2}{\pi i k_2(k_1 k_2 -1)}}
-
e^{-\frac{2(A+\pi i)^2}{\pi i k_2}} e^{-\frac{2(A+\pi i)^2}{\pi i k_2(k_1 k_2 -1)}}
\Big)
\nonumber\\
&\qquad=
\pi i a_+ + \frac{\pi i }{2}
\int \text{d}A \frac{\cosh(A)}{\sinh(A)} 
\Big(
e^{-\frac{2(A+\pi i)^2}{\pi i k_2}} e^{-\frac{2(A+\pi i)^2}{\pi i k_2(k_1 k_2 -1)}}
-
e^{-\frac{2(A-\pi i)^2}{\pi i k_2}} e^{-\frac{2(A-\pi i)^2}{\pi i k_2(k_1 k_2 -1)}}
\Big) \;.
\label{Simultaneous contour 1}
\end{align}
where $a_+ = \pi i e^{-\frac{2\pi i k_1}{k_1 k_2 -1}}$ is the residue of $\frac{A \cosh(A)}{\sinh(A)} e^{-\frac{2A^2}{\pi i k_2}} e^{-\frac{2A^2}{\pi i k_2(k_1 k_2 -1)}}$ at $A=\pi i$. Again, the last integral in \eqref{Simultaneous contour 1} can be evaluated from the below contour integral:
\begin{figure}[H]
    \centering
    \begin{tikzpicture}
[decoration={markings,
	mark=at position 0.13 with {\arrow[line width=1pt]{>}},
	mark=at position 0.33 with {\arrow[line width=1pt]{>}},
	mark=at position 0.53 with {\arrow[line width=1pt]{>}},
	mark=at position 0.8 with {\arrow[line width=1pt]{>}},
	mark=at position 0.93 with {\arrow[line width=1pt]{>}}
}
]
\draw[help lines,->] (-5,0) -- (5,0) coordinate (xaxis);
\draw[help lines,->] (0,-1.5) -- (0,1.5) coordinate (yaxis);

\path[draw,line width=0.8pt,postaction=decorate] 
(-4,1) -- (-0.3,1)  arc (-180:0:0.3) -- (4,1) -- (4,-1)  -- (0.3,-1) arc (0:180:0.3) -- (-4,-1) --(-4,1);

\node at (0,1) {$\bullet$};
\node at (0,0) {$\bullet$};
\node at (0,-1) {$\bullet$};

\node[below] at (xaxis) {Re$A$};
\node at (0,1.8) {Im$A$};
\node at (2,1.3) {$l'$};
\node at (0.3,1.3) {$\pi i$};
\node at (-0.5,-1.3) {$-\pi i$};
\node at (4.3,0.15) {$\infty$};
\node at (-4.4,0.15) {$-\infty$};
\end{tikzpicture}
    \caption{A path $l'$ for \eqref{Simultaneous Contour 2}. There are simple poles at $A=0,\pm \pi i$.}
    \label{Diagram for Simultaneous contour 2}
\end{figure}
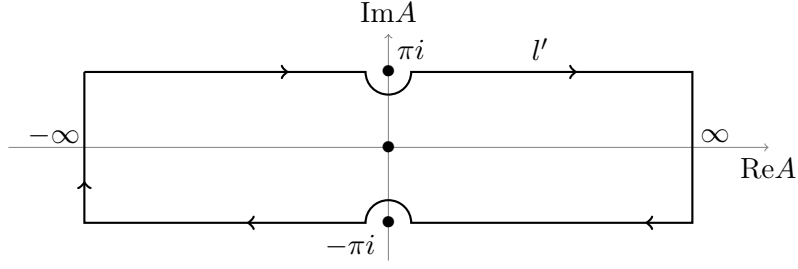
\begin{align}
&\oint_{l'} \text{d}A \,
\frac{\cosh(A)}{\sinh(A)} e^{-\frac{2A^2}{\pi i k_2}} e^{-\frac{2A^2}{\pi i k_2(k_1 k_2 -1)}} 
= -2\pi i \tilde{a}_0
\nonumber\\
&\qquad\rightarrow
\int \text{d}A \, \frac{\cosh(A)}{\sinh(A)}
\Big(
e^{-\frac{2(A+\pi i)^2}{\pi i k_2}} e^{-\frac{2(A+\pi i)^2}{\pi i k_2(k_1 k_2 -1)}}
-
e^{-\frac{2(A-\pi i)^2}{\pi i k_2}} e^{-\frac{2(A-\pi i)^2}{\pi i k_2(k_1 k_2 -1)}}
\Big)\nonumber
\\&\qquad=
-\pi i (\tilde{a}_- + 2\tilde{a}_0 + \tilde{a}_+)\;.
\label{Simultaneous Contour 2}
\end{align}
where $\tilde{a}_0=1$, $\tilde{a}_\pm = e^{-\frac{2\pi i k_1}{k_1 k_2 -1}}$ are the residues of $\frac{\cosh(A)}{\sinh(A)} e^{-\frac{2A^2}{\pi i k_2}} e^{-\frac{2A^2}{\pi i k_2(k_1 k_2 -1)}}$ at $A=0,\pm\pi i$ respectively. Plugging this into \eqref{Simultaneous contour 1}, we have
\begin{align}
\int \text{d}A\, \frac{A \cosh(A)}{\sinh(A)}
\Big(
e^{-\frac{2A^2}{\pi i k_2}} e^{-\frac{2A^2}{\pi i k_2(k_1 k_2 -1)}}
-
e^{-\frac{2(A+\pi i)^2}{\pi i k_2}} e^{-\frac{2(A+\pi i)^2}{\pi i k_2(k_1 k_2 -1)}}
\Big)
=
\pi^2 \;.
\label{simultaneous result1}
\end{align}
Likewise, the rest terms in \eqref{Simultaneous A integral} can also be evaluated in a similar way as
\begin{align}
\int \text{d}A\,
&\frac{A \cosh(A)}{\sinh(A)}
\Bigg[
e^{-\frac{2(A+\pi i)^2}{\pi i k_2}}
\Big(
e^{-\frac{2(A+\pi i + k_2 \pi i)^2}{\pi i k_2 (k_1 k_2 -1)}}
+
e^{-\frac{2(A+\pi i - k_2 \pi i)^2}{\pi i k_2 (k_1 k_2 -1)}}
\Big)
\nonumber\\
&\qquad
-e^{-\frac{2 A^2}{\pi i k_2}}
\Big(
e^{-\frac{2(A + k_2 \pi i)^2}{\pi i k_2 (k_1 k_2 -1)}}
+
e^{-\frac{2(A-k_2 \pi i)^2}{\pi i k_2 (k_1 k_2 -1)}}
\Big)
\Bigg]
=
-2\pi^2 e^{-\frac{2\pi i k_2}{k_1 k_2 -1}}\,.
\label{simultanetous result2}
\end{align}
Combining the two results \eqref{simultaneous result1} and \eqref{simultanetous result2}, and restoring the overall factor in \eqref{Simultaneous A integral}, we have
\begin{align}
\mathcal{Z}^{S_b^3}_{(k_1,k_2)}(b=1,m=0,\nu=1)
&=
\frac{e^{\frac{2\pi i}{3}}}{4\pi^2}
\frac{1}{\sqrt{k_1 k_2 -1}}
\Big(
2\pi^2
-
2\pi^2 e^{-\frac{2\pi i k_2}{k_1 k_2 -1}}
\Big)
\nonumber\\
&=
e^{\frac{7\pi i}{6}-\frac{k_2\pi i}{k_1 k_2 -1}}
\frac{1}{\sqrt{k_1 k_2 -1}} \sin\Big(\frac{k_2 \pi}{k_1 k_2 -1}\Big)\;.
\end{align}
For $\nu=-1$, thanks to \eqref{mirror property}, the only difference is nothing but an exchange of the role of $k_1$ and $k_2$:
\begin{align}
\mathcal{Z}^{S_b^3}_{(k_1,k_2)}(b=1,m=0,\nu=-1) 
=
e^{\frac{7\pi i}{6}-\frac{k_1\pi i}{k_1 k_2 -1}}
\frac{1}{\sqrt{k_1 k_2 -1}} \sin\Big(\frac{k_1 \pi}{k_1 k_2 -1}\Big)\;.
\end{align}
%
\subsection{\texorpdfstring{$\mathcal{Z}^{S_b^3}_{\text{diag}_k}(b=1,m=0,\nu=\pm 1)$}{Z(S3b)(diag k)(b=1,m=0,nu=+- 1)}}
\label{App : Diagonal nu=pm1}
By the mirror-symmetry property \eqref{mirror property} it is enough to consider the case of $\nu=1$. With the help of Appendix \ref{App : QDL}, \eqref{diverging adjoint expansion}, and \eqref{vanishing int}, the partition function \eqref{Diagona gauging partition function} for $b=1$, $\nu=1$ is simplified as
\begin{align}
\mathcal{Z}^{S_b^3}_{\text{diag}_k}(b=1,m=0,\nu=1)
=
-\frac{e^{\frac{5\pi i}{12}}}{4\pi^3}
\int \text{d}Z\text{d}X \sinh^2(X) e^{\frac{i Z X}{\pi}}e^{\frac{k X^2}{2\pi i}}
\Bigg(
\frac{X \sinh(X) - Z \sinh(Z)}{\cosh(X)-\cosh(Z)}
\Bigg)\;.
\end{align}
Changing the variables as $X\rightarrow A+B$, $Z\rightarrow A-B$, we have
\begin{align}
\mathcal{Z}^{S_b^3}_{\text{diag}_k}(b=1,m=0,\nu=1)
=
-\frac{e^{\frac{5\pi i}{12}}}{2\pi^3}
&\int \text{d}A\text{d}B \sinh^2(A+B) e^{\frac{k A B}{\pi i}}
\nonumber\\
&\times
e^{\frac{(k-2)A^2}{2\pi i}}e^{\frac{(k+2)B^2}{2\pi i}}
\Bigg(
\frac{A \cosh(A)}{\sinh(A)}
+
\frac{B \cosh(B)}{\sinh(B)}
\Bigg)\;.
\label{Diagonal decoupled integrals}
\end{align}
We first consider the first term in the integrand which is an Gaussian integral of $B$.
\begin{align}
&\int \text{d}A\text{d}B \, \sinh^2(A+B) e^{\frac{k A B}{\pi i}}
e^{\frac{(k-2)A^2}{2\pi i}}e^{\frac{(k+2)B^2}{2\pi i}}
\frac{A \cosh(A)}{\sinh(A)}
\nonumber\\
&\rightarrow
\frac{\pi}{\sqrt{8 i (k+2)}}
\int\text{d}A\,
\frac{A \cosh(A)}{\sinh(A)}
\Big(
e^{-\frac{2(A-\pi i)^2}{\pi i(k+2)}} + e^{-\frac{2(A+\pi i)^2}{\pi i(k+2)}} - 2 e^{-\frac{2A^2}{\pi i(k+2)}}
\Big)\;.
\end{align}
Now, consider contour integrals along the paths shown below:
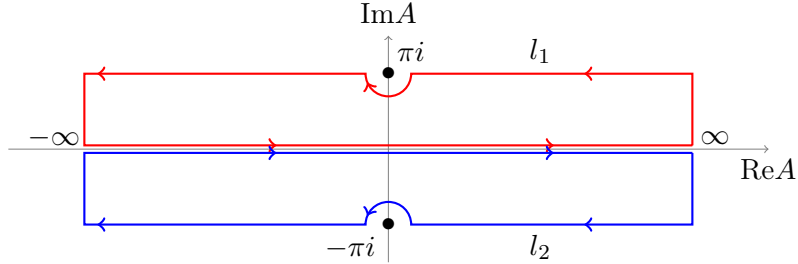
\begin{figure}[H]
    \centering
    \begin{tikzpicture}
[decoration={markings,
	mark=at position 0.13 with {\arrow[line width=1pt]{>}},
	mark=at position 0.3 with {\arrow[line width=1pt]{>}},
	mark=at position 0.5 with {\arrow[line width=1pt]{>}},
	mark=at position 0.7 with {\arrow[line width=1pt]{>}},
	mark=at position 0.9 with {\arrow[line width=1pt]{>}}
}
]
\draw[help lines,->] (-5,0) -- (5,0) coordinate (xaxis);
\draw[help lines,->] (0,-1.5) -- (0,1.5) coordinate (yaxis);

\path[draw,line width=0.8pt,red,postaction=decorate] 
(4,0.05) -- (4,1) -- (0.3,1)  arc (0:-180:0.3) -- (-4,1) -- (-4,0.05)  -- (4,0.05);

\path[draw,line width=0.8pt,blue,postaction=decorate] 
(4,-0.05) -- (4,-1) -- (0.3,-1)  arc (0:180:0.3) -- (-4,-1) -- (-4,-0.05)  -- (4,-0.05);

\node at (0,1) {$\bullet$};
\node at (0,-1) {$\bullet$};

\node[below] at (xaxis) {Re$A$};
\node at (0,1.8) {Im$A$};
\node at (2,1.3) {$l_1$};
\node at (2,-1.3) {$l_2$};
\node at (0.3,1.3) {$\pi i$};
\node at (-0.5,-1.3) {$-\pi i$};
\node at (4.3,0.15) {$\infty$};
\node at (-4.4,0.15) {$-\infty$};
\end{tikzpicture}
    \caption{Two paths $l_1$ and $l_2$ for \eqref{Diagonal contour 1}. There are simple poles at $A = \pm \pi i$.}
    \label{Diagram for Diagonal contour 1}
\end{figure}
\begin{align}
&\oint_{l_1} \text{d}A \,\frac{A\cosh(A)}{\sinh(A)}e^{-\frac{2 A^2}{\pi i(k+2)}}
+
\oint_{l_2} \text{d}A \,\frac{A\cosh(A)}{\sinh(A)}e^{-\frac{2 A^2}{\pi i(k+2)}}
= 0
\nonumber\\
&\rightarrow
\int\text{d}A\,
\frac{A \cosh(A)}{\sinh(A)}
\Big(
e^{-\frac{2(A-\pi i)^2}{\pi i(k+2)}} + e^{-\frac{2(A+\pi i)^2}{\pi i(k+2)}} - 2 e^{-\frac{2A^2}{\pi i(k+2)}}
\Big)
\nonumber\\
&\qquad\qquad=\pi i (u_- - u_+) - \pi i \int \text{d}A\, \frac{\cosh(A)}{\sinh(A)} 
\Big(
e^{-\frac{2(A+\pi i)^2}{\pi i(k+2)}}
-
e^{-\frac{2(A-\pi i)^2}{\pi i(k+2)}}
\Big) \;,
\label{Diagonal contour 1}
\end{align}
where $u_{\pm} = \pm \pi i e^{-\frac{2\pi i}{k+2}}$ are the residues of $\frac{A\cosh(A)}{\sinh(A)}e^{-\frac{2A^2}{\pi i (k+2)}}$ at $A=\pm \pi i$. Again, the integral at the last term of \eqref{Diagonal contour 1} can be evaluated by considering the following path:
\begin{figure}[H]
    \centering
    \begin{tikzpicture}
[decoration={markings,
	mark=at position 0.13 with {\arrow[line width=1pt]{>}},
	mark=at position 0.33 with {\arrow[line width=1pt]{>}},
	mark=at position 0.53 with {\arrow[line width=1pt]{>}},
	mark=at position 0.8 with {\arrow[line width=1pt]{>}},
	mark=at position 0.93 with {\arrow[line width=1pt]{>}}
}
]
\draw[help lines,->] (-5,0) -- (5,0) coordinate (xaxis);
\draw[help lines,->] (0,-1.5) -- (0,1.5) coordinate (yaxis);

\path[draw,line width=0.8pt,postaction=decorate] 
(-4,1) -- (-0.3,1)  arc (-180:0:0.3) -- (4,1) -- (4,-1)  -- (0.3,-1) arc (0:180:0.3) -- (-4,-1) --(-4,1);

\node at (0,1) {$\bullet$};
\node at (0,0) {$\bullet$};
\node at (0,-1) {$\bullet$};

\node[below] at (xaxis) {Re$A$};
\node at (0,1.8) {Im$A$};
\node at (2,1.3) {$l_3$};
\node at (0.3,1.3) {$\pi i$};
\node at (-0.5,-1.3) {$-\pi i$};
\node at (4.3,0.15) {$\infty$};
\node at (-4.4,0.15) {$-\infty$};
\end{tikzpicture}
    \caption{A path $l_3$ for \eqref{Diagonal Contour 2}. There are simple poles at $A=0,\pm \pi i$.}
    \label{Diagram for Diagonal contour 2}
\end{figure}
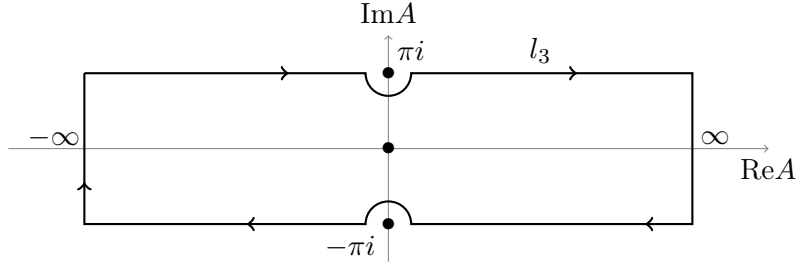
\begin{align}
&\oint_{l_3}\text{d}A\, \frac{\cosh(A)}{\sinh(A)} e^{-\frac{2A^2}{\pi i(k+2)}} = - 2\pi i v_0
\nonumber\\
&\rightarrow
\int \text{d}A\, \frac{\cosh(A)}{\sinh(A)} 
\Big(
e^{-\frac{2(A+\pi i)^2}{\pi i(k+2)}}
-
e^{-\frac{2(A-\pi i)^2}{\pi i(k+2)}}
\Big)
=
-\pi i (v_- + 2 v_0 + v_+)
=
-2\pi i(1 + e^{-\frac{2\pi i}{k+2}})\;.
\label{Diagonal Contour 2}
\end{align}
where $v_0 = 1$, $v_{\pm} = e^{-\frac{2\pi i}{k+2}}$ are the residues of $\frac{\cosh(A)}{\sinh(A)}e^{-\frac{2A^2}{\pi i(k+2)}}$ at $A=0,\pm \pi i$ respectively. Combining the results \eqref{Diagonal contour 1} and \eqref{Diagonal Contour 2}, we have
\begin{align}
\int \text{d}A\text{d}B \, \sinh^2(A+B) e^{\frac{k A B}{\pi i}}
e^{\frac{(k-2)A^2}{2\pi i}}e^{\frac{(k+2)B^2}{2\pi i}}
\frac{A \cosh(A)}{\sinh(A)}
=
-\frac{\pi^3}{\sqrt{2i(k+2)}}\;.
\label{diag integral result1}
\end{align}
Similarly, the second term in the integrand of \eqref{Diagonal decoupled integrals} can be evaluated as
\begin{align}
\int \text{d}A\text{d}B \sinh^2(A+B) e^{\frac{k A B}{\pi i}}
e^{\frac{(k-2)A^2}{2\pi i}}e^{\frac{(k+2)B^2}{2\pi i}}
\frac{B \cosh(B)}{\sinh(B)}
=
-\frac{\pi^3}{\sqrt{2i(k-2)}}\;.
\label{diag integral result2}
\end{align}
Finally, with the results \eqref{diag integral result1} and \eqref{diag integral result2}, and restoring the overall factor in \eqref{Diagonal decoupled integrals}, we have
\begin{align}
\mathcal{Z}^{S_b^3}_{\text{diag}_k}(b=1,m=0,\nu=1)
=
e^{\frac{\pi i}{6}}
\Bigg(
\sqrt{\frac{1}{8(k-2)}}
+
\sqrt{\frac{1}{8(k+2)}}
\Bigg)\;.
\end{align}

\section{Quantum dilogarithm function}  \label{App : QDL}
 
The quantum dilogarithm function (Q.D.L) $\psi_{\hbar} (Z) $ is defined by \cite{Faddeev:1993rs} $(\hbar = 2\pi i b^2)$ 
\begin{align}
\begin{split}
\psi_{\hbar}(Z) := 
\begin{cases}
\prod_{r=1}^{\infty} \frac{1 - q^r e^{-Z}}{1 - \tilde{q}^{-r+1} e^{-\tilde{Z} } }  \quad \text{if} \quad |q| < 1\;,
\\
\prod_{r=1}^{\infty} \frac{1 - \tilde{q}^r e^{-\tilde{Z}}}{1 - q^{-r+1} e^{-Z } }  \quad \text{if} \quad |q| > 1\;,
\end{cases}
\end{split}
\end{align}
with
\begin{align}
\begin{split}
q := e^{2\pi i b^2}
\,,\qquad
\tilde{q} := e^{2\pi i b^{-2}}
\,,\qquad
\tilde{Z} := \frac{1}{b^2}Z\,,
\end{split}
\end{align}
where $b$ is the squashing parameter. 
The function satisfies the following difference equations:
\begin{align}
\begin{split}
\psi_{\hbar}(Z+2\pi i b^2) = (1-e^{-Z}) \psi_{\hbar}(Z)
\,,\quad
\psi_{\hbar}(Z+2\pi i ) = (1-e^{-\frac{Z}{b^{2}}}) \psi_{\hbar}(Z)\;.
\end{split}
\end{align}
At the special value $b=1$, the Q.D.L simplifies as
\begin{align}
\begin{split}
\log \psi_{\hbar=2\pi i}(Z) =
\frac{-(2\pi + i Z)\log(1-e^{-Z})+i \text{Li}_2(e^{-Z})}{2\pi}\,,
\end{split}
\end{align}
and there is a special limit at $b=1$ 
\begin{align}
\begin{split}
\lim_{p\rightarrow 0} p \, \psi_{\hbar=2\pi i}(p) = e^{\frac{\pi i}{12}}\;.
\end{split}
\end{align}
On the other hand, the asymptotic expansion when $\hbar = 2\pi i b^2 \rightarrow 0$ is given by
\begin{align}
\begin{split}
\log \psi_{\hbar}(Z) \xrightarrow{b^{2}\rightarrow 0}
\sum_{n=0}^{\infty} \frac{B_n \hbar^{n-1}}{n !} \text{Li}_{2-n}(e^{-Z})\;.
\end{split}
\end{align}
Here $B_n$ is the $n$-th Bernoulli number with $B_1=1/2$.
For several computations in the main text, one needs to utilize the identity
\begin{align}
\begin{split}
\text{Li}_2 (u) + \text{Li}_2 (u^{-1}) = -\frac{\pi^2}{6} -\frac{1}{2}\big(\log(-u)\big)^2\;.
\end{split}
\end{align}

\newpage
\bibliographystyle{ytphys}
\bibliography{ref}

\end{document}